\documentclass[3p]{elsarticle}

\usepackage{amsmath}
\usepackage{amsthm}
\usepackage{amssymb}
\usepackage{graphicx}
\usepackage{esint}
\usepackage{color}

\PassOptionsToPackage{hyphens}{url}
\usepackage[pdfencoding=auto,psdextra,colorlinks=true]{hyperref}
\usepackage{bookmark}

\journal{Progress in Biophysics \& Molecular Biology}

\biboptions{authoryear}

\begin{document}

\begin{frontmatter}

\title{Quantum information theoretic approach to the mind--brain problem}

\author[address1]{Danko D. Georgiev\corref{mycorrespondingauthor}}
\ead{danko.georgiev@mail.bg}
\cortext[mycorrespondingauthor]{Corresponding author}

\address[address1]{Institute for Advanced Study, 30 Vasilaki Papadopulu Str., Varna 9010, Bulgaria}

\begin{abstract}
The brain is composed of electrically excitable neuronal networks regulated by the activity of voltage-gated ion channels. Further portraying the molecular composition of the brain, however, will not reveal anything remotely reminiscent of a feeling, a sensation or a conscious experience. In classical physics, addressing the mind--brain problem is a formidable task because no physical mechanism is able to explain how the brain generates the unobservable, inner psychological world of conscious experiences and how in turn those conscious experiences steer the underlying brain processes toward desired behavior. Yet, this setback does not establish that consciousness is non-physical. Modern quantum physics affirms the interplay between two types of physical entities in Hilbert space: unobservable quantum states, which are vectors describing what exists in the physical world, and quantum observables, which are operators describing what can be observed in quantum measurements. Quantum no-go theorems further provide a framework for studying quantum brain dynamics, which has to be governed by a physically admissible Hamiltonian. Comprising consciousness of unobservable quantum information integrated in quantum brain states explains the origin of the inner privacy of conscious experiences and revisits the dynamic timescale of conscious processes to picosecond conformational transitions of neural biomolecules. The observable brain is then an objective construction created from classical bits of information, which are bound by Holevo's theorem, and obtained through the measurement of quantum brain observables. Thus, quantum information theory clarifies the distinction between the unobservable mind and the observable brain, and supports a solid physical foundation for consciousness research.
\end{abstract}


\date{August 2, 2020}

\end{frontmatter}



\section*{Highlights}

\begin{itemize}
\item Psychological inner world remains private and unobservable from a third-person perspective.
\item Physiological brain activity due to electric excitations of neuronal networks is observable.
\item Quantum information theory makes a distinction between physical states and observables.
\item Unobservable quantum information built in quantum brain states comprises consciousness.
\item The observable brain is constructed from bits of information constrained by Holevo's theorem.
\end{itemize}

\pagebreak

\section{Introduction}

The essence of consciousness is \emph{experience} \citep{Nagel1974,Nagel1987,Nagel2012}.
Through conscious experiences such as the perceived colors of the
rainbow, the pleasant sound of a musical instrument, the fresh smell
of the sea breeze, or the wet touch of the water, we access the surrounding
physical world and become aware of our own bodies \citep{Georgiev2017}.
The introspective access to our conscious experiences is privately
reserved only for us from a subjective, phenomenal, first-person perspective,
and it is denied to others who happen to observe us from an objective,
third-person perspective \citep{Nagel1974,Nagel1987,Nagel2012}. It
is an empirical fact that the very process of observation of someone
else's brain does not elicit in us experiences that are identical
with those experienced by the observed brain \citep{Georgiev2020}.
Consequently, we do not have at our disposal an objective method to
determine whether any other living or non-living physical system is
conscious or not. The unobservability of conscious experiences does
not prevents us from being able to specify the particular subject
whose experiences we are talking about, or to characterize the physical
circumstances under which certain conscious experiences are elicited
\citep{Georgiev2017,Georgiev2020}; for example, dolphins' double sonar
experience of reflected ultrasound waves used for hunting prey or
orientation in their natural habitat \citep{Starkhammar2011,Branstetter2012,Jensen2013,Ridgway2015,Ladegaard2019}.
But this is the most we can do. We are unable to describe in words and
communicate to others what it is like to have those experiences. Thus,
conscious experiences are fundamentally \emph{unobservable} and their
phenomenal qualia are \emph{incommunicable} \citep{Georgiev2020}.
Since we have direct access to our own conscious experiences, we know
that there is at least one conscious entity in the physical universe.
From our shared evolutionary ancestry with other humans or animal
species \citep{Darwin2006,Dawkins2004,Stringer2017,Hublin2017,Chan2019}, we also have solid scientific grounds
to maintain that we are not the only conscious entity in existence.
Therefore, the primary aim of a physical theory of consciousness is
to provide criteria that will allow unambiguous specification of which
physical systems are conscious and which are not. Once the conscious
mind is physically identified, the physical laws will regulate how
the mind affects the world \citep{Georgiev2017}. It should be noted
that \emph{consciousness}, \emph{conscious experience}, \emph{conscious
state}, \emph{mental state} and \emph{mind} are used interchangeably
throughout this work. A \emph{mental process} (\emph{conscious process})
is a process that involves a sequence of mental states (i.e. dynamically changing
conscious experiences).

\begin{figure}[t]
\begin{centering}
\includegraphics[width=162mm]{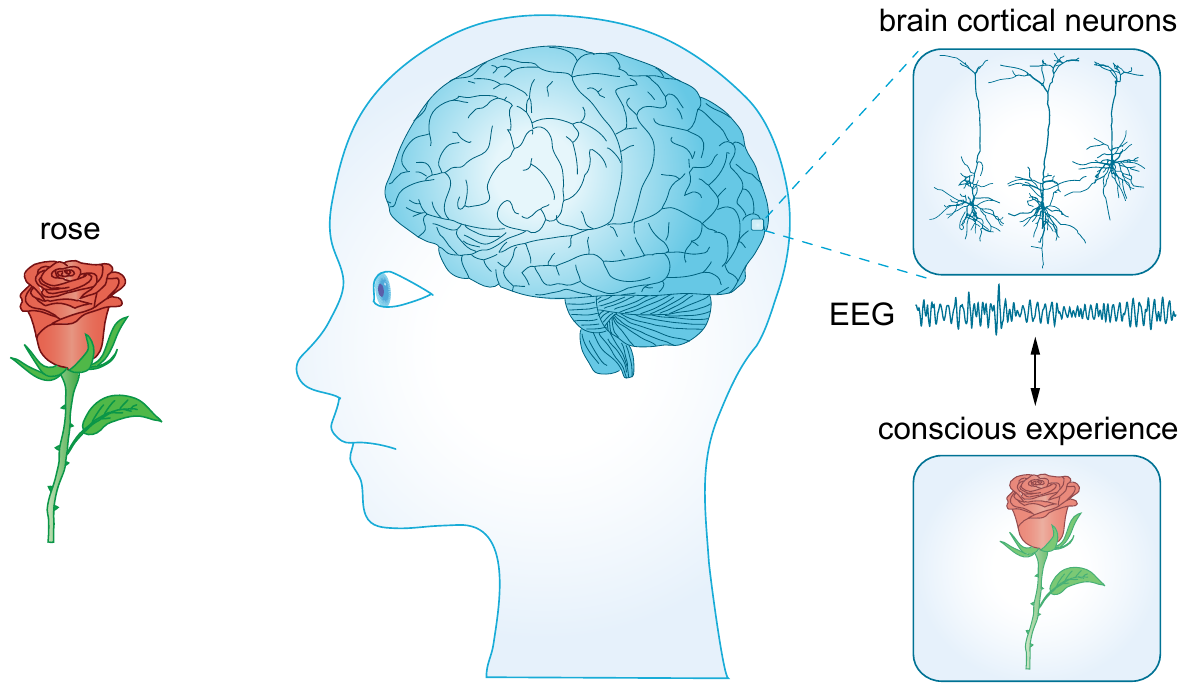}
\par\end{centering}
\caption{\label{fig:1}The mind--brain problem. Neither the brain cortex whose
anatomy can be observed during an open skull surgery, nor the cortical
electric activity recorded by electroencephalography (EEG) resemble
the visual conscious experience elicited by observation of a red rose.
Explaining the physical relationship between the observable brain
and the unobservable mind has troubled philosophers for centuries. Modified from \cite{Georgiev2020}.}
\end{figure}

The seat of the human mind is the \emph{brain cortex}. Cortical electric activity is mainly due to excitation of principal pyramidal neurons, which comprise over 70\% of all cortical neurons (Fig.~\ref{fig:1}). Pyramidal neurons were designated as the `psychic cells' of the brain
by the father of modern neuroscience Ram\'{o}n y Cajal since their electric activities instantiate feelings \citep{Goldman-Rakic2002}.
Substantial medical evidence supports a cohesive
relationship between the brain cortical electric excitation and the conscious
mind because direct electric stimulation of the
brain cortex elicits sensations \citep{Bosking2017a,Hiremath2017,Yoshor2007},
whereas discrete cortical lesions impair cognitive abilities or change
the way one experiences the world \citep{Chen2017,Hadid2017,Sajja2017,Lau2018}.
For example, direct electric stimulation of the visual cortex through
implanted electrodes that deliver digitized signals captured by a
camera is capable of restoring the vision in blind patients whose
eyes were injured by trauma \citep{Dobelle2000,Bosking2017b,Lewis2016},
while various injuries to the occipital lobe of the cortex produce
blindness \citep{Hadid2017,Chen2017,Sajja2017}. Apparently, the mind
and the brain are not identical, because anesthetized brains do not
generate conscious experiences. In the course of general anesthesia
consciousness is erased by the pharmacological action of the anesthetic
drug, yet the experimenter may stimulate with visible light the open
eyes of anesthetized animals and still evoke electric potentials by
pyramidal neurons located in the primary visual cortex \citep{Lamme1998,Imas2005,Sellers2015,Hudetz2007}.
Similar experiments in anesthetized human subjects showed evoked electroencephalographic
\citep{McNeer2009} or electrocorticographic \citep{Nourski2017} responses
under auditory stimulation. If mind states were related to brain states
through one-to-one correspondence (logical identity relation), it
should not have been possible to turn mental states on or off using
general anesthetics, because the brain states would have always remained
mental states. Thus, the mind--brain problem is to explain how the
unobservable conscious mind and the observable brain relate to each
other: do they interact or does one unilaterally generate the other?

The unobservability and incommunicability of conscious experiences
has been marshaled as evidence for the \emph{nonphysical} nature of
consciousness \citep{Robinson1976,Jackson1982,Jackson1986,Sprigge1994,Chalmers1995,Chalmers1996,Zhao2012}
and the alleged inadequacy of physics to answer questions related
to our mentality and sentience \citep{Nagel1965,Kim1998,Campbell2011}.
Such a view is often grounded in the principles of \emph{classical
physics} according to which everything inside the physical world is
observable, governed by deterministic physical laws, and causally
closed in regard to its time dynamics with respect to non-physical entities \citep{Susskind2013}.

Classical reductionism fails because reductive identification of unobservable
consciousness with observable physical properties is logically inconsistent.
According to the postulates of classical physics (including classical mechanics, electromagnetism, and Einstein's theory of relativity) all existing things are physical and all physical entities are observable.
In other words, by logical contraposition, it follows that if an entity is not observable, then it is not physical and does not exist. In the precise mathematical language of set theory, non-existing entities such as unicorns, centaurs or fairies, are all members of the empty set. Therefore, identifying the unobservable consciousness with any non-observable non-existing entity would be logically equivalent to classical eliminativism \citep{Dennett1991} according to which consciousness does not exist and is a member of the empty set. Here, our goal is not to ban consciousness from existence, but to incorporate it into the physical description of the world.

The obstacles faced by classical reductionism compel many philosophers and neuroscientists to reject the postulate that all existing things are physical thereby assuming that the brain generates existing, but non-physical, non-observable conscious experiences, and consciousness is a functional product emerging out of the underlying brain activity \citep{Fodor1981,Baars2005,Piccinini2010,Wenzel2019}.
Classical functionalism, however, also fails due to the closure of the physical world to non-physical entities.
The conscious mind, if viewed as a functional product or an emergent property of the brain,
can only be admitted as an \emph{epiphenomenon} without any causal influence
on physical events \citep{Jackson1982}, hence providing no basis for
natural selection and no evolutionary advantage to conscious organisms
in the fight for survival \citep{Georgiev2013,Georgiev2019a,Popper1983}.
The adequate match between our conscious experiences, the neural responses,
and the corresponding behavior, provides compelling evidence against
epiphenomenalism. For example, the neural responses to detrimental
factors are always associated with unpleasant feelings and avoiding
behavior. If conscious experiences are causally effective, the evolution
theory is able to explain the adequate matching between the unpleasant
experiences, the avoiding behavior and the negative influence of detrimental
factors upon the organism. Clearly, those animals that would have
enjoyed detrimental factors would not have avoided dangers, thereby
dying out in the competition with rival organisms. However, if conscious
experiences were causally ineffective, the evolutionary explanation
would fail since one would expect that in nature still there will be organisms
that experience pleasant feelings when being injured but
avoid detrimental factors due to the organization of their neural
processes \citep{James1879}. Introspectively, we could verify that
we never enjoy detrimental factors, hence epiphenomenalism and classical
functionalism have to be false.

To avoid the charge of epiphenomenalism, consciousness has to be physical, thereby entering directly
into the mathematical equations of the fundamental physical laws that
describe the dynamics of physical systems endowed with conscious experiences.
Certainly, this requires a fundamental revision of the principles
of classical physics and incorporation of consciousness into a modern, non-classical physical theory.
Fortunately, in 1920s the failure of classical
physics to describe faithfully the physical world was well-established:
it was unable to explain the stability of chemical atoms, the photoelectric
effect, the electron diffraction in crystals and the spectral curve
of blackbody radiation \citep{Fayngold2013}. The concerted efforts
of quantum physicists have replaced the inadequate classical physics
with a radically new, empirically successful, quantum theory based not only
on different physical equations, but also on conceptually distinct
quantum principles \citep{Susskind2014}. Among the newly introduced
concepts is the quantum indeterminism, which endows elementary particles
with the \emph{propensity to make choices} among different future
possibilities available for actualization, and a dichotomy between
what physically exists, described by unobservable quantum state vectors,
and what can be physically observed, described by observable quantum
operators. This dichotomy is crucial for addressing the
mind--brain problem. Identifying the unobservable consciousness with
the quantum information integrated in unobservable quantum brain state
vectors makes consciousness causally effective in determining the
probabilities for producing different quantum outcomes upon measurement.
The observable brain then is nothing but the classical record of observed
outcomes of brain quantum observables upon measurement with brain
imaging devices \citep{Georgiev2017}.

In this present work, after briefly reviewing some preliminary background on the mind--brain problem to make the exposition self-contained, we will focus on the quantum information theoretic differences between the unobservable mind and the observable brain, and will elaborate on the Holevo bound, including its explicit mathematical formulation and the physical conditions that maximize its value to $n$ bits of classical information for a system composed of $n$ qubits.

\section{Classical information-theoretic approach to the mind--brain problem}

Even though quantum physics superseded classical physics a century
ago, current neuroscience is still based solely on classical principles.
This conservative approach denies any essential role of quantum effects
in regard to consciousness and assumes that the brain processes related
to the input, processing, storage and output of classical information
are sufficient to explain consciousness. Limiting quantum theory to
a narrow domain where quantum physical systems exhibit classical behavior,
however, leads again to classical functionalism, epiphenomenalism,
and the infamous hard problem of consciousness \citep{Chalmers1995}.

The characteristic features of classical behavior are the observability
and communicability of classical information, and deterministic time
evolution of physical states \citep{Susskind2013}. Classical information
encoded onto a physical carrier can be \emph{read} and \emph{copied}
onto a new carrier. If in the process of copying the old copy is preserved
intact, classical information can be \emph{multiplied}. Changing the
nature of the physical carriers (e.g. from massive electrons to massless
photons) allows classical information to be \emph{broadcast} to a
distant receiver where it is \emph{recorded} and \emph{stored}. The
obtained classical information can be further \emph{processed} using
irreversible logic gates and/or \emph{erased}. An illustrative example
of classical information is the digital string of bits, 0s and 1s,
which encodes a text file on a computer hard disk drive. One can display
the text on a monitor and read it, copy the information contained
in the digital file multiple times, or even erase the file in order
to free hard disk memory space \citep{Georgiev2013}. Thus, the physical
properties of classical information are ideal for memory storage and
retrieval, namely, once memory traces are formed in the brain, they
can be read again and recalled at a later time. Observable classical
information, however, cannot lead to unobservable conscious experiences
without assuming some form of functional emergence.

\begin{figure}[t]
\begin{centering}
\includegraphics[width=155mm]{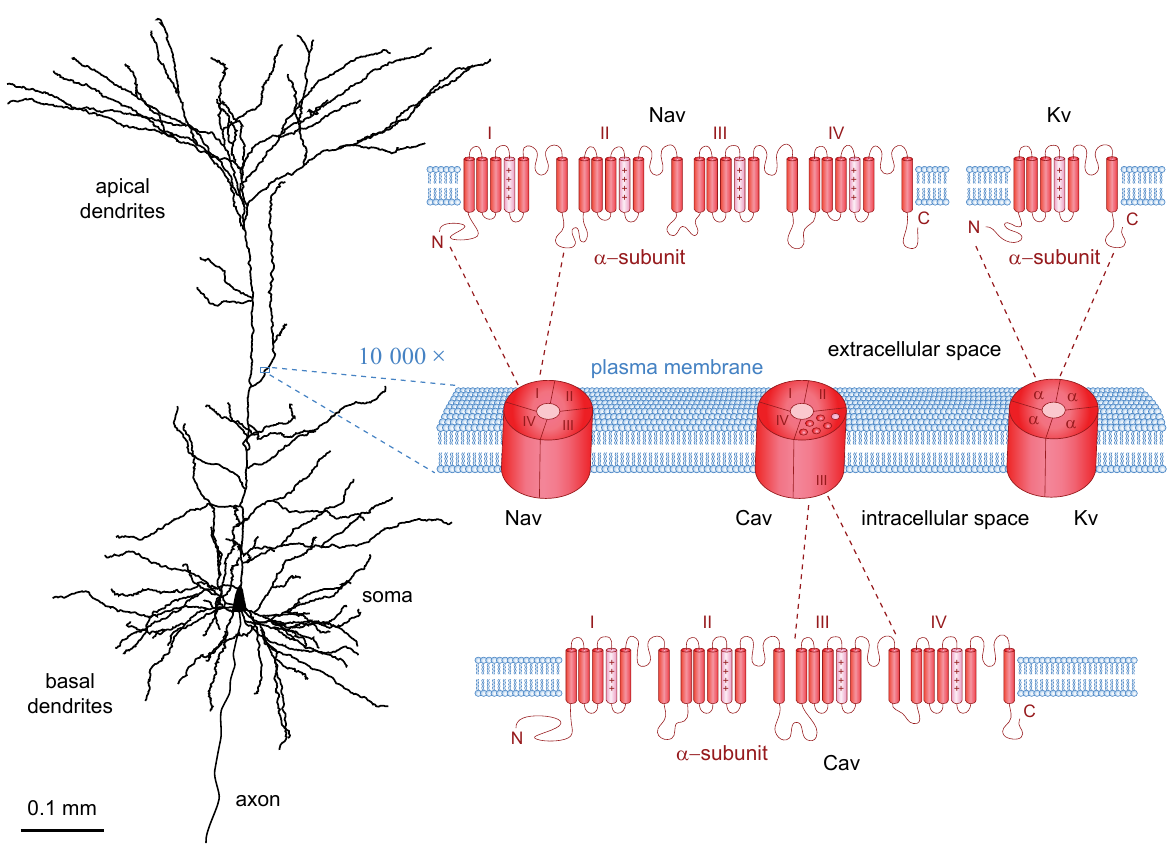}
\par\end{centering}
\caption{\label{fig:2}Morphology of a cortical pyramidal neuron (NeuroMorpho.org NMO\_09565)
with structural representation of voltage-gated ion channels in a
patch of the electrically excitable plasma membrane. The pyramidal
neuron receives excitatory or inhibitory synaptic inputs applied at
the apical and basal dendrites or the soma. If the summated synaptic
electric currents depolarize the axon initial segment over the threshold
value of $-55$ mV, the neuron discharges an action potential, which
propagates down the axon and initiates the release of neurotransmitter
from terminal axonal boutons onto target neurons. The electric activity
of the neuron is mainly driven by ionic fluxes through sodium (Nav),
potassium (Kv) and calcium (Cav) voltage-gated ion channels. Individual
Nav and Cav channels are composed of $\alpha$-subunits with four
protein domains (I--IV) each, whereas Kv channels have disjoint protein
domains into separate $\alpha$-subunits. Each transmembrane channel
domain is formed by six $\alpha$ helices bundled in parallel. The
electrically charged voltage sensor that gates the ion channel is
located within the 4th $\alpha$-helix of each domain. Modified from \cite{Georgiev2014}.}
\end{figure}

Functionally, neurons encode and transmit classical information in
terms of electric spikes. Neuronal electric activity is due to ionic
fluxes through excitatory or inhibitory ion channels incorporated
in the excitable plasma membrane. Instrumental for most neurophysiological
processes are sodium (Nav), potassium (Kv) and calcium (Cav) voltage-gated
ion channels, selectively conducting Na\textsuperscript{+}, K\textsuperscript{+}
or Ca\textsuperscript{2+} ions down the respective ion concentration
gradients \citep{Georgiev2014}. The voltage sensing is performed by
an electrically charged 4th $\alpha$-helix inside each domain of
the $\alpha$-subunit of ion channels (Fig.~\ref{fig:2}). Macroscopic
electric currents flow through a rich repertoire of neuronal voltage-gated
ion channels, whose opening is regulated by the local voltage across
the plasma membrane \citep{Georgiev2015}. As a result, the transmembrane
voltage of neurons undergoes dynamical changes in time. Pyramidal
neurons stay at rest if the transmembrane voltage in the soma and the axon
initial segment does not exceed a threshold value of about $-55$
mV \citep{Gasparini2004}. When the voltage threshold is reached, the
neuron fires a brief electric spike that propagates down the axon
in order to activate synapses innervating other target neurons.
Glial cells, including astrocytes and oligodendrocytes, maintain homeostasis of electrolytes and other biologically active substances in the brain, thereby nourishing and nurturing the easily vulnerable neurons \citep{Verkhratsky2017}.
With the use of electric spikes propagating within the neural network,
the brain is able to perform a variety of computational tasks. Yet,
the hard problem of consciousness is to explain why neuronal computation
in the brain generates any conscious experiences at all \citep{Chalmers1995}.

The hard problem of consciousness is a hallmark of functional theories
of consciousness, in which conscious experiences are assumed to be
generated by the brain in the process of performing a certain kind
of classical function. Once the function is precisely specified, e.g.
neuronal computation, it becomes impossible to explain why it is the
case that the brain does not operate in a mindless, nonconscious mode
where the neurons perform the specified function without any generation
of conscious experiences. Thus, the hard problem is an excellent test
for epiphenomenal consciousness, namely, if the dynamics can be fully
specified in advance without any reference to consciousness, then
the generated conscious experiences have to be causally ineffective \citep{Georgiev2017}.
Noteworthy, the hard problem does not occur for reductive theories
of consciousness, in which conscious experiences are identified with
physical states. Indeed, if the logical identity relation makes the
mind equivalent to some physical state $\Psi$, it will be inconsistent
to define alternative physical worlds in which the state
$\Psi$ is not a mind.

\section{Quantum information-theoretic approach to the mind--brain problem}

Quantum information is a novel kind of information that is held in
the quantum states of quantum physical systems. Quantum information
cannot be converted to classical information, which means that it is not contained in the mathematical description of a quantum physical state $\Psi$, but it is held in the physically existing
substrate signified by $\Psi$.
In other words, exactly as the \emph{map} is not the \emph{territory}, the quantum physical state $\Psi$ of, say, an electron written as a mathematical symbol on a sheet of paper is not the same as the quantum state of the electron in the quantum physical reality. This needs to be properly understood if one is to overcome the possible discomfort resulting from staring at quantum mechanical expressions for~$\Psi$ while deriving mathematically, for example, that $\Psi$ is not observable \citep{Georgiev2017}.

Quantum information differs from classical
information in a number of striking ways, which we will briefly review using Dirac's bra-ket notation \citep{Dirac1967}. The two main quantum physical laws are given by the \emph{Schr\"{o}dinger
equation} \citep{Hayashi2015}, which governs what physically exists and how it changes in time, and the
\emph{Born rule} \citep{Busch2016}, which governs what can be observed or measured with physical devices.

\emph{Schr\"{o}dinger equation}.
The fabric of physical reality is woven from quantum probability amplitudes,
which define the physical state of closed quantum physical systems
obeying the Schr\"{o}dinger equation
\begin{equation}
\imath\hbar\frac{\partial}{\partial t}|\Psi(\mathbf{r},t)\rangle=\hat{H}\,|\Psi(\mathbf{r},t)\rangle
\end{equation}
where $\imath$ is the imaginary unit, 
$\hbar$ is the reduced Planck constant,
$\frac{\partial}{\partial t}$ is the partial time derivative
operator, $|\Psi(\mathbf{r},t)\rangle$ is the
quantum wavefunction, $\mathbf{r}=\left(x,y,z\right)$ is the vector
of position coordinates, $t$ is time, and $\hat{H}$ is the Hamiltonian
operator corresponding to the total energy of the quantum system \citep{Georgiev2017,Georgiev2020}.

The quantum wavefunction $|\Psi(\mathbf{r},t)\rangle$ of the physical
system, which solves the Schr\"{o}dinger equation, is a continuous distribution
in three-dimensional space. As a consequence of the linearity of the
Schr\"{o}dinger equation, it follows that any two solutions $|\Psi_{1}(\mathbf{r},t)\rangle$
and $|\Psi_{2}(\mathbf{r},t)\rangle$ can be linearly superposed to
form a new solution 
\begin{equation}
|\Psi_{s}(\mathbf{r},t)\rangle=\alpha_{1}|\Psi_{1}(\mathbf{r},t)\rangle+\alpha_{2}|\Psi_{2}(\mathbf{r},t)\rangle
\end{equation}
where $\alpha_{1}$ and $\alpha_{2}$ are complex numbers satisfying
the normalization condition $|\alpha_{1}|^{2}+|\alpha_{2}|^{2}=1$.
Due to the principle of quantum superposition, the quantum wavefunction
$|\Psi(\mathbf{r},t)\rangle$ has the properties of a vector (and behaves like
a vector) in an abstract Hilbert space \citep{Georgiev2018a}.

\emph{Born rule}.
The value of the quantum wavefunction at a certain location $(\mathbf{r},t)$
in space and time is a complex number $\Psi(\mathbf{r},t)$ known
as quantum probability amplitude \citep{Feynman1948,Feynman2013,Feynman2014}.
According to the Born rule, the absolute square of the quantum probability
amplitude $|\Psi(\mathbf{r},t)|^{2}$ determines the quantum probability
for some physical event involving the quantum system of interest to
occur at the location with coordinates $(\mathbf{r},t)$. Importantly,
the behavior of the quantum system is inherently indeterministic.
If the exact state $|\Psi(\mathbf{r},t)\rangle$ of the quantum system
is known through meticulous preparation (post-selection), the quantum
probabilities for occurrence of different physical events (observations)
will still arise, not because of our ignorance of what the quantum
state is, but due to the inherent propensity of the quantum system
to generate the observed outcomes under experimental measurement.
Even though the quantum physical laws may preclude
prediction with absolute certainty of the future state (or event)
of a quantum system, they allow calculation of the probability for
a given future state (or event) to be actualized by the system. As
a consequence of indeterminism, consciousness does not have to be
epiphenomenal in a quantum world and the origin of free will could
be recognized in the process of actualization of physical events compliant
with the Born rule \citep{Georgiev2013,Georgiev2017}.

The main characteristic property of the quantum wavefunction $|\Psi\rangle$
is that it is not observable. Of course, here we do not mean that
the mathematical symbol $|\Psi\rangle$ typeset on paper is unobservable,
but rather that the physical entity to which $|\Psi\rangle$ refers
to in reality is unobservable. Exactly because the quantum information
is fundamentally tied to its physical carrier, it is impossible for
classical computers to replicate the behavior of quantum physical
systems. While classical waves could be easily observed as ripples
in a water tank, quantum waves are comprised from a contrastingly
different fabric, namely unobservable quantum probability amplitudes
\citep{Georgiev2020}. Observable physical quantities during a quantum
measurement are the eigenvalues of some quantum operator (observable)
$\hat{A}$, which is represented by a matrix \citep{Dirac1967,Fayngold2013,Susskind2014,Hayashi2015}. The
quantum observable $\hat{A}$ may operate upon any quantum wavefunction
and generate another quantum wavefunction \citep{Holevo2001}. Of special
physical interest is the set of eigenvectors of $\hat{A}$ such that
the action of $\hat{A}$ on an eigenvector $|\Phi\rangle$ returns
the same eigenvector $|\Phi\rangle$ multiplied by a number $\lambda$
referred to as an eigenvalue, namely $\hat{A}|\Phi\rangle=\lambda|\Phi\rangle$
\citep{Strang2016}. In $n$-dimensional Hilbert space, the quantum
observable $\hat{A}$ will have $n$ orthogonal eigenvectors $|\Phi_{1}\rangle,|\Phi_{2}\rangle,\ldots,|\Phi_{n}\rangle$,
each of which will be associated with a corresponding eigenvalue,
$\lambda_{1},\lambda_{2},\ldots,\lambda_{n}$.
The eigenvectors and eigenvalues of quantum observables play a major role
in the process of quantum measurement as they are exhibited as measurement outcomes.
Suppose that at time $t$ we measure a quantum system whose quantum state is
\begin{equation}
|\Psi(t)\rangle=\hat{I}|\Psi(t)\rangle=\sum\limits _{n}|\Phi_{n}\rangle\langle\Phi_{n}||\Psi(t)\rangle=\sum\limits _{n}\alpha_{n}|\Phi_{n}\rangle
\end{equation}
where $\hat{I}$ is the identity operator and $\alpha_{n}=\langle\Phi_{n}|\Psi(t)\rangle$
are complex coefficients \citep{Georgiev2020}. For the quantum measurement
of the observable $\hat{A}$, the outcome could only be one of the
eigenvalues present in the set $\{\lambda_{1},\lambda_{2},\ldots,\lambda_{n}\}$,
after which the quantum system will collapse into the eigenvector
corresponding to the obtained eigenvalue. For example, if the eigenvalue
$\lambda_{n}$ is obtained, immediately after the measurement at time
$t'$ the quantum system will turn into the state $|\Psi(t')\rangle=|\Phi_{n}\rangle$.
The probability to obtain the particular eigenvalue $\lambda_{n}$
is determined by the Born rule as the absolute square of the inner
product between the initial quantum state $|\Psi(t)\rangle$ and the
final quantum state given by the corresponding eigenvector $|\Phi_{n}\rangle$,
namely
\begin{equation}
\text{Prob}(\lambda_{n})=|\langle\Phi_{n}|\Psi(t)\rangle|^{2}
\end{equation}
In other words, the very act of quantum measurement is intrusive and
forces the measured system to react. This abrupt collapse $|\Psi(t)\rangle\to|\Phi_{n}\rangle$
provides some insight into the origin of unobservability of the quantum
wavefunction $|\Psi(t)\rangle$, namely the measurement of a quantum
observable $\hat{A}$ transforms probabilistically the measured system
into an eigenvector $|\Phi_{n}\rangle$ of the measured observable.
As a result, the external observer may learn the quantum state $|\Phi_{n}\rangle$
after the measurement, but is unable to reconstruct with certainty
what the quantum state $|\Psi(t)\rangle$ of the measured system was
before the measurement \citep{Georgiev2017}.

As a highlight of our brief excursion into quantum foundations, we could
say that quantum states are vectors $|\Psi\rangle$ whereas quantum
observables are operators $\hat{A}$ in Hilbert space. This mathematical
distinction results in a \emph{schism} between what physically exists
and what can be physically observed. The failure of naive realism,
which identifies what is observed with what exists, allows quantum
theory to accommodate conscious experiences that are subjective, private
and inaccessible to observation.

Thomas Nagel has been able to nicely illustrate in a thought experiment
the inner privacy of conscious experiences by examining the taste
of chocolate \citep{Nagel1987}: suppose that an awake patient is undergoing
an open skull neurosurgery with local anesthesia while eating chocolate.
The surgeon will be able to directly see the soft, spongy substance
of the patient's brain, and if a microscope equipped with sophisticated
patch-clamp device is used, electrical recordings from individual
neurons would reveal a chain of complicated physicochemical processes.
Nonetheless, the surgeon would find nowhere the taste of chocolate,
because the patient's conscious experiences are unobservable. In fact,
conscious experiences are inside the mind with a kind of insideness
that is quite unlike how the brain is inside the skull \citep{Nagel1987}.

In classical physics, the reductive approach to consciousness is unsuccessful
because the unobservable conscious mind cannot be identified with
an observable physical subsystem of the brain. In quantum physics,
however, the conscious mind could be reductively identified with the
quantum information integrated in some quantum state of the brain
$|\Psi\rangle$ because both are unobservable. In this case, the unobservable
conscious mind $|\Psi\rangle$ will not be a functional product of
the observable brain $\hat{A}$ that could be investigated with brain
imaging devices. Instead, the situation will be reversed---the observable brain $\hat{A}$ will be
the product (classical record) of actualized mind decisions (choices) \citep{Georgiev2017}.

The temporal dynamics of quantum states governed by the Schr\"{o}dinger
equation has an important implication for the dynamic timescale of
conscious processes. The total energy of the quantum system is given
by the eigenvalues $E_{n}$ and eigenstates $\left|E_{n}\right\rangle $
of the Hamiltonian operator $\hat{H}$, namely $\hat{H}\left|E_{n}\right\rangle =E_{n}\left|E_{n}\right\rangle $.
For each eigenstate with definite energy, the Schr\"{o}dinger equation
generates solutions of the form
\begin{equation}
\Psi_{n}(t)\left|E_{n}\right\rangle =\Psi_{n}(0)e^{-\imath\,\omega_{n}\,t}\left|E_{n}\right\rangle 
\end{equation}
The angular frequency $\omega_{n}=E_{n}/\hbar$ of each solution $\Psi_{n}(t)\left|E_{n}\right\rangle $
establishes the quantum dynamic scale (period) of the process $t_{n}={2\pi}/{\omega_{n}}$ with which the wavefunction $\Psi_{n}(t)$ rotates inside the Hilbert space.
For biomolecular energies exceeding the energy of thermal fluctuations
$E\ge k_{B}T$, the dynamic timescale is
\begin{equation}
t\le\frac{2\pi\hbar}{k_{B}T}
\end{equation}
where $k_B$ is the Boltzmann constant, and $T$ is the temperature.
At physiological temperature $T=310$~K, the dynamic timescale is faster than 0.15 picoseconds, which is in the realm of quantum
chemistry. Thus, the dynamic timescale of conscious processes is consistent
with picosecond conformational transitions of neural biomolecules
\citep{Georgiev2007}.
Examples of picosecond protein dynamics, which is directly related to the neuronal processing of information, include regulation of conductance of voltage-gated ion channels \citep{Callahan2018}, activation of ionotropic glutamate receptors by neurotransmitter binding \citep{Kubo2004}, or vibrational
motions of the $\alpha$-helix backbone involved in the conformational flexibility of SNARE proteins that drive exocytosis of synaptic vesciles \citep{Stelzer2008}.
In contrast, the observable brain dynamics of
electric spikes propagating along neuronal projections at a millisecond
timescale describes transfer of classical information, which is triggered
by quantum processes, but sets no lower bound on how fast these quantum
processes are.

\section{Physical properties and theoretical utility of quantum information}

Quantum information theory has been able to distill the main properties
of quantum information into a number of no-go theorems \citep{Nielsen2010,Pathak2013,Hayashi2015}.
Unknown quantum states that are not prepared by us cannot be read
(cannot be unambiguously reconstructed from measurements) \citep{Busch1997},
cannot be cloned (multiplied) \citep{Wootters1982}, cannot be deleted
\citep{Pati2000}, cannot be broadcast \citep{Barnum1996}, and cannot
be converted into a string of bits of classical information, even
if an infinite string of classical bits were allowed \citep{Pathak2013}.
Bell's theorem further shows that quantum entangled states of composite
quantum systems exhibit nonlocality, which can only be explained by
the admission of superluminal action at a distance between the spatially
separated quantum components \citep{Aspect1999,Georgescu2014,Rosenfeld2017}.
These physical properties of quantum information are able to address different
aspects of consciousness.

The \emph{unobservability} of quantum information is able to protect the inner
privacy of conscious experiences against external peering with alleged
mind-reading devices \citep{Georgiev2017}. This means that we may
keep secrets in our mind insofar we do not verbalize them in words.
Because our inner monologue is expressible as a string of classical
bits of information, we should be aware that the neural electric signals
corresponding to individual words are subject to eavesdropping, even
before the words are spoken in the form of audible acoustic waves.

The \emph{nonclonability} of quantum information is able to protect the identity
of the self against possible duplication at a different location in
space \citep{Georgiev2017}. Being macroscopically localized ensures
the historicity of the self and underlies our perception of being
embodied. Knowing that we cannot be clones of another conscious being
also provides us with an insurance against external blackmailing with
alleged torture of multiple brain-in-a-vat clones of ourselves, which
are supposed to have conscious experiences indistinguishable from
ours \citep{Elga2004}. Classically, we may doubt whether we are the
genuine original or a clone, whereas in a quantum world, we may rest
assured that we are physically unique.

The \emph{inconvertibility} of quantum information into classical
information is able to explain why we cannot communicate the phenomenal qualia
of our experiences (what it is like to feel what we experience) to
others \citep{Georgiev2017}. Suppose that in an imaginary world our
conscious experiences were communicable. In such a world, we would
have been able to restore missing senses through transmission of classical
strings of bits, 0s and 1s, which comprise the digital files of textual
or audiovisual material. For example, we would have been able to make
a blind person see a visual scenery by simply describing it with our
words \citep{Georgiev2017,Georgiev2020}. Effectively, blindness would
have been cured through genuine visual experiences elicited by words.
Unfortunately, this is not our world.
Nonetheless, existing regularities within our conscious experiences can be communicated
to other people. Conscious experiences can be categorized into distinct
categories: familiar or unfamiliar, pleasant or unpleasant, etc. Such
categorization is expressible in classical bits of information. In
fact, classical information can be used for the preparation of a set
of orthogonal (distinguishable) quantum states. This may provide a possible
explanation of why we do not continuously experience stored memories
but need a recall in order for certain memories to be consciously
relived, namely our memories are nothing but classical instructions
of how to bring back the quantum brain in a certain quantum state
selected from an orthogonal set of states.

The \emph{inerasability} of quantum information is able to protect against external
hindering with one's causal potency or free will \citep{Georgiev2017}.
Classical deletion works by taking different input classical states
and deterministically preparing them in a certain output state designated
to be the empty state \citep{Shen2011}. Different quantum states,
however, produce probabilistic outcomes upon measurement, and there
is no general unitary operation that could transform all possible
input quantum states into the same empty quantum state. This means
that the preparation of a desirable quantum state is always a post-selection,
namely, the quantum system chooses a measurement outcome in accordance
with inherent propensities given by the Born rule, and then the external
observer chooses to work with only those quantum systems that have
produced a certain desirable outcome. It should also be noted, that
future quantum choices are not dependent on past quantum choices.
For example, a single electron can be prepared multiple times in an
initial state $\left|\uparrow_{z}\right\rangle $ such that the electron
spin points up along the $z$-axis, after which the electron spin
can be measured along the $x$-axis. For each of these measurements,
there will be an equal probability for the spin to point up $\left|\uparrow_{x}\right\rangle $
or down $\left|\downarrow_{x}\right\rangle $ along the $x$-axis,
regardless of what the sequence of previous measurement outcomes for
the $x$ component of spin is.

The \emph{nonlocality} of quantum information due to quantum entanglement
is able to explain the combination problem (binding problem) of conscious experiences
\citep{Georgiev2017}. Indeed, while the reductive solution to the
mind--brain problem avoids the charge of epiphenomenal emergence by
attributing elementary conscious experiences to all matter, it requires
a mechanism by which elementary conscious experiences could combine
together and unite into larger, more complex, and ultimately human-like
conscious minds \citep{Basile2010,Coleman2012,Morris2017}. Quantum
entanglement provides the physical mechanism by which the quantum
probability amplitudes of component quantum systems become inseparable
\citep{Horodecki2009,Peled2020,Gudder2}. As an example, consider the singlet state composed
of two entangled spin-$\frac{1}{2}$ particles given by a composite
state vector
\begin{equation}
\left|\Psi\right\rangle _{AB}=\frac{1}{\sqrt{2}}\left(\left|\uparrow\right\rangle _{A}\left|\downarrow\right\rangle _{B}-\left|\downarrow\right\rangle _{A}\left|\uparrow\right\rangle _{B}\right)
\end{equation}
where neither the particle $A$, nor the particle $B$ has its own
separable state vector. However, if the composite state of particles
$A$ and $B$ can be expressed as a tensor product state
\begin{equation}
\left|\Psi\right\rangle _{AB}=\left|\Psi\right\rangle _{A}\otimes\left|\Psi\right\rangle _{B}
\end{equation}
then both particles $A$ and $B$ will have their own individual state
vectors such that
\begin{align}
\left|\Psi\right\rangle _{A} & =a_{1}\left|\uparrow\right\rangle _{A}+a_{2}\left|\downarrow\right\rangle _{A}\\
\left|\Psi\right\rangle _{B} & =b_{1}\left|\uparrow\right\rangle _{B}+b_{2}\left|\downarrow\right\rangle _{B}
\end{align}
where the complex quantum probability amplitudes are normalized
$\sum\left|a_{i}\right|^{2}=\sum\left|b_{i}\right|^{2}=1$.

In the tensor product state $\left|\Psi\right\rangle _{A}\otimes\left|\Psi\right\rangle _{B}$,
it is not only that the particles $A$ and $B$ have their own state
vectors $\left|\Psi\right\rangle _{A}$ and $\left|\Psi\right\rangle _{B}$,
but the composite system also has a state vector $\left|\Psi\right\rangle _{AB}$.

If the quantum information contained in the state vector is to be
identified with the mind of the system, to avoid paradoxical existence
of minds within other minds it has to be specified that only state
vectors that cannot be expressed as tensor products correspond to
\emph{individual minds}. If a state vector can be expressed as a tensor
product, then it will correspond to a \emph{collection of separate
minds}. Thus, by decomposing the quantum state $\left|\Psi\right\rangle _{U}$
of the universe into a tensor product form
\begin{equation}
\left|\Psi\right\rangle _{U}=\left|\Psi\right\rangle _{1}\otimes\left|\Psi\right\rangle _{2}\otimes\ldots\otimes\left|\Psi\right\rangle _{k} \label{eq:minds}
\end{equation}
quantum information theory provides a theoretical rule that explicitly
specifies the boundaries of individual minds $\left|\Psi\right\rangle _{k}$
in a universal complex-valued Hilbert space \citep{Georgiev2017}.
The combination of conscious experiences through quantum entanglement
also elaborates on the free will theorem by Conway and Specker \citep{Conway2006,Conway2009}
by showing that elementary quantum particles possess free will that
allows them to produce contextual outcomes upon different quantum
measurements but only as long as they are in a separable tensor product
state \citep{Georgiev2017}. Otherwise, if the elementary quantum particles
enter into a composite entangled state, it is the composite system
that manifests free will, not the individual components whose measurement
outcomes have to be nonlocally correlated.
In order to avoid quantum entanglement of the whole universe into a single universal cosmic mind, the reductive quantum information theoretic approach to consciousness requires the existence of a non-unitary physical process of objective reduction with associated disentanglement of mascroscopic quantum physical systems that have attained a certain energy threshold [for details see Chapter~6 in \cite{Georgiev2017}].

The reductive identification of the quantum information contained in the quantum probability amplitudes~$\Psi$ with consciousness leads to quantum panpsychism.
This means that mentality in the form of primitive subjective conscious experiences is attributed to all matter in the quantum universe. In other words, consciousness does not emerge at some later stage in the history of the universe, but is already present in the quantum probability amplitudes $\Psi$ of all quantum particles at the very origin of the universe. Here, it is important to emphasize that there is no single cosmic conscious mind permeating the universe, but a stochastic collection of fleeting minds popping in and out of existence. Consider as an example, liquid water near thermodynamic equilibrium. Thermal noise leads to vibrational motion of water molecules, but does not prevent the formation of water clusters such as hexamer prisms or cages (H$_2$O)$_6$ through hydrogen bonding \citep{Foley2013,Richardson2016}. The hydrogen bonds share quantum entanglement \citep{Pusuluk2018} between water molecules, which would then imply the existence of fleeting minds for each entangled cluster $|\Psi_k\rangle$ present in Eq.~\eqref{eq:minds}. These brief moments of subjectivity in the individual entangled water clusters would last only for a few picoseconds \citep{Elgabarty2020} before the water clusters are destroyed by the thermal noise. Because such fleeting minds in nonliving matter possess no memory and no means for communication with the surrounding world, they lack many of the features that we attribute to human consciousness. In the presence of interfaces provided by lipid membranes or protein biomolecules, however, water molecules can form extended hydrogen bond networks \citep{Paciaroni2008,Fayer2012,Stohr2019} whose interfacial orientational relaxation timescale of 18~ps is almost an order of magnitude larger compared with the relaxation time of 2.6~ps in bulk water \citep{Fayer2012}. This already illustrates how the presence of biomolecules supplies a form of ultrafast memory for thin layers of interfacial water. To build a fully functional mental unit for the human mind, natural selection had to assemble electrically excitable neurons from phospholipid membranes, proteins, signaling biomolecules, interfacial water and electrolytes, which collectively not only support complex entangled quantum states, but could also store long-term memories of those quantum states and use past memories to react to environmental stimuli.

\section{Molecular mechanisms of human consciousness}

The proposed quantum information-theoretic approach to the mind--brain problem is reductive in nature and identifies quantum information contained in quantum probability amplitudes with \emph{experiences}. This endorses a form of quantum \emph{panpsychism} or \emph{panexperientialism}, namely all quantum particles posses some elementary experiences that need to be organized through natural evolutionary processes in the proper way to compose a human mind that has a sense of ``self'' and is capable of rational thinking.
In humans, the brain cortex is the seat of consciousness and normally there is only one dominant personality, which we experiences as our conscious ``I''. In certain abnormal cases, however, a single brain can contain several minds, which may or may not be aware of each other's existence \citep{Georgiev2017}. For example, in split-brain patients who had their corpus callosum surgically severed in order to treat refractory epilepsy, each of the two cortical hemispheres hosts a separate mind that could communicate individually and control a half of the body apparently unaware of the other mind in the opposite hemisphere \citep{Gazzaniga1962,Sperry1966,Gazzaniga1967,Sperry1982,Gazzaniga2002,Wolman2012}. In psychiatric disease, such as dissociative identity disorder (also known as multiple personality disorder) a single brain may contain a number of different minds that take control over the body in succession \citep{Gillig2009}. The possible existence of multiple minds inside a single brain is hard to explain classically, but it is a natural prediction of quantum panpsychism. In normal conditions, a healthy dominant conscious ``I'' residing in the brain cortex will not be aware of the existence of conscious experiences in other subcortical areas of the brain. Thus, from the viewpoint of the conscious ``I'', what is going on in the subcortical areas could be called ``subconsciousness'', but qualitatively these will be experiences to which the conscious ``I'' has no concurrent access. In the rest of this section, we will focus on molecular mechanisms whose quantum dynamics in the brain cortex may be characteristic for the realization of human consciousness expressed through the conscious ``I''.

The main thrust of quantum information theory is to conceptualize
the differences between classical and quantum physical behavior and
distill those differences into explicit no-go theorems \citep{Pathak2013,Nielsen2010}.
The application of quantum information-theoretic results to brain
physiology, however, requires underlying biomolecular substrates that
exhibit quantum behavior. Recent advances in the methods for solving
numerically the Schr\"{o}dinger equation allow \emph{ab initio} studies
of quantum physical systems composed of several thousand atoms \citep{Sholl2009,Ullrich2012,Helgaker2013,Su2017}.
So far, computational quantum chemistry has provided evidence for
dynamic quantum effects in biomolecule-ion interaction \citep{Kolev2013,Kolev2018},
enzyme catalysis \citep{Ishida2006,Ranaghan2017,Sousa2016}, ion channel
gating \citep{Bucher2010a,Bucher2010b,Kariev2007,Kariev2009,Kariev2012,Kariev2014,Kariev2019,Roy2009,Maffeo2012,Flood2019},
and protein-induced remodeling of phospholipid membranes \citep{Ingolfsson2016}.
Quantum tunneling in the gating of voltage-gated ion channels \citep{Chancey1992,Vaziri2010,Kariev2019}
or zipping of SNARE (soluble N-ethylmaleimide-sensitive factor attachment
protein receptor) proteins in neurotransmitter release \citep{Georgiev2018b}
may act as a quantum trigger whose effects are amplified into macroscopic
patterns of electric activity of the cortical neural network.

Quantum effects inside the pores of ion channels confer selectivity
for passage of a certain type of ion \citep{Bucher2010a,Bucher2010b,Kariev2007,Kariev2009,Kariev2012,Kariev2014,Kariev2019}.
Ion selectivity divides channels into excitatory or inhibitory. Sodium
and calcium channels excite the neuron because they let positively
charged Na\textsuperscript{+} and Ca\textsuperscript{2+} ions, respectively,
enter into the cytosol. Conversely, potassium channels inhibit the
neuron, as they let positively charged K\textsuperscript{+} ions
escape from the cytosol toward the extracellular space. At places
where positively charged ions enter the neuron, the electric voltage
across the membrane increases and the membrane depolarizes. Alternatively,
if positively charged ions leave the neuron, the electric voltage
decreases and the membrane hyperpolarizes. Both depolarizations and
hyperpolarizations spread along the neuronal projections and summate
at the axonal hillock where electric spikes are generated. The quantum
behavior of individual voltage-gated ion channels is manifested in
the binary dynamic change of their electric conductance: the open
channel selectively conducts ions with a characteristic picosiemens
single channel conductance, whereas the closed channel does not conduct
at all. At a given value of the transmembrane voltage of the neuron,
individual voltage-gated ion channels undergo stochastic (probabilistic)
transitions between closed and open states \citep{Sakmann1995}. When
the voltage across the neuronal plasma membrane is far away from the
threshold for generation of an electric spike, the neuronal activity
is not particularly sensitive to the tiny stochastic fluctuations
in the transmembrane potential due to single channel transitions between
closed and open conformations \citep{Georgiev2015}. When the voltage
is near to the threshold value of $-55$ mV, however, closing or opening
of a single ion channel may influence the generation of electric spike
\citep{Destexhe2006}. In the human brain cortex, there are $\approx1.6\times10^{10}$
neurons \citep{Azevedo2009} which could attain firing frequencies
of $\approx40$ Hz. Thus, each millisecond thousands of cortical neurons
may sense the closing or opening of a single channel and amplify its
quantum dynamics into a macroscopically distinct electric firing pattern
of the cortical neuronal network.

\begin{figure}[t]
\begin{centering}
\includegraphics[width=162mm]{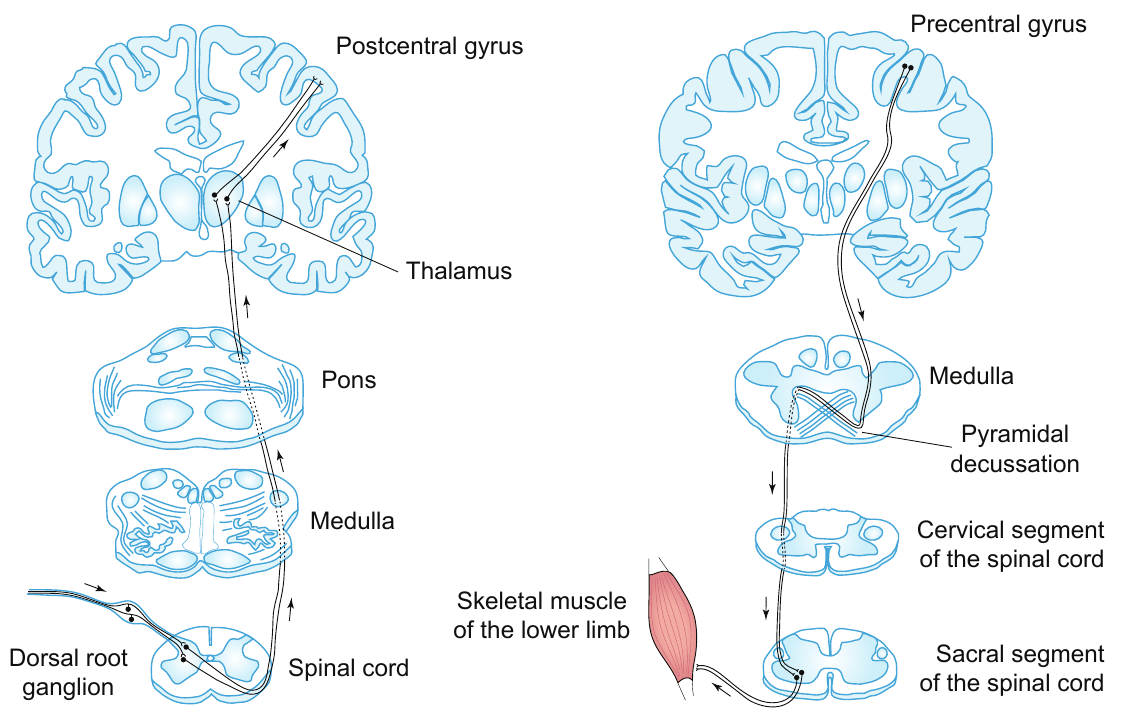}
\par\end{centering}
\caption{\label{fig:3}Classical communication through electric signals between
the brain cortex and the body. The somatosensory pathway (left) delivers
sensory information from the body to the somatosensory cortex in the
postcentral gyrus, whereas the somatomotor pathway (right) delivers
motor information from the motor cortex in the precentral gyrus to
the body muscles. The spinal cord segments, medulla and pons are represented
with their transversal sections, whereas thalamus and cortex are shown
in frontal slice. Modified from \cite{Georgiev2017}.}
\end{figure}

Sensory and somatomotor information encoded in electric spikes is
reliably transmitted across the synapses of the sensory pathways from
the sensory organs toward the brain cortex \citep{Kim2013,Singer2007,Glowatzki2002,Magistretti2015}
or the somatomotor pathways from the motor cortex toward the muscles
\citep{Kuno1971}. To achieve reliability of transmission, the chemical
synapses in these pathways release multiple synaptic vesicles upon
depolarization of the presynaptic axonal boutons \citep{Rudolph2015}.
The multivesicular release of neurotransmitter molecules then generates
large postsynaptic currents in the target neuron \citep{Rudolph2015}.
The reliable transmission of electric signals between the brain cortex
and the body (Fig.~\ref{fig:3}) ensures the survival of the organism
through the execution of fight-or-flight responses.

\begin{figure}[t]
\begin{centering}
\includegraphics[width=162mm]{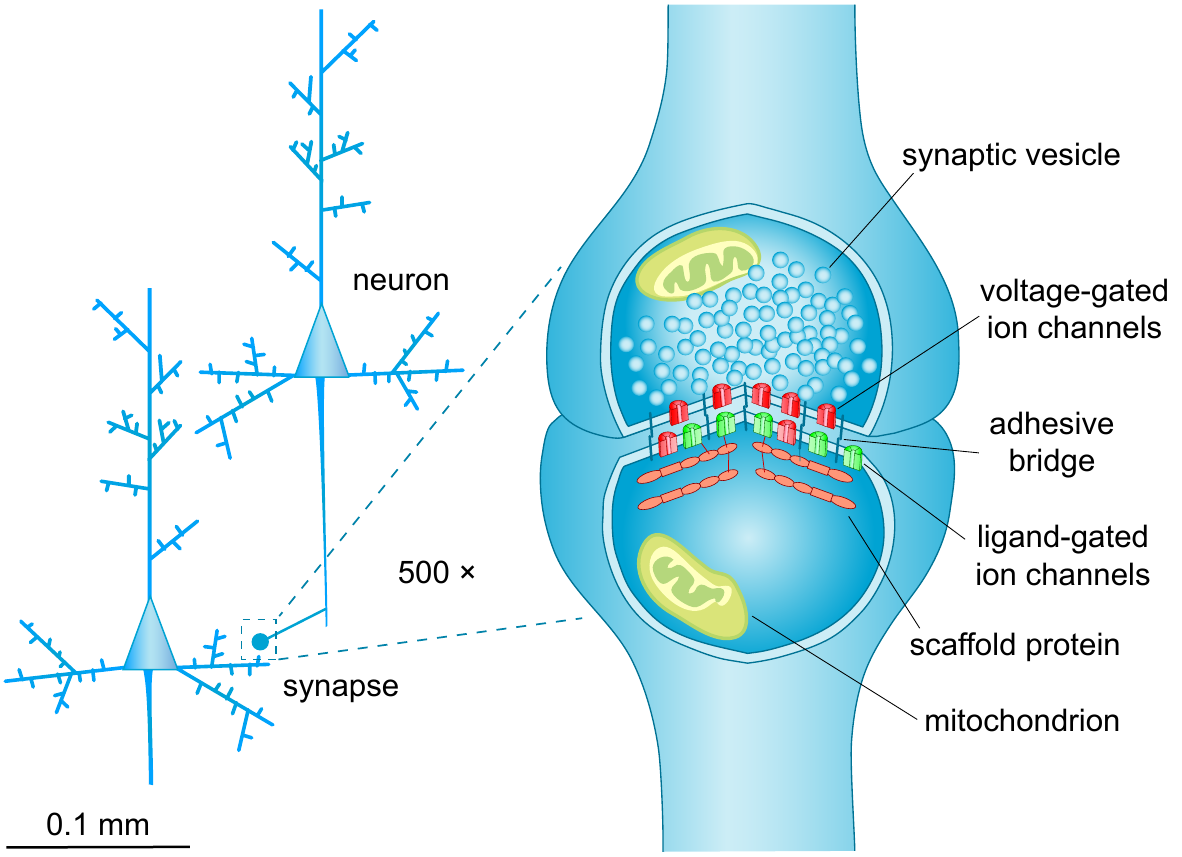}
\par\end{centering}
\caption{\label{fig:4}Excitatory synaptic contact between pyramidal cortical
neurons. The presynaptic axonal bouton has a pool of synaptic vesicles
that contain neurotransmitter (glutamate). During an electric spike, the activation
of presynaptic voltage-gated calcium channels initiates Ca\protect\textsuperscript{2+}
influx at the active zone, which may trigger fusion of a single synaptic
vesicle with the plasma membrane. The released neurotransmitter (glutamate) acts
on postsynaptic ligand-gated ion channels, such as AMPA ($\alpha$-amino-3-hydroxy-5-methyl-4-isoxazolepropionic acid) or NMDA ($N$-methyl-$_D$-aspartate) receptors, to induce postsynaptic electric currents in the target neuron. Structural support for the synapse is provided
by adhesive bridges and scaffold proteins, whereas mitochondria ensure
robust energy supply for synaptic neurotransmission. Modified from \cite{Georgiev2017}.}
\end{figure}

In contrast to extracortical synapses, individual synapses inside
the brain cortex and the hippocampus were found to release either
a single synaptic vesicle or none. Thus, each cortical synapse (Fig.~\ref{fig:4})
appears to possess only one functional release site at a given time
\citep{Stevens1995}. The probability for release of neurotransmitter
through synaptic vesicle exocytosis at intracortical synapses is $0.35\pm0.23$
per axonal spike \citep{Dobrunz1997}. This means that an electrically
excited axonal bouton is twice more likely to fail than succeed in
releasing neurotransmitter. If it is conservatively estimated that
each cortical neuron has only $n=1000$ axonal boutons, and on average
only $k=350$ of these boutons release neurotransmitter per electric
spike, the number of possible combinations given by the binomial coefficient
$\frac{n!}{k!(n-k)!}$ for exocytosis per neuron is over $10^{279}$.
If the mind were not in control of synaptic vesicle release, the cortical
neural network would have been disorganized within seconds \citep{Georgiev2018b}.
Fortunately, recent advances in molecular neuroscience have revealed
an elaborate protein machinery that regulates synaptic vesicle fusion
with the active patch of presynaptic plasma membrane \citep{Branco2009,Sudhof2013}.

\begin{figure}[t!]
\begin{centering}
\includegraphics[width=162mm]{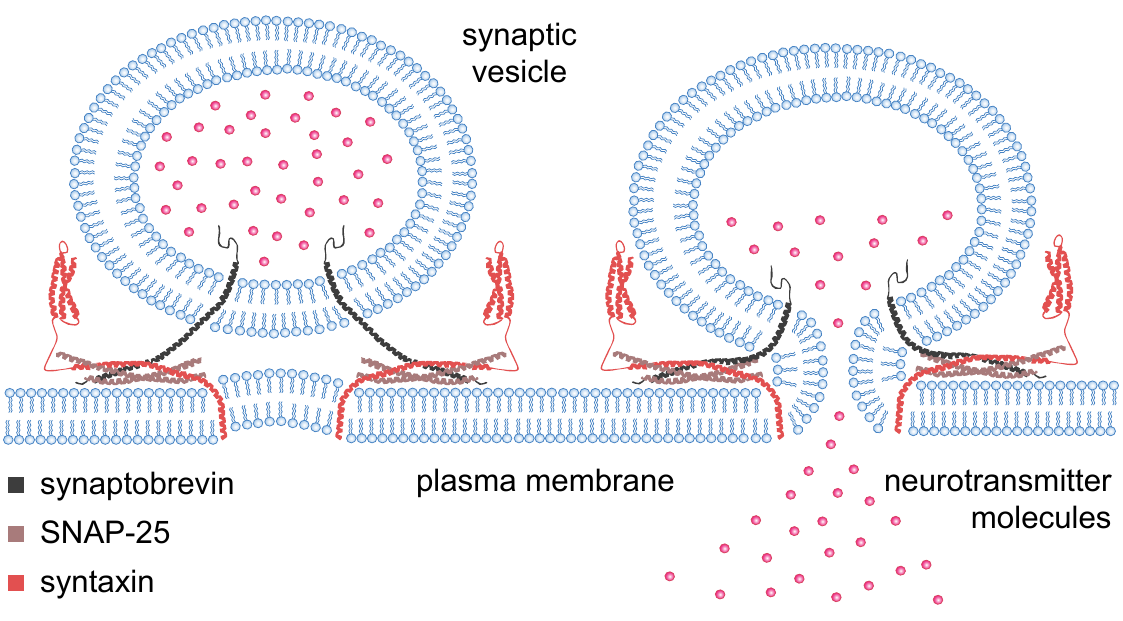}
\par\end{centering}
\caption{\label{fig:5}Synaptic vesicle exocytosis driven by cooperative zipping
of SNARE protein complexes. Initial docking of the synaptic vesicles
at the active zone of the synapse is actuated by partially zipped
SNARE protein complexes (left). Electric activity of the presynaptic
terminal with its accompanying Ca\protect\textsuperscript{2+} influx
triggers full zipping of the SNARE complexes in a docked vesicle with
subsequent opening of the fusion pore and extrusion of neurotransmitter
from the synaptic vesicle into the synaptic cleft (right). Modified from \cite{Georgiev2014}.}
\end{figure}

The minimal molecular machinery capable of driving synaptic vesicle
fusion is comprised of only three SNARE proteins: synaptobrevin, syntaxin
and SNAP-25 (Fig.~\ref{fig:5}) \citep{Weber1998}. These three SNARE
proteins zip together to form a bundle of four $\alpha$-helices referred
to as the core SNARE complex. The twisting of the 4-$\alpha$-helix
bundle inside the core SNARE complex applies a traction force that
drives the fusion of the opposing phospholipid bilayers of the synaptic
vesicle and the plasma membrane \citep{Weber1998,Risselada2012,Sudhof2012,Sudhof2013,Zhou2015}.
The zippering of the core SNARE complex is potent enough to drive
synaptic vesicle exocytosis even in neurons expressing artificially
engineered lipid-anchored synaptobrevin and syntaxin molecules that
lack their transmembrane regions \citep{Zhou2013}.

In different neuron types, the process of exocytosis is regulated
by different sets of SNARE master proteins \citep{Zhou2015,Rizo2012,Sudhof2009,Giraudo2006}
that effectively set the potential energy barrier for vesicle fusion.
The potential energy $V(\mathbf{r})$ enters into the Hamiltonian
\begin{equation}
\hat{H}=-\frac{\hbar^{2}}{2m}\nabla^{2}+V(\mathbf{r})
\end{equation}
for the Schr\"{o}dinger equation, where it constrains the motion of massive
quantum particles with mass $m$. The motion of quantum particles
is free inside spatial regions with zero potential energy, analogously
to the classical case. However, there is a substantial difference
between classical and quantum behavior for regions in which the potential
energy is non-zero. Classical particles are forbidden from entering
spatial regions where the particle energy is less than the potential
energy, $E_{0}<V(\mathbf{r})$, whereas quantum particles are not \citep{Landau1965}.
In fact, the quantum wavefunction $\Psi(\mathbf{r})$ needs to be
continuous throughout space, which allows the quantum particle to
tunnel through the potential energy barrier with height $V_{0}>E_{0}$
and appear on the other side. In protein $\alpha$-helices,
quantum quasi-particles called Davydov solitons, which are composed
of amide I excitation self-trapped in the lattice distortion of hydrogen
bonded peptide groups, are able to transport energy along the protein
and could trigger conformational transitions \citep{Georgiev2019b,Georgiev2019c,Georgiev2020b}.
Because the probability of release in intracortical synapses is less
than one, the quantum mechanical description of the process involves
quantum tunneling through potential barrier whose height is higher
than the energy of quantum quasi-particle, presumably Davydov soliton,
which triggers the release \citep{Beck1992,Georgiev2012,Georgiev2019c}.

\begin{figure}[t!]
\begin{centering}
\includegraphics[width=160mm]{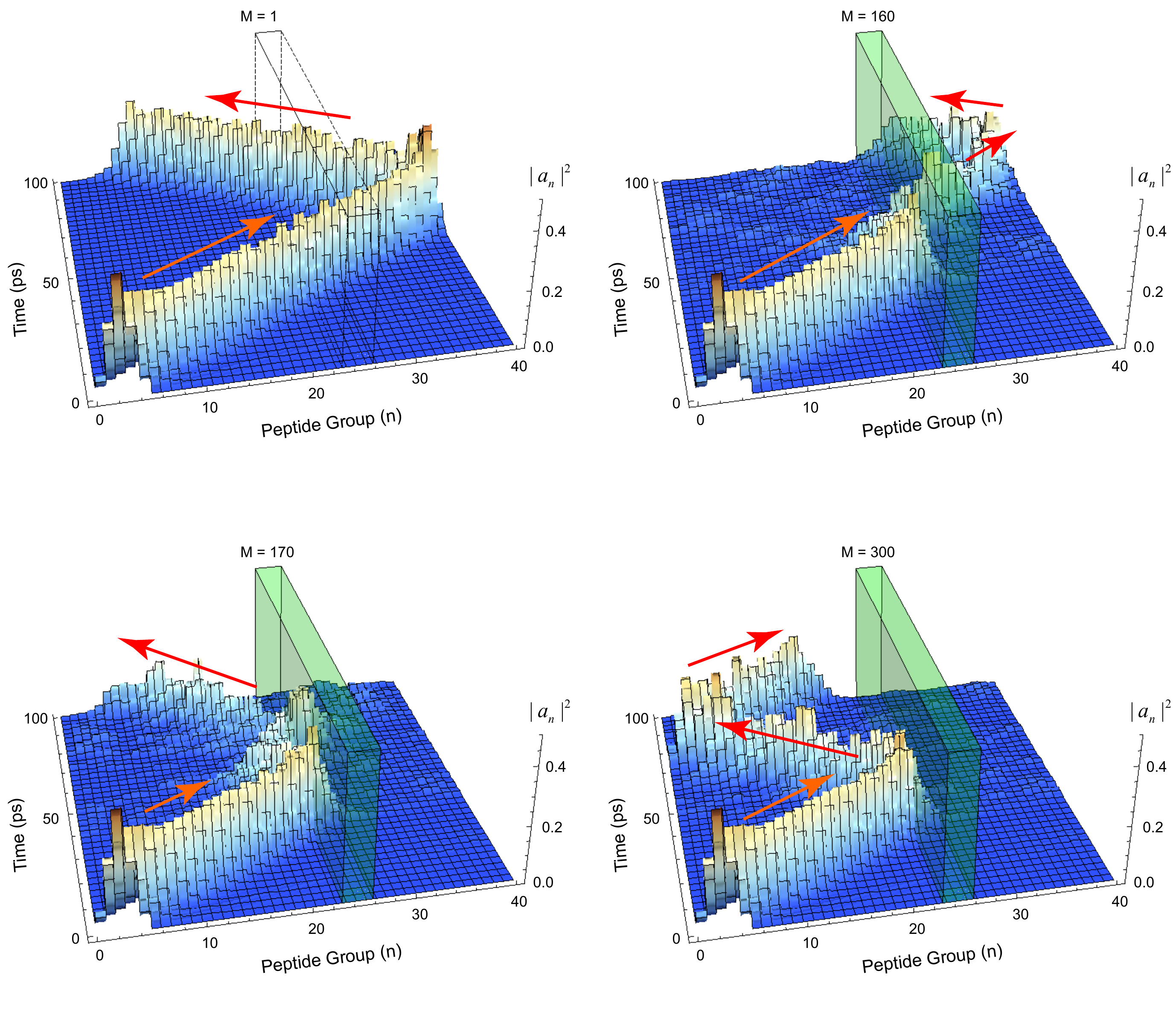}
\par\end{centering}
\caption{\label{fig:6}Quantum dynamics of Davydov soliton visualized by the probability $|a_n|^2$ of finding the exciton at the peptide group~$n$ inside a hydrogen bonded protein $\alpha$-helix spine simulated for 100~ps. In the absence of massive barrier ($M=1$), the evelope of excition probabilities $|a_n|^2$ propagates as a solitary wave (soliton), which reflects from the ends of the protein $\alpha$-helix. In the presence of massive barrier ($M=160$) extending over 3 peptide groups $n=22-25$, each of which with effective mass 160 times greater than the mass of an average amino acid residue, the soliton tunnels through the barrier with picosecond time delay. When the barrier mass is increased by only $6.25\%$ to $M=170$, the tunneling is drastically suppressed due to exponential decay of excition quantum probability amplitudes inside the barrier and the soliton cannot penetrate to the other side. For barrier with $M=300$, the soliton bounces off the barrier resembling the elastic collision of a classical billiard ball with a wall. Red arrows help indicate the direction of motion.}
\end{figure}

The mass of the Davydov soliton is $\approx 5\%$ of the proton mass and readily undergoes quantum tunneling \citep{Georgiev2020c}. The Davydov soliton is a type of acoustic polaron formed by the interaction of excitons (C=O bond vibrations) with phonons (deformations of the hydrogen bonded lattice of peptide groups) inside protein $\alpha$-helices. Because the protein backbone is quite massive (the average mass on an amino acid residue is 114 proton masses), the quantum motion of peptide groups (given by the generalized Ehrenfest theorem) is virtually indistinguishable from classical motion. Consequently, the protein backbone could be viewed as providing a potential energy landscape for the motion of the much lighter excitons. Through nonlinear feedback effect called \emph{self-trapping}, the presence of the exciton, in turn, influences the motion of the hydrogen bonded peptide groups inducing \emph{phonon dressing}. The combination of an exciton together with its phonon dressing is referred to as Davydov soliton. When a portion of the protein $\alpha$-helix is embedded inside a phospholipid membrane or interacts with another protein, there is a resulting clamping action that constrains the motion of the $\alpha$-helix thereby increasing locally its effective mass. The result of external clamps is that an exciton/soliton propagating along the $\alpha$-helix would meet a massive barrier. Depending on the mass of the barrier, the exciton may be able to pass through the barrier employing quantum tunneling or may reflect from the barrier (Fig.~\ref{fig:6}). In the case of SNARE zipping, the role of the barrier is played by the Ca$^{2+}$ sensor synaptotagmin, which clamps the SNARE complex in partially zipped conformation. Quantum tunneling of Davydov soliton through the barrier may induce full zipping of the SNARE complex and trigger exocytosis. In essence, massive proteins do not quantum tunnel, whereas quantum excitations propagating along the proteins do. Quantum tunneling of such excitations could act as a trigger that steers the overall protein motion at points of bifurcation into one of two alternative classical paths.
Thus, cortical neurons that have surpassed with certainty the voltage
threshold for the generation of electric spike, are able to amplify
the quantum dynamics of SNARE proteins at individual axonal buttons
into a macroscopic pattern of active synapses that release neurotransmitter
molecules.

In summary, the dominant conscious ``I'' in the brain cortex is to be identified with the unobservable quantum information contained in the quantum state of quantum entangled membrane-bound proteins, including voltage-gated ion channels, ligand-gated ion channels and SNARE proteins, residing in the plasma membranes of millions of cortical neurons. The dynamic timescale of the underlying quantum processes is on the order of picoseconds, which addresses effectively previously noted issues with decoherence \citep{Tegmark2000}. The power of the thermal noise is both frequency and temperature dependent as given by the Johnson--Nyquist formula \citep{Johnson1928,Nyquist1928}
\begin{equation}
P(f)=\int_{f_1}^{f_2}\frac{h f}{e^{\frac{h f}{k_B T}}-1} df
\label{eq:JN}
\end{equation}
where $h$ is the Planck constant, $f$ is the frequency, $\Delta f = f_2 - f_1$ is the bandwidth, $k_B$ is the Boltzmann constant and $T$ is the temperature. Thus, quantum effects readily persist not only for very low temperatures near absolute zero but also for very high frequencies over 1 THz. Psychophysical and clinical evidence from patients with time agnosia, which supports a submillisecond timescale of consciousness, has been extensively discussed in previous works \citep{Georgiev2013,Georgiev2017}.

To highlight the novel features of the quantum information theoretic approach to the mind--brain problem, it would be helpful to compare it with the orchestrated objective reduction (Orch OR) theory \citep{Hameroff2014} that also attempts to connect quantum mechanics to consciousness. Because the points of disagreement are substantial and somewhat involved, we will discuss each conceptual difference separately.

First, the two theories use the word ``quantum'' in mutually exclusive senses. In quantum information theory, the ``quantum'' characteristics of physical systems (wave-particle duality, superposition, interference, entanglement, tunneling) originate from compliance with continuous unitary dynamics according to the Schr\"{o}dinger equation, whose solution is the quantum wavefunction $\Psi$. In the Orch OR theory, the main focus is on discrete non-unitary, non-computable violations of the Schr\"{o}dinger equation induced by ``quantum gravity''. Thus, the two theories disagree on whether conscious experiences are present during the continuous period of unitary dynamics according to the Schr\"{o}dinger equation: the quantum information theory says yes, the Orch OR theory says no.
Also, consciousness in the quantum information theory originates in the unitary dynamics, whereas consciousness in the Orch OR theory originates in the non-unitary dynamics.

Second, the two theories assess the need of ``emergence'' with contrasting evaluations. In quantum information theory, the quantum wavefunction $\Psi$ is reductively identified with the mind of the physical system. Therefore, conscious experiences originate as physical solutions $\Psi$ of the Schr\"{o}dinger equation and need no further physical laws to miraculously pop into existence emergent sentience from insentient matter. In the Orch OR theory, conscious experiences emerge in the form of discrete subjective ``flashes'', ``bings'', or ``nows'' from non-unitary, non-computable objective reduction events in the fundamental space-time geometry. The main problem with ``emergence'' is that it only postulates violation of the physical laws, but does not inform us about the properties of the emergent phenomenon. In contrast, the reductive identification of consciousness with $\Psi$ implies that consciousness should necessarily possess all the physical characteristics that can be deduced from the Schr\"{o}dinger equation.

Third, the two theories attribute distinct ``causal efficacy'' to consciousness. In quantum information theory, the absolute square of the quantum wavefunction $|\Psi|^2$ determines the probabilities for future courses of action. Therefore, if consciousness is $\Psi$, then it causally introduces biases in what future physical events can happen and how probable they are. Reduction of the wavefunction $\Psi \equiv \sum_i \Psi_i$ to a particular outcome~$\Psi_i$ upon quantum measurement, is then viewed as an act of choice making and manifestation of inherent free will by the quantum system, which already possesses consciousness. In the Orch OR theory, conscious experiences are generated by the non-computable objective reduction process, $\sum_i \Psi_i \to \Psi_i$, yet due to the physical closure of the world, the emergent consciousness becomes an epiphenomenon that only witnesses what has happened, but is causally ineffective to determine what can happen or will happen next. Natural selection and evolution of epiphenomenal consciousness is impossible.

Fourth, the two theories operate at different timescales and rely on different underlying biomolecular substrates. In quantum information theory, human consciousness operates at picosecond timescale and is supported by voltage-gated ion channels and other membrane-bound proteins incorporated in excitable neuronal membranes, which is consistent with the neural delivery of sensory information and output of motor information in the form of electric spikes. In the Orch OR theory, human consciousness operates at millisecond timescale and is supported by cytoskeletal microtubules, which are hypothesized to be isolated from the surrounding neural electric activities to prevent quantum decoherence, yet also somehow capable to control the generation of electric spikes at the axonal hillock.
Fortunately, the discovery of general anesthetics provides an experimental framework for studying the molecular mechanisms of human consciousness through attempts to reversibly turn it off and on in clinical settings with human subjects.

\section{The SNARE protein complex is main molecular target for volatile anesthetics}

Pharmacological studies have shown that volatile anesthetics activate
ligand-gated GABA\textsubscript{A} chloride ion channels \citep{Garcia2010},
and block voltage-gated sodium \citep{Purtell2015} or calcium \citep{Joksovic2009}
ion channels. The increased sensitivity to volatile anesthetics conferred
by the expression of mutant voltage-gated sodium channels \citep{Pal2015},
however, provides only indirect link to consciousness, and experimental
attempts to introduce resistance to anesthesia in mice expressing
artificially engineered mutant voltage-gated calcium channels \citep{Petrenko2007}
or volatile anesthetic-resistant GABA\textsubscript{A} chloride ion
channels \citep{Werner2011} have been unsuccessful. In contrast, several
lines of experimental evidence point to direct involvement of SNARE
proteins in volatile anesthetic-induced unconsciousness \citep{Georgiev2010}.
Firstly, general anesthesia with clinical concentrations of volatile
anesthetics, such as isoflurane or halothane, inhibits excitatory
neurotransmitter release \citep{MacIver1996,Herring2009,Wu2004}, and
selectively erases consciousness, but not all cortical electric responses.
In particular, electric potentials evoked by applied visual stimuli
are routinely recorded from the visual cortex of anesthetized animals
\citep{Lamme1998,Imas2005,Sellers2015}. Secondly, volatile anesthetics
bind with high affinity to syntaxin either as a purified protein or
as a component of the core SNARE complex \citep{Nagele2005,Johansson2000}.
Thirdly, resistance to volatile anesthetics is conferred by a genetic
mutation that results in the translation of a truncated form of syntaxin
\citep{vanSwinderen1999}. Animals with truncated syntaxin require
higher doses of isoflurane for induction of anesthesia and also recover
faster from anesthesia in comparison with wild-type animals \citep{Troup2019}.
\emph{In vitro} experiments further show that the expression of the truncated
syntaxin in PC12 cells completely blocks the effects of isoflurane
on the neurotransmitter release machinery despite the presence of
endogenous syntaxin \citep{Herring2009}. Taken together, these findings
support the aforementioned quantum model of neurotransmitter release
in which volatile anesthetics inhibit quantum tunneling by raising
the height of the potential energy barrier for SNARE protein zipping.
Hence, the quantum quasiparticles (Davydov solitons) required for
exocytosis are effectively isolated from their environment by being
trapped in deep potential wells and their quantum states remain separable
from the rest of the brain. According to the quantum information-theoretic
approach to consciousness, it is quantum entanglement (as opposed
to quantum separability) of component systems that binds conscious
experiences together \citep{Georgiev2017}. Therefore, anesthesia works
not by turning conscious experiences on or off, but rather by interfering
with the combination of elementary conscious experiences into a unitary
composite conscious experience. In essence, by changing the potential
energy barrier landscape in the brain, volatile anesthetics temporarily
split the conscious mind into myriad pieces, analogously to how physical
severing of the two brain hemispheres generates two separate minds
in split-brain patients \citep{Wolman2012}. The only difference is
that the anesthetic action is reversible, whereas the physical severing
of the brain is not.

\section{Holevo's bound on accessible classical information from quantum measurements}

The phenomenal nature of conscious experiences is incommunicable as established by Locke's inverted qualia thought experiment [for a detailed discussion see \citet{Georgiev2017,Georgiev2020}]. This means that our introspective access to our own feelings and conscious experiences is fundamentally different from a scientific ``observation'' whose outcomes can be communicated to others in the form of classical bits of information. In fact, a consistency argument based on indistinguishability of non-orthogonal quantum states also establishes that conscious experiences cannot have the status of observations: if in reductive quantum theories of mind non-orthogonal quantum states comprise different conscious experiences, then we should not be able to report what exactly it is to be in one of the states instead of the other, or which of two non-orthogonal states we have been in, otherwise external observers would have been able to use our reports to distinguish non-orthogonal quantum brain states, i.e. a well-known no-go theorem by Busch \citep{Busch1997} would have been violated.

The incommunicability of consciousness has troubled philosophers for
centuries. Whereas we cannot express what the phenomenal nature of
qualia is \citep{Nagel1974,Nagel1987,Chalmers1995}, Wittgenstein has
argued that we can express some of our feelings in words such as ``pain,''
``anger,'' or ``love,'' which we all understand \citep{Wittgenstein2009,Tang2014,Baker1998}.
Quantum physics provides a valuable insight into why we can say something
meaningful about our consciousness, but not all there is about it.
As we have already elaborated in some detail, the quantum information
comprising the quantum state of a quantum physical system is unobservable
and cannot be fully converted into classical information. However,
Alexander Holevo has been able to show that each quantum system can
carry a certain amount of accessible classical information, which
does not exceed Holevo's bound \citep{Holevo1973,Holevo1998} 
\begin{equation}
\chi=S\left(\sum\limits _{k}p_{k}\hat{\rho}_{k}\right)-\sum\limits _{k}p_{k}S\left(\hat{\rho}_{k}\right)
\end{equation}
where $\left\{ \hat{\rho}_{1},\hat{\rho}_{2},\ldots,\hat{\rho}_{k}\right\} $
is a set of quantum states (described by their \emph{density matrix
operators}, cf. \citep{Belinfante1980,Hughston1993,deGosson2018})
that are drawn from the probability distribution $\left\{ p_{1},p_{2},\ldots,p_{k}\right\} $,
and $S\left(\hat{\rho}\right)$ is the \emph{von Neumann entropy}
\citep{vonNeumann1955,Ohya2011} of the density matrix $\hat{\rho}$
measured in bits
\begin{equation}
S\left(\hat{\rho}\right)=-\sum\limits _{j}\lambda_{j}\log_{2}\lambda_{j}
\end{equation}
with $\lambda_{j}$ denoting the eigenvalues of the density matrix
operator $\hat{\rho}$. Noteworthy, the von Neumann entropy is a concave functional on the space of density matrices, $S\left(\hat{\rho}\right)\geq 0$.
Holevo's accessible information is maximal
when: (i) the quantum states $\hat{\rho}_{k}$ are pure, that is $\hat{\rho}_{k}=|\psi_{k}\rangle\langle\psi_{k}|$,
since they have zero von Neumann entropy $S\left(\hat{\rho}_{k}\right)=0$
leading to
\begin{equation}
-\sum\limits _{k}p_{k}S\left(\hat{\rho}_{k}\right)=0\label{eq:14}
\end{equation}
(ii) the quantum states $\hat{\rho}_{k}$ are orthogonal, $\langle\psi_{k}|\psi_{k'}\rangle=0$
for $k\ne k'$ and $\langle\psi_{k}|\psi_{k'}\rangle=1$ for $k=k'$,
and (iii) the probabilities $p_{k}$ in the probability distribution
are equal. A composite quantum system consisting of $n$~qubits evolves
in a Hilbert space with $2^{n}$ dimensions. This provides $2^{n}$
orthogonal quantum basis states $\hat{\rho}_{k}$ available for encoding
of classical information. If those $2^{n}$ states $\hat{\rho}_{k}$
are equiprobable, we get
\begin{equation}
\sum\limits _{k}p_{k}\hat{\rho}_{k}=\frac{1}{2^{n}}\left(\begin{array}{cccc}
1 & 0 & \cdots & 0\\
0 & 1 & \cdots & 0\\
\vdots & \vdots & \ddots & \vdots\\
0 & 0 & \cdots & 1
\end{array}\right)
\end{equation}
with von Neumann entropy
\begin{equation}
S\left(\sum\limits _{k}p_{k}\hat{\rho}_{k}\right)=-\sum\limits _{k=1}^{2^{n}}\frac{1}{2^{n}}\log_{2}\left(\frac{1}{2^{n}}\right)=n\label{eq:16}
\end{equation}
Combining Eqs.~\eqref{eq:14} and \eqref{eq:16} together implies
that for a quantum system of $n$ qubits the maximal value of Holevo's
bound is $\chi=n$ bits. In other words, a composite quantum system
that contains $n$ qubits can carry up to $n$ bits of accessible
classical information, which can be extracted through measurement
by an external observer \citep{Holevo1973,Holevo1998}.

Quantum physics allows orthogonal quantum states to be distinguished
even though the same cannot be done unambiguously for
non-orthogonal quantum states. Therefore, we may be able to communicate
something meaningful in the form of words (bits of classical
information) about those of our conscious experiences that correspond
to orthogonal quantum brain states \citep{Georgiev2020}. If consciousness
is comprised of quantum information, it will not be completely inaccessible for others,
but rather it will have an accessible part which does not exceed the
classical bits of information allowed by Holevo's theorem. 

Returning back to Wittgenstein's example, we are able to conclude
that we do not really communicate the phenomenal qualia of ``pain,''
``anger,'' or ``love,'' but rather describe distinguishable classical
situations under which we expect other humans to consciously experience
feelings that are qualitatively similar to ours. For example, one
way to explain what the ``pain'' is could be to note that it is the
subjective feeling that we experience when we hit our finger with
a hammer. Such an explanation will work for all people who are capable
of experiencing pain, but will fail for people with syringomyelia
who are unable to feel pain in their extremities due to a cyst that
damages their spinal cord \citep{Klekamp2002}. Thus, consciousness
is not directly observable or directly accessible. We are able to
say some meaningful facts about our conscious experiences, but not
everything there is about these experiences \citep{Georgiev2017,Georgiev2020}.
Holevo's theorem links the orthogonality of quantum states in Hilbert
space with the possibility of a partially accessible consciousness by others.
Conscious experiences can be private in regard to their phenomenal
content (qualia) insofar we can communicate as a string of bits of classical
information only the distinguishable physical circumstances for their
occurrence. The latter is sufficient to differentiate between conscious
experiences with perceptibly distinct phenomenal content and allows
us to label these experiences with words such as ``pain,'' ``anger,''
or ``love.''

Introspectively, we do not perceive ourselves as being built of atoms,
molecules or neurons. When we decide to close our eyes or to move
one of our limbs, we do not have any idea which neuron in our brain
is firing to deliver the appropriate electric signal to our muscles.
Yet, if we undergo open skull neurosurgery, our brain can be observed
and its electric activity recorded. Thus, the brain may appear to
be what the conscious mind looks like from a third-person, objective
perspective. But according to quantum information theory this cannot
be exactly right. If consciousness comprises the quantum information
in the existing quantum brain state, then it cannot be observed. Instead,
upon measurement of some quantum observable of the brain, the quantum
state has to choose a possible quantum outcome among the possible
eigenvalues of the observable. Thus, the observable brain is the classical
record of observable eigenvalues of past mind choices (decisions).
The irreversibility of mind choices ensures that the observable brain
is an objective construct for all possible observers. Because the
anatomical brain is accessible classical information, it can be observed, shared,
communicated, copied and analyzed by multiple observers. The observability
of the brain supports objectivity in neuroscience as it grants multiple
scientists with simultaneous access to multiple identical copies of
classical information about the same brain, which can then be analyzed
independently.

\section{Concluding remarks}

Our conscious minds exist in the physical universe where they appear
to be causally potent agents capable of controlling our behavior and
transforming the surrounding world \citep{Yablo1992,Crane1995,Jackson1996}. Therefore,
if we are to have a scientific theory of consciousness, conscious
experiences should be represented in physical equations and need to
be governed by physical laws \citep{Georgiev2013,Georgiev2017}. In
a classical world, all physical quantities are observable and communicable.
This severely restrains the scope of classical physical theories because
an unobservable and incommunicable consciousness cannot be reduced
to anything already present in the physical equations. Instead consciousness
needs to somehow emerge as a functional product of the observable
brain, which will unfortunately turn the emergent consciousness into
a useless, causally ineffective epiphenomenon overrun by the deterministic
laws of classical physics \citep{Georgiev2019a}. In a quantum universe
comprised of unobservable and incommunicable quantum information,
however, epiphenomenal consciousness is avoidable thanks to quantum
indeterminism, and the perplexing inner privacy of consciousness could
be seen to originate in the physical properties of quantum information
integrated in brain cortical quantum states \citep{Georgiev2017,Georgiev2020,Melkikh2015,Melkikh2019}.

In the quantum reductive approach to consciousness, conscious experiences are identified with quantum brain states, which are in some relevant sense private or noncommunicable. When we introspect our quantum conscious states, we are able to report only a certain amount of classical information that does not exceed Holevo's bound.
The incommunicable qualia of conscious experiences, which are privately accessed through introspection, then correspond to quantum information that cannot be converted into classical information.
This incommunicable quantum information is inaccessible to external observers.
Thus, introspection \emph{is not} equivalent to quantum measurement. We continuously experience the contents of our conscious minds, which means that the introspective access to our inner mental world, provided through identity relation between consciousness and quantum information contained in the quantum brain state $\Psi$, is continuous and not discrete. The unitary quantum evolution of the quantum brain state $\Psi$, which among other things also leads to entanglement of different brain subcomponents, describes changes in unobservable conscious experiences and their binding or composition. The quantum measurements of the brain are discrete events performed by effector organs such as muscles or glands, which react to neuronal electric impulses outputted by efferent axon terminals. The glial cells, which nourish the neurons, also perform quantum measurements upon brain neurons for the purposes of maintaining proper homeostasis of brain electrolytes or other chemicals. The measurement of quantum brain observables by glial cells, effector organs or physical devices leads to decoherence of the quantum brain state $\Psi$ and extracts accessible classical bits of information from which is constructed the ``observable brain.'' Thus, quantum information theory addresses the mind--brain problem by utilizing the dichotomy between quantum state vectors and quantum observables.

The quantum physical laws, expressed in the Schr\"{o}dinger
equation and the Born rule, introduce a paradigm shift in consciousness
research by making all physical statements about the mind or the brain
subject to mathematical precision. General statements valid for all
physically admissible Hamiltonians, which govern the dynamics of the
quantum system, are subject to quantum no-go theorems. This can resolve
philosophical problems only by considering the physical properties
of quantum information without the need of explicitly solving any
equations. Specific statements valid for a concrete biomolecular Hamiltonian,
however, require quantum chemical methods for solving numerically
the many-body Schr\"{o}dinger equation. Currently available supercomputers
allow the derivation and experimental testing of quantum predictions
for physical systems consisting of several thousand atoms. Future
developments of faster supercomputers and better techniques for numerically
solving the Schr\"{o}dinger equation will enable even tighter integration
of quantum physics in consciousness research.

\section*{Conflict of interest}

The author declares that he has no conflict of interest.


\begin{thebibliography}{189}
\expandafter\ifx\csname natexlab\endcsname\relax\def\natexlab#1{#1}\fi
\providecommand{\url}[1]{\texttt{#1}}
\providecommand{\href}[2]{#2}
\providecommand{\path}[1]{#1}
\providecommand{\DOIprefix}{doi:}
\providecommand{\ArXivprefix}{arXiv:}
\providecommand{\URLprefix}{URL: }
\providecommand{\Pubmedprefix}{pmid:}
\providecommand{\doi}[1]{\href{http://dx.doi.org/#1}{\path{#1}}}
\providecommand{\Pubmed}[1]{\href{pmid:#1}{\path{#1}}}
\providecommand{\bibinfo}[2]{#2}
\ifx\xfnm\relax \def\xfnm[#1]{\unskip,\space#1}\fi
\bibitem[{Aspect(1999)}]{Aspect1999}
\bibinfo{author}{Aspect, A.} (\bibinfo{year}{1999}).
\newblock \bibinfo{title}{Bell's inequality test: more ideal than ever}.
\newblock {\it \bibinfo{journal}{Nature}\/},  {\it \bibinfo{volume}{398}\/},
  \bibinfo{pages}{189--190}. \DOIprefix\doi{10.1038/18296}.
\bibitem[{Azevedo et~al.(2009)Azevedo, Carvalho, Grinberg, Farfel, Ferretti,
  Leite, Filho, Lent \& Herculano-Houzel}]{Azevedo2009}
\bibinfo{author}{Azevedo, F. A.~C.}, \bibinfo{author}{Carvalho, L. R.~B.},
  \bibinfo{author}{Grinberg, L.~T.}, \bibinfo{author}{Farfel, J.~M.},
  \bibinfo{author}{Ferretti, R. E.~L.}, \bibinfo{author}{Leite, R. E.~P.},
  \bibinfo{author}{Filho, W.~J.}, \bibinfo{author}{Lent, R.}, \&
  \bibinfo{author}{Herculano-Houzel, S.} (\bibinfo{year}{2009}).
\newblock \bibinfo{title}{Equal numbers of neuronal and nonneuronal cells make
  the human brain an isometrically scaled-up primate brain}.
\newblock {\it \bibinfo{journal}{Journal of Comparative Neurology}\/},  {\it
  \bibinfo{volume}{513}\/}, \bibinfo{pages}{532--541}.
  \DOIprefix\doi{10.1002/cne.21974}.
\bibitem[{Baars(2005)}]{Baars2005}
\bibinfo{author}{Baars, B.~J.} (\bibinfo{year}{2005}).
\newblock \bibinfo{title}{Global workspace theory of consciousness: toward a
  cognitive neuroscience of human experience}.
\newblock {\it \bibinfo{journal}{Progress in Brain Research}\/},  {\it
  \bibinfo{volume}{150}\/}, \bibinfo{pages}{45--53}.
  \DOIprefix\doi{10.1016/S0079-6123(05)50004-9}.
\bibitem[{Baker(1998)}]{Baker1998}
\bibinfo{author}{Baker, G.} (\bibinfo{year}{1998}).
\newblock \bibinfo{title}{The private language argument}.
\newblock {\it \bibinfo{journal}{Language and Communication}\/},  {\it
  \bibinfo{volume}{18}\/}, \bibinfo{pages}{325--356}.
  \DOIprefix\doi{10.1016/S0271-5309(98)00010-X}.
\bibitem[{Barnum et~al.(1996)Barnum, Caves, Fuchs, Jozsa \&
  Schumacher}]{Barnum1996}
\bibinfo{author}{Barnum, H.}, \bibinfo{author}{Caves, C.~M.},
  \bibinfo{author}{Fuchs, C.~A.}, \bibinfo{author}{Jozsa, R.}, \&
  \bibinfo{author}{Schumacher, B.} (\bibinfo{year}{1996}).
\newblock \bibinfo{title}{Noncommuting mixed states cannot be broadcast}.
\newblock {\it \bibinfo{journal}{Physical Review Letters}\/},  {\it
  \bibinfo{volume}{76}\/}, \bibinfo{pages}{2818--2821}.
  \DOIprefix\doi{10.1103/PhysRevLett.76.2818}.
\bibitem[{Basile(2010)}]{Basile2010}
\bibinfo{author}{Basile, P.} (\bibinfo{year}{2010}).
\newblock \bibinfo{title}{It must be true -- but how can it be? {S}ome remarks
  on panpsychism and mental composition}.
\newblock {\it \bibinfo{journal}{Royal Institute of Philosophy Supplement}\/},
  {\it \bibinfo{volume}{67}\/}, \bibinfo{pages}{93--112}.
  \DOIprefix\doi{10.1017/S1358246110000044}.
\bibitem[{Beck \& Eccles(1992)}]{Beck1992}
\bibinfo{author}{Beck, F.}, \& \bibinfo{author}{Eccles, J.~C.}
  (\bibinfo{year}{1992}).
\newblock \bibinfo{title}{Quantum aspects of brain activity and the role of
  consciousness}.
\newblock {\it \bibinfo{journal}{Proceedings of the National Academy of
  Sciences}\/},  {\it \bibinfo{volume}{89}\/}, \bibinfo{pages}{11357--11361}.
  \DOIprefix\doi{10.1073/pnas.89.23.11357}.
\bibitem[{Belinfante(1980)}]{Belinfante1980}
\bibinfo{author}{Belinfante, F.~J.} (\bibinfo{year}{1980}).
\newblock \bibinfo{title}{Density matrix formulation of quantum theory and its
  physical interpretation}.
\newblock {\it \bibinfo{journal}{International Journal of Quantum
  Chemistry}\/},  {\it \bibinfo{volume}{17}\/}, \bibinfo{pages}{1--24}.
  \DOIprefix\doi{10.1002/qua.560170102}.
\bibitem[{Bosking et~al.(2017{\natexlab{a}})Bosking, Beauchamp \&
  Yoshor}]{Bosking2017b}
\bibinfo{author}{Bosking, W.~H.}, \bibinfo{author}{Beauchamp, M.~S.}, \&
  \bibinfo{author}{Yoshor, D.} (\bibinfo{year}{2017}{\natexlab{a}}).
\newblock \bibinfo{title}{Electrical stimulation of visual cortex: relevance
  for the development of visual cortical prosthetics}.
\newblock {\it \bibinfo{journal}{Annual Review of Vision Science}\/},  {\it
  \bibinfo{volume}{3}\/}, \bibinfo{pages}{141--166}.
  \DOIprefix\doi{10.1146/annurev-vision-111815-114525}.
\bibitem[{Bosking et~al.(2017{\natexlab{b}})Bosking, Sun, Ozker, Pei, Foster,
  Beauchamp \& Yoshor}]{Bosking2017a}
\bibinfo{author}{Bosking, W.~H.}, \bibinfo{author}{Sun, P.},
  \bibinfo{author}{Ozker, M.}, \bibinfo{author}{Pei, X.},
  \bibinfo{author}{Foster, B.~L.}, \bibinfo{author}{Beauchamp, M.~S.}, \&
  \bibinfo{author}{Yoshor, D.} (\bibinfo{year}{2017}{\natexlab{b}}).
\newblock \bibinfo{title}{Saturation in phosphene size with increasing current
  levels delivered to human visual cortex}.
\newblock {\it \bibinfo{journal}{Journal of Neuroscience}\/},  {\it
  \bibinfo{volume}{37}\/}, \bibinfo{pages}{7188--7197}.
  \DOIprefix\doi{10.1523/jneurosci.2896-16.2017}.
\bibitem[{Branco \& Staras(2009)}]{Branco2009}
\bibinfo{author}{Branco, T.}, \& \bibinfo{author}{Staras, K.}
  (\bibinfo{year}{2009}).
\newblock \bibinfo{title}{The probability of neurotransmitter release:
  variability and feedback control at single synapses}.
\newblock {\it \bibinfo{journal}{Nature Reviews Neuroscience}\/},  {\it
  \bibinfo{volume}{10}\/}, \bibinfo{pages}{373--383}.
  \DOIprefix\doi{10.1038/nrn2634}.
\bibitem[{Branstetter et~al.(2012)Branstetter, Finneran, Fletcher, Weisman \&
  Ridgway}]{Branstetter2012}
\bibinfo{author}{Branstetter, B.~K.}, \bibinfo{author}{Finneran, J.~J.},
  \bibinfo{author}{Fletcher, E.~A.}, \bibinfo{author}{Weisman, B.~C.}, \&
  \bibinfo{author}{Ridgway, S.~H.} (\bibinfo{year}{2012}).
\newblock \bibinfo{title}{Dolphins can maintain vigilant behavior through
  echolocation for 15 days without interruption or cognitive impairment}.
\newblock {\it \bibinfo{journal}{PloS One}\/},  {\it \bibinfo{volume}{7}\/},
  \bibinfo{pages}{e47478}. \DOIprefix\doi{10.1371/journal.pone.0047478}.
\bibitem[{Bucher et~al.(2010)Bucher, Guidoni, Carloni \&
  Rothlisberger}]{Bucher2010a}
\bibinfo{author}{Bucher, D.}, \bibinfo{author}{Guidoni, L.},
  \bibinfo{author}{Carloni, P.}, \& \bibinfo{author}{Rothlisberger, U.}
  (\bibinfo{year}{2010}).
\newblock \bibinfo{title}{Coordination numbers of {K}$^+$ and {N}a$^+$ ions
  inside the selectivity filter of {KcsA} potassium channel: insights from
  first principles molecular dynamics}.
\newblock {\it \bibinfo{journal}{Biophysical Journal}\/},  {\it
  \bibinfo{volume}{98}\/}, \bibinfo{pages}{L47--L49}.
  \DOIprefix\doi{10.1016/j.bpj.2010.01.064}.
\bibitem[{Bucher \& Rothlisberger(2010)}]{Bucher2010b}
\bibinfo{author}{Bucher, D.}, \& \bibinfo{author}{Rothlisberger, U.}
  (\bibinfo{year}{2010}).
\newblock \bibinfo{title}{Molecular simulations of ion channels: a quantum
  chemist's perspective}.
\newblock {\it \bibinfo{journal}{Journal of General Physiology}\/},  {\it
  \bibinfo{volume}{135}\/}, \bibinfo{pages}{549--554}.
  \DOIprefix\doi{10.1085/jgp.201010404}.
\bibitem[{Busch(1997)}]{Busch1997}
\bibinfo{author}{Busch, P.} (\bibinfo{year}{1997}).
\newblock \bibinfo{title}{Is the quantum state (an) observable?}
\newblock {\it \bibinfo{journal}{Boston Studies in the Philosophy of
  Science}\/},  {\it \bibinfo{volume}{194}\/}, \bibinfo{pages}{61--70}.
  \DOIprefix\doi{10.1007/978-94-017-2732-7_5}.
\bibitem[{Busch et~al.(2016)Busch, Lahti, Pellonp\"{a}\"{a} \&
  Ylinen}]{Busch2016}
\bibinfo{author}{Busch, P.}, \bibinfo{author}{Lahti, P.},
  \bibinfo{author}{Pellonp\"{a}\"{a}, J.-P.}, \& \bibinfo{author}{Ylinen, K.}
  (\bibinfo{year}{2016}).
\newblock {\it \bibinfo{title}{Quantum Measurement}\/}.
\newblock Theoretical and Mathematical Physics.
\newblock \bibinfo{address}{Cham}: \bibinfo{publisher}{Springer}.
\newblock \DOIprefix\doi{10.1007/978-3-319-43389-9}.
\bibitem[{Callahan \& Roux(2018)}]{Callahan2018}
\bibinfo{author}{Callahan, K.~M.}, \& \bibinfo{author}{Roux, B.}
  (\bibinfo{year}{2018}).
\newblock \bibinfo{title}{Molecular dynamics of ion conduction through the
  selectivity filter of the {N}a$_{V}${A}b sodium channel}.
\newblock {\it \bibinfo{journal}{Journal of Physical Chemistry B}\/},  {\it
  \bibinfo{volume}{122}\/}, \bibinfo{pages}{10126--10142}.
  \DOIprefix\doi{10.1021/acs.jpcb.8b09678}.
\bibitem[{Campbell \& Bickhard(2011)}]{Campbell2011}
\bibinfo{author}{Campbell, R.~J.}, \& \bibinfo{author}{Bickhard, M.~H.}
  (\bibinfo{year}{2011}).
\newblock \bibinfo{title}{Physicalism, emergence and downward causation}.
\newblock {\it \bibinfo{journal}{Axiomathes}\/},  {\it \bibinfo{volume}{21}\/},
  \bibinfo{pages}{33--56}. \DOIprefix\doi{10.1007/s10516-010-9128-6}.
\bibitem[{Chalmers(1995)}]{Chalmers1995}
\bibinfo{author}{Chalmers, D.~J.} (\bibinfo{year}{1995}).
\newblock \bibinfo{title}{Facing up to the problem of consciousness}.
\newblock {\it \bibinfo{journal}{Journal of Consciousness Studies}\/},  {\it
  \bibinfo{volume}{2}\/}, \bibinfo{pages}{200--219}.
\bibitem[{Chalmers(1996)}]{Chalmers1996}
\bibinfo{author}{Chalmers, D.~J.} (\bibinfo{year}{1996}).
\newblock {\it \bibinfo{title}{The Conscious Mind: In Search of a Fundamental
  Theory}\/}.
\newblock Philosophy of Mind.
\newblock \bibinfo{address}{Oxford}: \bibinfo{publisher}{Oxford University
  Press}.
\bibitem[{Chan et~al.(2019)Chan, Timmermann, Baldi, Moore, Lyons, Lee,
  Kalsbeek, Petersen, Rautenbach, F\"{o}rtsch, Bornman \& Hayes}]{Chan2019}
\bibinfo{author}{Chan, E. K.~F.}, \bibinfo{author}{Timmermann, A.},
  \bibinfo{author}{Baldi, B.~F.}, \bibinfo{author}{Moore, A.~E.},
  \bibinfo{author}{Lyons, R.~J.}, \bibinfo{author}{Lee, S.-S.},
  \bibinfo{author}{Kalsbeek, A. M.~F.}, \bibinfo{author}{Petersen, D.~C.},
  \bibinfo{author}{Rautenbach, H.}, \bibinfo{author}{F\"{o}rtsch, H. E.~A.},
  \bibinfo{author}{Bornman, M. S.~R.}, \& \bibinfo{author}{Hayes, V.~M.}
  (\bibinfo{year}{2019}).
\newblock \bibinfo{title}{Human origins in a southern {A}frican palaeo-wetland
  and first migrations}.
\newblock {\it \bibinfo{journal}{Nature}\/},  {\it \bibinfo{volume}{575}\/},
  \bibinfo{pages}{185--189}. \DOIprefix\doi{10.1038/s41586-019-1714-1}.
\bibitem[{Chancey et~al.(1992)Chancey, George \& Marshall}]{Chancey1992}
\bibinfo{author}{Chancey, C.~C.}, \bibinfo{author}{George, S.~A.}, \&
  \bibinfo{author}{Marshall, P.~J.} (\bibinfo{year}{1992}).
\newblock \bibinfo{title}{Calculations of quantum tunnelling between closed and
  open states of sodium channels}.
\newblock {\it \bibinfo{journal}{Journal of Biological Physics}\/},  {\it
  \bibinfo{volume}{18}\/}, \bibinfo{pages}{307--321}.
  \DOIprefix\doi{10.1007/bf00419427}.
\bibitem[{Chen et~al.(2017)Chen, Chen \& Tseng}]{Chen2017}
\bibinfo{author}{Chen, C.~Y.}, \bibinfo{author}{Chen, C.~J.}, \&
  \bibinfo{author}{Tseng, Y.~C.} (\bibinfo{year}{2017}).
\newblock \bibinfo{title}{A case report of transient cortical blindness after
  angiography}.
\newblock {\it \bibinfo{journal}{The Neurologist}\/},  {\it
  \bibinfo{volume}{22}\/}, \bibinfo{pages}{82--84}.
  \DOIprefix\doi{10.1097/nrl.0000000000000115}.
\bibitem[{Coleman(2012)}]{Coleman2012}
\bibinfo{author}{Coleman, S.} (\bibinfo{year}{2012}).
\newblock \bibinfo{title}{Mental chemistry: combination for panpsychists}.
\newblock {\it \bibinfo{journal}{Dialectica}\/},  {\it \bibinfo{volume}{66}\/},
  \bibinfo{pages}{137--166}.
  \DOIprefix\doi{doi:10.1111/j.1746-8361.2012.01293.x}.
\bibitem[{Conway \& Kochen(2006)}]{Conway2006}
\bibinfo{author}{Conway, J.~H.}, \& \bibinfo{author}{Kochen, S.~B.}
  (\bibinfo{year}{2006}).
\newblock \bibinfo{title}{The free will theorem}.
\newblock {\it \bibinfo{journal}{Foundations of Physics}\/},  {\it
  \bibinfo{volume}{36}\/}, \bibinfo{pages}{1441--1473}.
  \DOIprefix\doi{10.1007/s10701-006-9068-6}.
\bibitem[{Conway \& Kochen(2009)}]{Conway2009}
\bibinfo{author}{Conway, J.~H.}, \& \bibinfo{author}{Kochen, S.~B.}
  (\bibinfo{year}{2009}).
\newblock \bibinfo{title}{The strong free will theorem}.
\newblock {\it \bibinfo{journal}{Notices of the AMS}\/},  {\it
  \bibinfo{volume}{56}\/}, \bibinfo{pages}{226--232}.
\bibitem[{Crane \& Brewer(1995)}]{Crane1995}
\bibinfo{author}{Crane, T.}, \& \bibinfo{author}{Brewer, B.}
  (\bibinfo{year}{1995}).
\newblock \bibinfo{title}{Mental causation}.
\newblock {\it \bibinfo{journal}{Proceedings of the Aristotelian Society,
  Supplementary Volumes}\/},  {\it \bibinfo{volume}{69}\/},
  \bibinfo{pages}{211--253}.
\bibitem[{Darwin(2006)}]{Darwin2006}
\bibinfo{author}{Darwin, C.} (\bibinfo{year}{2006}).
\newblock {\it \bibinfo{title}{From So Simple a Beginning: The Four Great Books
  of Charles Darwin (The Voyage of the Beagle, On the Origin of Species, The
  Descent of Man, The Expression of the Emotions in Man and Animals)}\/}.
\newblock \bibinfo{address}{New York}: \bibinfo{publisher}{W. W. Norton \&
  Company}.
\bibitem[{Dawkins(2004)}]{Dawkins2004}
\bibinfo{author}{Dawkins, R.} (\bibinfo{year}{2004}).
\newblock {\it \bibinfo{title}{The Ancestor's Tale: A Pilgrimage to the Dawn of
  Life}\/}.
\newblock \bibinfo{address}{London}: \bibinfo{publisher}{Weidenfeld \&
  Nicolson}.
\bibitem[{Dennett(1991)}]{Dennett1991}
\bibinfo{author}{Dennett, D.~C.} (\bibinfo{year}{1991}).
\newblock {\it \bibinfo{title}{Consciousness Explained}\/}.
\newblock \bibinfo{address}{New York}: \bibinfo{publisher}{Back Bay Books}.
\bibitem[{Destexhe \& Contreras(2006)}]{Destexhe2006}
\bibinfo{author}{Destexhe, A.}, \& \bibinfo{author}{Contreras, D.}
  (\bibinfo{year}{2006}).
\newblock \bibinfo{title}{Neuronal computations with stochastic network
  states}.
\newblock {\it \bibinfo{journal}{Science}\/},  {\it \bibinfo{volume}{314}\/},
  \bibinfo{pages}{85--90}. \DOIprefix\doi{10.1126/science.1127241}.
\bibitem[{Dirac(1967)}]{Dirac1967}
\bibinfo{author}{Dirac, P. A.~M.} (\bibinfo{year}{1967}).
\newblock {\it \bibinfo{title}{The Principles of Quantum Mechanics}\/}.
\newblock (\bibinfo{edition}{4th} ed.).
\newblock \bibinfo{address}{Oxford}: \bibinfo{publisher}{Oxford University
  Press}.
\bibitem[{Dobelle(2000)}]{Dobelle2000}
\bibinfo{author}{Dobelle, W.~H.} (\bibinfo{year}{2000}).
\newblock \bibinfo{title}{Artificial vision for the blind by connecting a
  television camera to the visual cortex}.
\newblock {\it \bibinfo{journal}{ASAIO Journal}\/},  {\it
  \bibinfo{volume}{46}\/}, \bibinfo{pages}{3--9}.
  \DOIprefix\doi{10.1097/00002480-200001000-00002}.
\bibitem[{Dobrunz \& Stevens(1997)}]{Dobrunz1997}
\bibinfo{author}{Dobrunz, L.~E.}, \& \bibinfo{author}{Stevens, C.~F.}
  (\bibinfo{year}{1997}).
\newblock \bibinfo{title}{Heterogeneity of release probability, facilitation,
  and depletion at central synapses}.
\newblock {\it \bibinfo{journal}{Neuron}\/},  {\it \bibinfo{volume}{18}\/},
  \bibinfo{pages}{995--1008}. \DOIprefix\doi{10.1016/S0896-6273(00)80338-4}.
\bibitem[{Elga(2004)}]{Elga2004}
\bibinfo{author}{Elga, A.} (\bibinfo{year}{2004}).
\newblock \bibinfo{title}{Defeating {D}r. {E}vil with self-locating belief}.
\newblock {\it \bibinfo{journal}{Philosophy and Phenomenological Research}\/},
  {\it \bibinfo{volume}{69}\/}, \bibinfo{pages}{383--396}.
  \DOIprefix\doi{10.1111/j.1933-1592.2004.tb00400.x}.
\bibitem[{Elgabarty et~al.(2020)Elgabarty, Kampfrath, Bonthuis, Balos,
  Kaliannan, Loche, Netz, Wolf, K\"{u}hne \& Sajadi}]{Elgabarty2020}
\bibinfo{author}{Elgabarty, H.}, \bibinfo{author}{Kampfrath, T.},
  \bibinfo{author}{Bonthuis, D.~J.}, \bibinfo{author}{Balos, V.},
  \bibinfo{author}{Kaliannan, N.~K.}, \bibinfo{author}{Loche, P.},
  \bibinfo{author}{Netz, R.~R.}, \bibinfo{author}{Wolf, M.},
  \bibinfo{author}{K\"{u}hne, T.~D.}, \& \bibinfo{author}{Sajadi, M.}
  (\bibinfo{year}{2020}).
\newblock \bibinfo{title}{Energy transfer within the hydrogen bonding network
  of water following resonant terahertz excitation}.
\newblock {\it \bibinfo{journal}{Science Advances}\/},  {\it
  \bibinfo{volume}{6}\/}, \bibinfo{pages}{eaay7074}.
  \DOIprefix\doi{10.1126/sciadv.aay7074}.
\bibitem[{Fayer(2012)}]{Fayer2012}
\bibinfo{author}{Fayer, M.~D.} (\bibinfo{year}{2012}).
\newblock \bibinfo{title}{Dynamics of water interacting with interfaces,
  molecules, and ions}.
\newblock {\it \bibinfo{journal}{Accounts of Chemical Research}\/},  {\it
  \bibinfo{volume}{45}\/}, \bibinfo{pages}{3--14}.
  \DOIprefix\doi{10.1021/ar2000088}.
\bibitem[{Fayngold \& Fayngold(2013)}]{Fayngold2013}
\bibinfo{author}{Fayngold, M.}, \& \bibinfo{author}{Fayngold, V.}
  (\bibinfo{year}{2013}).
\newblock {\it \bibinfo{title}{Quantum Mechanics and Quantum Information}\/}.
\newblock \bibinfo{address}{Weinheim, Germany}: \bibinfo{publisher}{Wiley-VCH}.
\bibitem[{Feynman(1948)}]{Feynman1948}
\bibinfo{author}{Feynman, R.~P.} (\bibinfo{year}{1948}).
\newblock \bibinfo{title}{Space-time approach to non-relativistic quantum
  mechanics}.
\newblock {\it \bibinfo{journal}{Reviews of Modern Physics}\/},  {\it
  \bibinfo{volume}{20}\/}, \bibinfo{pages}{367--387}.
  \DOIprefix\doi{10.1103/RevModPhys.20.367}.
\bibitem[{Feynman(2014)}]{Feynman2014}
\bibinfo{author}{Feynman, R.~P.} (\bibinfo{year}{2014}).
\newblock {\it \bibinfo{title}{QED: The Strange Theory of Light and Matter}\/}.
\newblock \bibinfo{address}{Princeton}: \bibinfo{publisher}{Princeton
  University Press}.
\bibitem[{Feynman et~al.(2013)Feynman, Leighton \& Sands}]{Feynman2013}
\bibinfo{author}{Feynman, R.~P.}, \bibinfo{author}{Leighton, R.~B.}, \&
  \bibinfo{author}{Sands, M.~L.} (\bibinfo{year}{2013}).
\newblock {\it \bibinfo{title}{The Feynman Lectures on Physics. Volume III.
  Quantum Mechanics}\/}.
\newblock \bibinfo{address}{Pasadena, California}:
  \bibinfo{publisher}{California Institute of Technology}.
\newblock \URLprefix \url{http://www.feynmanlectures.caltech.edu/}.
\bibitem[{Flood et~al.(2019)Flood, Boiteux, Lev, Vorobyov \& Allen}]{Flood2019}
\bibinfo{author}{Flood, E.}, \bibinfo{author}{Boiteux, C.},
  \bibinfo{author}{Lev, B.}, \bibinfo{author}{Vorobyov, I.}, \&
  \bibinfo{author}{Allen, T.~W.} (\bibinfo{year}{2019}).
\newblock \bibinfo{title}{Atomistic simulations of membrane ion channel
  conduction, gating, and modulation}.
\newblock {\it \bibinfo{journal}{Chemical Reviews}\/},  {\it
  \bibinfo{volume}{119}\/}, \bibinfo{pages}{7737--7832}.
  \DOIprefix\doi{10.1021/acs.chemrev.8b00630}.
\bibitem[{Fodor(1981)}]{Fodor1981}
\bibinfo{author}{Fodor, J.~A.} (\bibinfo{year}{1981}).
\newblock \bibinfo{title}{The mind--body problem}.
\newblock {\it \bibinfo{journal}{Scientific American}\/},  {\it
  \bibinfo{volume}{244}\/}, \bibinfo{pages}{114--123}.
  \DOIprefix\doi{10.1038/scientificamerican0181-114}.
\bibitem[{Foley \& Mazziotti(2013)}]{Foley2013}
\bibinfo{author}{Foley, J.~J.}, \& \bibinfo{author}{Mazziotti, D.~A.}
  (\bibinfo{year}{2013}).
\newblock \bibinfo{title}{Cage versus prism: electronic energies of the water
  hexamer}.
\newblock {\it \bibinfo{journal}{Journal of Physical Chemistry A}\/},  {\it
  \bibinfo{volume}{117}\/}, \bibinfo{pages}{6712--6716}.
  \DOIprefix\doi{10.1021/jp405739d}.
\bibitem[{Garcia et~al.(2010)Garcia, Kolesky \& Jenkins}]{Garcia2010}
\bibinfo{author}{Garcia, P.~S.}, \bibinfo{author}{Kolesky, S.~E.}, \&
  \bibinfo{author}{Jenkins, A.} (\bibinfo{year}{2010}).
\newblock \bibinfo{title}{General anesthetic actions on {GABA}$_\textrm{A}$
  receptors}.
\newblock {\it \bibinfo{journal}{Current Neuropharmacology}\/},  {\it
  \bibinfo{volume}{8}\/}, \bibinfo{pages}{2--9}.
  \DOIprefix\doi{10.2174/157015910790909502}.
\bibitem[{Gasparini et~al.(2004)Gasparini, Migliore \& Magee}]{Gasparini2004}
\bibinfo{author}{Gasparini, S.}, \bibinfo{author}{Migliore, M.}, \&
  \bibinfo{author}{Magee, J.~C.} (\bibinfo{year}{2004}).
\newblock \bibinfo{title}{On the initiation and propagation of dendritic spikes
  in {CA1} pyramidal neurons}.
\newblock {\it \bibinfo{journal}{Journal of Neuroscience}\/},  {\it
  \bibinfo{volume}{24}\/}, \bibinfo{pages}{11046--11056}.
  \DOIprefix\doi{10.1523/jneurosci.2520-04.2004}.
\bibitem[{Gazzaniga(2002)}]{Gazzaniga2002}
\bibinfo{author}{Gazzaniga, M.~S.} (\bibinfo{year}{2002}).
\newblock \bibinfo{title}{The split brain revisited}.
\newblock {\it \bibinfo{journal}{Scientific American}\/},  {\it
  \bibinfo{volume}{12}\/}, \bibinfo{pages}{26--31}.
\bibitem[{Gazzaniga et~al.(1962)Gazzaniga, Bogen \& Sperry}]{Gazzaniga1962}
\bibinfo{author}{Gazzaniga, M.~S.}, \bibinfo{author}{Bogen, J.~E.}, \&
  \bibinfo{author}{Sperry, R.~W.} (\bibinfo{year}{1962}).
\newblock \bibinfo{title}{Some functional effects of sectioning the cerebral
  commissures in man}.
\newblock {\it \bibinfo{journal}{Proceedings of the National Academy of
  Sciences}\/},  {\it \bibinfo{volume}{48}\/}, \bibinfo{pages}{1765--1769}.
\bibitem[{Gazzaniga \& Sperry(1967)}]{Gazzaniga1967}
\bibinfo{author}{Gazzaniga, M.~S.}, \& \bibinfo{author}{Sperry, R.~W.}
  (\bibinfo{year}{1967}).
\newblock \bibinfo{title}{Language after section of the cerebral commissures}.
\newblock {\it \bibinfo{journal}{Brain}\/},  {\it \bibinfo{volume}{90}\/},
  \bibinfo{pages}{131--148}. \DOIprefix\doi{10.1093/brain/90.1.131}.
\bibitem[{Georgescu(2014)}]{Georgescu2014}
\bibinfo{author}{Georgescu, I.} (\bibinfo{year}{2014}).
\newblock \bibinfo{title}{Bell's theorem: Closing the loopholes}.
\newblock {\it \bibinfo{journal}{Nature Physics}\/},  {\it
  \bibinfo{volume}{10}\/}, \bibinfo{pages}{248}.
  \DOIprefix\doi{10.1038/nphys2945}.
\bibitem[{Georgiev(2013)}]{Georgiev2013}
\bibinfo{author}{Georgiev, D.~D.} (\bibinfo{year}{2013}).
\newblock \bibinfo{title}{Quantum no-go theorems and consciousness}.
\newblock {\it \bibinfo{journal}{Axiomathes}\/},  {\it \bibinfo{volume}{23}\/},
  \bibinfo{pages}{683--695}. \DOIprefix\doi{10.1007/s10516-012-9204-1}.
\bibitem[{Georgiev(2015)}]{Georgiev2015}
\bibinfo{author}{Georgiev, D.~D.} (\bibinfo{year}{2015}).
\newblock \bibinfo{title}{Monte {C}arlo simulation of quantum {Z}eno effect in
  the brain}.
\newblock {\it \bibinfo{journal}{International Journal of Modern Physics B}\/},
   {\it \bibinfo{volume}{29}\/}, \bibinfo{pages}{1550039}.
  \DOIprefix\doi{10.1142/S0217979215500393}.
\bibitem[{Georgiev(2017)}]{Georgiev2017}
\bibinfo{author}{Georgiev, D.~D.} (\bibinfo{year}{2017}).
\newblock {\it \bibinfo{title}{Quantum Information and Consciousness: A Gentle
  Introduction}\/}.
\newblock \bibinfo{address}{Boca Raton}: \bibinfo{publisher}{CRC Press}.
\newblock \DOIprefix\doi{10.1201/9780203732519}.
\bibitem[{Georgiev(2019)}]{Georgiev2019a}
\bibinfo{author}{Georgiev, D.~D.} (\bibinfo{year}{2019}).
\newblock \bibinfo{title}{Chalmers' principle of organizational invariance
  makes consciousness fundamental but meaningless spectator of its own drama}.
\newblock {\it \bibinfo{journal}{Activitas Nervosa Superior}\/},  {\it
  \bibinfo{volume}{61}\/}, \bibinfo{pages}{159--164}.
  \DOIprefix\doi{10.1007/s41470-019-00062-z}.
\bibitem[{Georgiev(2020)}]{Georgiev2020}
\bibinfo{author}{Georgiev, D.~D.} (\bibinfo{year}{2020}).
\newblock \bibinfo{title}{Inner privacy of conscious experiences and quantum
  information}.
\newblock {\it \bibinfo{journal}{Biosystems}\/},  {\it
  \bibinfo{volume}{187}\/}, \bibinfo{pages}{104051}.
  \DOIprefix\doi{10.1016/j.biosystems.2019.104051}.
\bibitem[{Georgiev \& Cohen(2018)}]{Georgiev2018a}
\bibinfo{author}{Georgiev, D.~D.}, \& \bibinfo{author}{Cohen, E.}
  (\bibinfo{year}{2018}).
\newblock \bibinfo{title}{Probing finite coarse-grained virtual {F}eynman
  histories with sequential weak values}.
\newblock {\it \bibinfo{journal}{Physical Review A}\/},  {\it
  \bibinfo{volume}{97}\/}, \bibinfo{pages}{052102}.
  \DOIprefix\doi{10.1103/PhysRevA.97.052102}.
\bibitem[{Georgiev \& Glazebrook(2007)}]{Georgiev2007}
\bibinfo{author}{Georgiev, D.~D.}, \& \bibinfo{author}{Glazebrook, J.~F.}
  (\bibinfo{year}{2007}).
\newblock \bibinfo{title}{Subneuronal processing of information by solitary
  waves and stochastic processes}.
\newblock In \bibinfo{editor}{S.~E. Lyshevski} (Ed.), {\it
  \bibinfo{booktitle}{Nano and Molecular Electronics Handbook}\/} Nano and
  Microengineering Series chapter~\bibinfo{chapter}{17}. (pp.
  \bibinfo{pages}{1--41}).
\newblock \bibinfo{address}{Boca Raton}: \bibinfo{publisher}{CRC Press}.
\newblock \DOIprefix\doi{10.1201/9781315221670}.
\bibitem[{Georgiev \& Glazebrook(2010)}]{Georgiev2010}
\bibinfo{author}{Georgiev, D.~D.}, \& \bibinfo{author}{Glazebrook, J.~F.}
  (\bibinfo{year}{2010}).
\newblock \bibinfo{title}{{SNARE} proteins as molecular masters of
  interneuronal communication}.
\newblock {\it \bibinfo{journal}{Biomedical Reviews}\/},  {\it
  \bibinfo{volume}{21}\/}, \bibinfo{pages}{17--23}.
  \DOIprefix\doi{10.14748/bmr.v21.43}.
\bibitem[{Georgiev \& Glazebrook(2012)}]{Georgiev2012}
\bibinfo{author}{Georgiev, D.~D.}, \& \bibinfo{author}{Glazebrook, J.~F.}
  (\bibinfo{year}{2012}).
\newblock \bibinfo{title}{Quasiparticle tunneling in neurotransmitter release}.
\newblock In \bibinfo{editor}{W.~A. Goddard~III}, \bibinfo{editor}{D.~Brenner},
  \bibinfo{editor}{S.~E. Lyshevski}, \& \bibinfo{editor}{G.~J. Iafrate} (Eds.),
  {\it \bibinfo{booktitle}{Handbook of Nanoscience, Engineering, and
  Technology}\/} Electrical Engineering Handbook chapter~\bibinfo{chapter}{30}.
  (pp. \bibinfo{pages}{983--1016}).
\newblock \bibinfo{address}{Boca Raton}: \bibinfo{publisher}{CRC Press}.
  (\bibinfo{edition}{3rd} ed.).
\newblock \DOIprefix\doi{10.1201/b11930-37}.
\bibitem[{Georgiev \& Glazebrook(2014)}]{Georgiev2014}
\bibinfo{author}{Georgiev, D.~D.}, \& \bibinfo{author}{Glazebrook, J.~F.}
  (\bibinfo{year}{2014}).
\newblock \bibinfo{title}{Quantum interactive dualism: From {B}eck and {E}ccles
  tunneling model of exocytosis to molecular biology of {SNARE} zipping}.
\newblock {\it \bibinfo{journal}{Biomedical Reviews}\/},  {\it
  \bibinfo{volume}{25}\/}, \bibinfo{pages}{15--24}.
  \DOIprefix\doi{10.14748/bmr.v25.1038}.
\bibitem[{Georgiev \& Glazebrook(2018)}]{Georgiev2018b}
\bibinfo{author}{Georgiev, D.~D.}, \& \bibinfo{author}{Glazebrook, J.~F.}
  (\bibinfo{year}{2018}).
\newblock \bibinfo{title}{The quantum physics of synaptic communication via the
  {SNARE} protein complex}.
\newblock {\it \bibinfo{journal}{Progress in Biophysics and Molecular
  Biology}\/},  {\it \bibinfo{volume}{135}\/}, \bibinfo{pages}{16--29}.
  \DOIprefix\doi{10.1016/j.pbiomolbio.2018.01.006}.
\bibitem[{Georgiev \& Glazebrook(2019{\natexlab{a}})}]{Georgiev2019b}
\bibinfo{author}{Georgiev, D.~D.}, \& \bibinfo{author}{Glazebrook, J.~F.}
  (\bibinfo{year}{2019}{\natexlab{a}}).
\newblock \bibinfo{title}{On the quantum dynamics of {D}avydov solitons in
  protein $\alpha$-helices}.
\newblock {\it \bibinfo{journal}{Physica A: Statistical Mechanics and its
  Applications}\/},  {\it \bibinfo{volume}{517}\/}, \bibinfo{pages}{257--269}.
  \DOIprefix\doi{10.1016/j.physa.2018.11.026}.
\bibitem[{Georgiev \& Glazebrook(2019{\natexlab{b}})}]{Georgiev2019c}
\bibinfo{author}{Georgiev, D.~D.}, \& \bibinfo{author}{Glazebrook, J.~F.}
  (\bibinfo{year}{2019}{\natexlab{b}}).
\newblock \bibinfo{title}{Quantum tunneling of {D}avydov solitons through
  massive barriers}.
\newblock {\it \bibinfo{journal}{Chaos, Solitons and Fractals}\/},  {\it
  \bibinfo{volume}{123}\/}, \bibinfo{pages}{275--293}.
  \DOIprefix\doi{10.1016/j.chaos.2019.04.013}.
\bibitem[{Georgiev \& Glazebrook(2020{\natexlab{a}})}]{Georgiev2020c}
\bibinfo{author}{Georgiev, D.~D.}, \& \bibinfo{author}{Glazebrook, J.~F.}
  (\bibinfo{year}{2020}{\natexlab{a}}).
\newblock \bibinfo{title}{Launching of {D}avydov solitons in protein
  $\alpha$-helix spines}.
\newblock {\it \bibinfo{journal}{Physica E: Low-dimensional Systems and
  Nanostructures}\/},  {\it \bibinfo{volume}{124}\/}, \bibinfo{pages}{114332}.
  \DOIprefix\doi{10.1016/j.physe.2020.114332}.
\bibitem[{Georgiev \& Glazebrook(2020{\natexlab{b}})}]{Georgiev2020b}
\bibinfo{author}{Georgiev, D.~D.}, \& \bibinfo{author}{Glazebrook, J.~F.}
  (\bibinfo{year}{2020}{\natexlab{b}}).
\newblock \bibinfo{title}{Quantum transport and utilization of free energy in
  protein $\alpha$-helices}.
\newblock {\it \bibinfo{journal}{Advances in Quantum Chemistry}\/},  {\it
  \bibinfo{volume}{82}\/}, \bibinfo{pages}{253--300}. \DOIprefix\doi{10.1016/bs.aiq.2020.02.001}.
\bibitem[{Gillig(2009)}]{Gillig2009}
\bibinfo{author}{Gillig, P.~M.} (\bibinfo{year}{2009}).
\newblock \bibinfo{title}{Dissociative identity disorder: a controversial
  diagnosis}.
\newblock {\it \bibinfo{journal}{Psychiatry}\/},  {\it \bibinfo{volume}{6}\/},
  \bibinfo{pages}{24--29}.
\bibitem[{Giraudo et~al.(2006)Giraudo, Eng, Melia \& Rothman}]{Giraudo2006}
\bibinfo{author}{Giraudo, C.~G.}, \bibinfo{author}{Eng, W.~S.},
  \bibinfo{author}{Melia, T.~J.}, \& \bibinfo{author}{Rothman, J.~E.}
  (\bibinfo{year}{2006}).
\newblock \bibinfo{title}{A clamping mechanism involved in {SNARE}-dependent
  exocytosis}.
\newblock {\it \bibinfo{journal}{Science}\/},  {\it \bibinfo{volume}{313}\/},
  \bibinfo{pages}{676--680}. \DOIprefix\doi{10.1126/science.1129450}.
\bibitem[{Glowatzki \& Fuchs(2002)}]{Glowatzki2002}
\bibinfo{author}{Glowatzki, E.}, \& \bibinfo{author}{Fuchs, P.~A.}
  (\bibinfo{year}{2002}).
\newblock \bibinfo{title}{Transmitter release at the hair cell ribbon synapse}.
\newblock {\it \bibinfo{journal}{Nature Neuroscience}\/},  {\it
  \bibinfo{volume}{5}\/}, \bibinfo{pages}{147}. \DOIprefix\doi{10.1038/nn796}.
\bibitem[{Goldman-Rakic(2002)}]{Goldman-Rakic2002}
\bibinfo{author}{Goldman-Rakic, P.~S.} (\bibinfo{year}{2002}).
\newblock \bibinfo{title}{The ``psychic cell'' of {R}am\'{o}n y {C}ajal}.
\newblock {\it \bibinfo{journal}{Progress in Brain Research}\/},  {\it
  \bibinfo{volume}{136}\/}, \bibinfo{pages}{427--434}.
  \DOIprefix\doi{10.1016/S0079-6123(02)36035-7}.
\bibitem[{de~Gosson(2018)}]{deGosson2018}
\bibinfo{author}{de~Gosson, M.~A.} (\bibinfo{year}{2018}).
\newblock \bibinfo{title}{Quantum harmonic analysis of the density matrix}.
\newblock {\it \bibinfo{journal}{Quanta}\/},  {\it \bibinfo{volume}{7}\/},
  \bibinfo{pages}{74--110}. \DOIprefix\doi{10.12743/quanta.v7i1.74}.
\bibitem[{Gudder(2020)}]{Gudder2}
\bibinfo{author}{Gudder, S.} (\bibinfo{year}{2020}).
\newblock \bibinfo{title}{A theory of entanglement}.
\newblock {\it \bibinfo{journal}{Quanta}\/},  {\it \bibinfo{volume}{9}\/},
  \bibinfo{pages}{7--15}. \DOIprefix\doi{10.12743/quanta.v9i1.115}.
\bibitem[{Hadid \& Lepore(2017)}]{Hadid2017}
\bibinfo{author}{Hadid, V.}, \& \bibinfo{author}{Lepore, F.}
  (\bibinfo{year}{2017}).
\newblock \bibinfo{title}{From cortical blindness to conscious visual
  perception: theories on neuronal networks and visual training strategies}.
\newblock {\it \bibinfo{journal}{Frontiers in Systems Neuroscience}\/},  {\it
  \bibinfo{volume}{11}\/}, \bibinfo{pages}{64}.
  \DOIprefix\doi{10.3389/fnsys.2017.00064}.
\bibitem[{Hameroff \& Penrose(2014)}]{Hameroff2014}
\bibinfo{author}{Hameroff, S.~R.}, \& \bibinfo{author}{Penrose, R.}
  (\bibinfo{year}{2014}).
\newblock \bibinfo{title}{Consciousness in the universe: a review of the `{Orch
  OR}' theory}.
\newblock {\it \bibinfo{journal}{Physics of Life Reviews}\/},  {\it
  \bibinfo{volume}{11}\/}, \bibinfo{pages}{39--78}.
  \DOIprefix\doi{10.1016/j.plrev.2013.08.002}.
\bibitem[{Hayashi et~al.(2015)Hayashi, Ishizaka, Kawachi, Kimura \&
  Ogawa}]{Hayashi2015}
\bibinfo{author}{Hayashi, M.}, \bibinfo{author}{Ishizaka, S.},
  \bibinfo{author}{Kawachi, A.}, \bibinfo{author}{Kimura, G.}, \&
  \bibinfo{author}{Ogawa, T.} (\bibinfo{year}{2015}).
\newblock {\it \bibinfo{title}{Introduction to Quantum Information Science}\/}.
\newblock Graduate Texts in Physics.
\newblock \bibinfo{address}{Berlin}: \bibinfo{publisher}{Springer}.
\newblock \DOIprefix\doi{10.1007/978-3-662-43502-1}.
\bibitem[{Helgaker et~al.(2013)Helgaker, Jorgensen \& Olsen}]{Helgaker2013}
\bibinfo{author}{Helgaker, T.}, \bibinfo{author}{Jorgensen, P.}, \&
  \bibinfo{author}{Olsen, J.} (\bibinfo{year}{2013}).
\newblock {\it \bibinfo{title}{Molecular Electronic-Structure Theory}\/}.
\newblock \bibinfo{address}{Chichester}: \bibinfo{publisher}{John Wiley \&
  Sons}.
\bibitem[{Herring et~al.(2009)Herring, Xie, Marks \& Fox}]{Herring2009}
\bibinfo{author}{Herring, B.~E.}, \bibinfo{author}{Xie, Z.},
  \bibinfo{author}{Marks, J.}, \& \bibinfo{author}{Fox, A.~P.}
  (\bibinfo{year}{2009}).
\newblock \bibinfo{title}{Isoflurane inhibits the neurotransmitter release
  machinery}.
\newblock {\it \bibinfo{journal}{Journal of Neurophysiology}\/},  {\it
  \bibinfo{volume}{102}\/}, \bibinfo{pages}{1265--1273}.
  \DOIprefix\doi{10.1152/jn.00252.2009}.
\bibitem[{Hiremath et~al.(2017)Hiremath, Tyler-Kabara, Wheeler, Moran, Gaunt,
  Collinger, Foldes, Weber, Chen, Boninger \& Wang}]{Hiremath2017}
\bibinfo{author}{Hiremath, S.~V.}, \bibinfo{author}{Tyler-Kabara, E.~C.},
  \bibinfo{author}{Wheeler, J.~J.}, \bibinfo{author}{Moran, D.~W.},
  \bibinfo{author}{Gaunt, R.~A.}, \bibinfo{author}{Collinger, J.~L.},
  \bibinfo{author}{Foldes, S.~T.}, \bibinfo{author}{Weber, D.~J.},
  \bibinfo{author}{Chen, W.}, \bibinfo{author}{Boninger, M.~L.}, \&
  \bibinfo{author}{Wang, W.} (\bibinfo{year}{2017}).
\newblock \bibinfo{title}{Human perception of electrical stimulation on the
  surface of somatosensory cortex}.
\newblock {\it \bibinfo{journal}{PLOS ONE}\/},  {\it \bibinfo{volume}{12}\/},
  \bibinfo{pages}{e0176020}. \DOIprefix\doi{10.1371/journal.pone.0176020}.
\bibitem[{Holevo(1973)}]{Holevo1973}
\bibinfo{author}{Holevo, A.~S.} (\bibinfo{year}{1973}).
\newblock \bibinfo{title}{Bounds for the quantity of information transmitted by
  a quantum communication channel}.
\newblock {\it \bibinfo{journal}{Problems of Information Transmission}\/},
  {\it \bibinfo{volume}{9}\/}, \bibinfo{pages}{177--183}.
\bibitem[{Holevo(1998)}]{Holevo1998}
\bibinfo{author}{Holevo, A.~S.} (\bibinfo{year}{1998}).
\newblock \bibinfo{title}{Quantum coding theorems}.
\newblock {\it \bibinfo{journal}{Russian Mathematical Surveys}\/},  {\it
  \bibinfo{volume}{53}\/}, \bibinfo{pages}{1295--1331}.
  \DOIprefix\doi{10.1070/rm1998v053n06abeh000091}.
\bibitem[{Holevo(2001)}]{Holevo2001}
\bibinfo{author}{Holevo, A.~S.} (\bibinfo{year}{2001}).
\newblock {\it \bibinfo{title}{Statistical Structure of Quantum Theory}\/}.
\newblock Lecture Notes in Physics.
\newblock \bibinfo{publisher}{Springer}.
\newblock \DOIprefix\doi{10.1007/3-540-44998-1}.
\bibitem[{Horodecki et~al.(2009)Horodecki, Horodecki, Horodecki \&
  Horodecki}]{Horodecki2009}
\bibinfo{author}{Horodecki, R.}, \bibinfo{author}{Horodecki, P.},
  \bibinfo{author}{Horodecki, M.}, \& \bibinfo{author}{Horodecki, K.}
  (\bibinfo{year}{2009}).
\newblock \bibinfo{title}{Quantum entanglement}.
\newblock {\it \bibinfo{journal}{Reviews of Modern Physics}\/},  {\it
  \bibinfo{volume}{81}\/}, \bibinfo{pages}{865--942}.
  \DOIprefix\doi{10.1103/RevModPhys.81.865}.
\bibitem[{Hublin et~al.(2017)Hublin, Ben-Ncer, Bailey, Freidline, Neubauer,
  Skinner, Bergmann, Le~Cabec, Benazzi, Harvati \& Gunz}]{Hublin2017}
\bibinfo{author}{Hublin, J.-J.}, \bibinfo{author}{Ben-Ncer, A.},
  \bibinfo{author}{Bailey, S.~E.}, \bibinfo{author}{Freidline, S.~E.},
  \bibinfo{author}{Neubauer, S.}, \bibinfo{author}{Skinner, M.~M.},
  \bibinfo{author}{Bergmann, I.}, \bibinfo{author}{Le~Cabec, A.},
  \bibinfo{author}{Benazzi, S.}, \bibinfo{author}{Harvati, K.}, \&
  \bibinfo{author}{Gunz, P.} (\bibinfo{year}{2017}).
\newblock \bibinfo{title}{New fossils from {J}ebel {I}rhoud, {M}orocco and the
  pan-{A}frican origin of {H}omo sapiens}.
\newblock {\it \bibinfo{journal}{Nature}\/},  {\it \bibinfo{volume}{546}\/},
  \bibinfo{pages}{289--292}. \DOIprefix\doi{10.1038/nature22336}.
\bibitem[{Hudetz \& Imas(2007)}]{Hudetz2007}
\bibinfo{author}{Hudetz, A.~G.}, \& \bibinfo{author}{Imas, O.~A.}
  (\bibinfo{year}{2007}).
\newblock \bibinfo{title}{Burst activation of the cerebral cortex by flash
  stimuli during isoflurane anesthesia in rats}.
\newblock {\it \bibinfo{journal}{Anesthesiology}\/},  {\it
  \bibinfo{volume}{107}\/}, \bibinfo{pages}{983--991}.
  \DOIprefix\doi{10.1097/01.anes.0000291471.80659.55}.
\bibitem[{Hughston et~al.(1993)Hughston, Jozsa \& Wootters}]{Hughston1993}
\bibinfo{author}{Hughston, L.~P.}, \bibinfo{author}{Jozsa, R.}, \&
  \bibinfo{author}{Wootters, W.~K.} (\bibinfo{year}{1993}).
\newblock \bibinfo{title}{A complete classification of quantum ensembles having
  a given density matrix}.
\newblock {\it \bibinfo{journal}{Physics Letters A}\/},  {\it
  \bibinfo{volume}{183}\/}, \bibinfo{pages}{14--18}.
  \DOIprefix\doi{10.1016/0375-9601(93)90880-9}.
\bibitem[{Imas et~al.(2005)Imas, Ropella, Ward, Wood \& Hudetz}]{Imas2005}
\bibinfo{author}{Imas, O.~A.}, \bibinfo{author}{Ropella, K.~M.},
  \bibinfo{author}{Ward, B.~D.}, \bibinfo{author}{Wood, J.~D.}, \&
  \bibinfo{author}{Hudetz, A.~G.} (\bibinfo{year}{2005}).
\newblock \bibinfo{title}{Volatile anesthetics enhance flash-induced $\gamma$
  oscillations in rat visual cortex}.
\newblock {\it \bibinfo{journal}{Anesthesiology}\/},  {\it
  \bibinfo{volume}{102}\/}, \bibinfo{pages}{937--947}.
\bibitem[{Ing\'{o}lfsson et~al.(2016)Ing\'{o}lfsson, Arnarez, Periole \&
  Marrink}]{Ingolfsson2016}
\bibinfo{author}{Ing\'{o}lfsson, H.~I.}, \bibinfo{author}{Arnarez, C.},
  \bibinfo{author}{Periole, X.}, \& \bibinfo{author}{Marrink, S.~J.}
  (\bibinfo{year}{2016}).
\newblock \bibinfo{title}{Computational `microscopy' of cellular membranes}.
\newblock {\it \bibinfo{journal}{Journal of Cell Science}\/},  {\it
  \bibinfo{volume}{129}\/}, \bibinfo{pages}{257--268}.
  \DOIprefix\doi{10.1242/jcs.176040}.
\bibitem[{Ishida et~al.(2006)Ishida, Fedorov \& Kitaura}]{Ishida2006}
\bibinfo{author}{Ishida, T.}, \bibinfo{author}{Fedorov, D.~G.}, \&
  \bibinfo{author}{Kitaura, K.} (\bibinfo{year}{2006}).
\newblock \bibinfo{title}{All electron quantum chemical calculation of the
  entire enzyme system confirms a collective catalytic device in the chorismate
  mutase reaction}.
\newblock {\it \bibinfo{journal}{Journal of Physical Chemistry B}\/},  {\it
  \bibinfo{volume}{110}\/}, \bibinfo{pages}{1457--1463}.
  \DOIprefix\doi{10.1021/jp0557159}.
\bibitem[{Jackson(1982)}]{Jackson1982}
\bibinfo{author}{Jackson, F.} (\bibinfo{year}{1982}).
\newblock \bibinfo{title}{Epiphenomenal qualia}.
\newblock {\it \bibinfo{journal}{The Philosophical Quarterly}\/},  {\it
  \bibinfo{volume}{32}\/}, \bibinfo{pages}{127--136}.
  \DOIprefix\doi{10.2307/2960077}.
\bibitem[{Jackson(1986)}]{Jackson1986}
\bibinfo{author}{Jackson, F.} (\bibinfo{year}{1986}).
\newblock \bibinfo{title}{What {M}ary didn't know}.
\newblock {\it \bibinfo{journal}{Journal of Philosophy}\/},  {\it
  \bibinfo{volume}{83}\/}, \bibinfo{pages}{291--295}.
  \DOIprefix\doi{10.2307/2026143}.
\bibitem[{Jackson(1996)}]{Jackson1996}
\bibinfo{author}{Jackson, F.} (\bibinfo{year}{1996}).
\newblock \bibinfo{title}{Mental causation}.
\newblock {\it \bibinfo{journal}{Mind}\/},  {\it \bibinfo{volume}{105}\/},
  \bibinfo{pages}{377--413}.
\bibitem[{James(1879)}]{James1879}
\bibinfo{author}{James, W.} (\bibinfo{year}{1879}).
\newblock \bibinfo{title}{Are we automata?}
\newblock {\it \bibinfo{journal}{Mind}\/},  {\it \bibinfo{volume}{4}\/},
  \bibinfo{pages}{1--22}. \DOIprefix\doi{10.1093/mind/os-4.13.1}.
\bibitem[{Jensen et~al.(2013)Jensen, Rocco, Mansur, Smith, Janik \&
  Madsen}]{Jensen2013}
\bibinfo{author}{Jensen, F.~H.}, \bibinfo{author}{Rocco, A.},
  \bibinfo{author}{Mansur, R.~M.}, \bibinfo{author}{Smith, B.~D.},
  \bibinfo{author}{Janik, V.~M.}, \& \bibinfo{author}{Madsen, P.~T.}
  (\bibinfo{year}{2013}).
\newblock \bibinfo{title}{Clicking in shallow rivers: short-range echolocation
  of {I}rrawaddy and {G}anges river dolphins in a shallow, acoustically complex
  habitat}.
\newblock {\it \bibinfo{journal}{PLoS One}\/},  {\it \bibinfo{volume}{8}\/},
  \bibinfo{pages}{e59284}. \DOIprefix\doi{10.1371/journal.pone.0059284}.
\bibitem[{Johansson et~al.(2000)Johansson, Scharf, Davies, Reddy \&
  Eckenhoff}]{Johansson2000}
\bibinfo{author}{Johansson, J.~S.}, \bibinfo{author}{Scharf, D.},
  \bibinfo{author}{Davies, L.~A.}, \bibinfo{author}{Reddy, K.~S.}, \&
  \bibinfo{author}{Eckenhoff, R.~G.} (\bibinfo{year}{2000}).
\newblock \bibinfo{title}{A designed four-$\alpha$-helix bundle that binds the
  volatile general anesthetic halothane with high affinity}.
\newblock {\it \bibinfo{journal}{Biophysical Journal}\/},  {\it
  \bibinfo{volume}{78}\/}, \bibinfo{pages}{982--993}.
  \DOIprefix\doi{10.1016/S0006-3495(00)76656-2}.
\bibitem[{Johnson(1928)}]{Johnson1928}
\bibinfo{author}{Johnson, J.~B.} (\bibinfo{year}{1928}).
\newblock \bibinfo{title}{Thermal agitation of electricity in conductors}.
\newblock {\it \bibinfo{journal}{Physical Review}\/},  {\it
  \bibinfo{volume}{32}\/}, \bibinfo{pages}{97--109}.
  \DOIprefix\doi{10.1103/PhysRev.32.97}.
\bibitem[{Joksovic et~al.(2009)Joksovic, Weiergr\"{a}ber, Lee, Struck,
  Schneider \& Todorovic}]{Joksovic2009}
\bibinfo{author}{Joksovic, P.~M.}, \bibinfo{author}{Weiergr\"{a}ber, M.},
  \bibinfo{author}{Lee, W.~Y.}, \bibinfo{author}{Struck, H.},
  \bibinfo{author}{Schneider, T.}, \& \bibinfo{author}{Todorovic, S.~M.}
  (\bibinfo{year}{2009}).
\newblock \bibinfo{title}{Isoflurane-sensitive presynaptic {R}-type calcium
  channels contribute to inhibitory synaptic transmission in the rat thalamus}.
\newblock {\it \bibinfo{journal}{Journal of Neuroscience}\/},  {\it
  \bibinfo{volume}{29}\/}, \bibinfo{pages}{1434--1445}.
  \DOIprefix\doi{10.1523/jneurosci.5574-08.2009}.
\bibitem[{Kariev \& Green(2009)}]{Kariev2009}
\bibinfo{author}{Kariev, A.~M.}, \& \bibinfo{author}{Green, M.~E.}
  (\bibinfo{year}{2009}).
\newblock \bibinfo{title}{Quantum calculations on water in the {KcsA} channel
  cavity with permeant and non-permeant ions}.
\newblock {\it \bibinfo{journal}{Biochimica et Biophysica Acta (BBA) -
  Biomembranes}\/},  {\it \bibinfo{volume}{1788}\/},
  \bibinfo{pages}{1188--1192}. \DOIprefix\doi{10.1016/j.bbamem.2008.12.015}.
\bibitem[{Kariev \& Green(2012)}]{Kariev2012}
\bibinfo{author}{Kariev, A.~M.}, \& \bibinfo{author}{Green, M.~E.}
  (\bibinfo{year}{2012}).
\newblock \bibinfo{title}{Voltage gated ion channel function: gating,
  conduction, and the role of water and protons}.
\newblock {\it \bibinfo{journal}{International Journal of Molecular
  Sciences}\/},  {\it \bibinfo{volume}{13}\/}, \bibinfo{pages}{1680--1709}.
  \DOIprefix\doi{10.3390/ijms13021680}.
\bibitem[{Kariev \& Green(2019)}]{Kariev2019}
\bibinfo{author}{Kariev, A.~M.}, \& \bibinfo{author}{Green, M.~E.}
  (\bibinfo{year}{2019}).
\newblock \bibinfo{title}{Quantum calculation of proton and other charge
  transfer steps in voltage sensing in the {K}v1.2 channel}.
\newblock {\it \bibinfo{journal}{Journal of Physical Chemistry B}\/},  {\it
  \bibinfo{volume}{123}\/}, \bibinfo{pages}{7984--7998}.
  \DOIprefix\doi{10.1021/acs.jpcb.9b05448}.
\bibitem[{Kariev et~al.(2014)Kariev, Njau \& Green}]{Kariev2014}
\bibinfo{author}{Kariev, A.~M.}, \bibinfo{author}{Njau, P.}, \&
  \bibinfo{author}{Green, M.~E.} (\bibinfo{year}{2014}).
\newblock \bibinfo{title}{The open gate of the {K}v1.2 channel: quantum
  calculations show the key role of hydration}.
\newblock {\it \bibinfo{journal}{Biophysical Journal}\/},  {\it
  \bibinfo{volume}{106}\/}, \bibinfo{pages}{548--555}.
  \DOIprefix\doi{10.1016/j.bpj.2013.11.4495}.
\bibitem[{Kariev et~al.(2007)Kariev, Znamenskiy \& Green}]{Kariev2007}
\bibinfo{author}{Kariev, A.~M.}, \bibinfo{author}{Znamenskiy, V.~S.}, \&
  \bibinfo{author}{Green, M.~E.} (\bibinfo{year}{2007}).
\newblock \bibinfo{title}{Quantum mechanical calculations of charge effects on
  gating the {KcsA} channel}.
\newblock {\it \bibinfo{journal}{Biochimica et Biophysica Acta (BBA) -
  Biomembranes}\/},  {\it \bibinfo{volume}{1768}\/},
  \bibinfo{pages}{1218--1229}. \DOIprefix\doi{10.1016/j.bbamem.2007.01.021}.
\bibitem[{Kim(1998)}]{Kim1998}
\bibinfo{author}{Kim, J.} (\bibinfo{year}{1998}).
\newblock {\it \bibinfo{title}{Mind in a Physical World: An Essay on the
  Mind--Body Problem and Mental Causation}\/}.
\newblock \bibinfo{address}{Cambridge, Massachusetts}: \bibinfo{publisher}{MIT
  Press}.
\bibitem[{Kim et~al.(2013)Kim, Li \& von Gersdorff}]{Kim2013}
\bibinfo{author}{Kim, M.-H.}, \bibinfo{author}{Li, G.-L.}, \&
  \bibinfo{author}{von Gersdorff, H.} (\bibinfo{year}{2013}).
\newblock \bibinfo{title}{Single {C}a$^{2+}$ channels and exocytosis at sensory
  synapses}.
\newblock {\it \bibinfo{journal}{Journal of Physiology}\/},  {\it
  \bibinfo{volume}{591}\/}, \bibinfo{pages}{3167--3178}.
  \DOIprefix\doi{10.1113/jphysiol.2012.249482}.
\bibitem[{Klekamp \& Samii(2002)}]{Klekamp2002}
\bibinfo{author}{Klekamp, J.}, \& \bibinfo{author}{Samii, M.}
  (\bibinfo{year}{2002}).
\newblock {\it \bibinfo{title}{Syringomyelia: Diagnosis and Treatment}\/}.
\newblock \bibinfo{address}{Berlin}: \bibinfo{publisher}{Springer}.
\newblock \DOIprefix\doi{10.1007/978-3-642-56023-1}.
\bibitem[{Kolev et~al.(2013)Kolev, Petkov, Rangelov \& Vayssilov}]{Kolev2013}
\bibinfo{author}{Kolev, S.~K.}, \bibinfo{author}{Petkov, P.~S.},
  \bibinfo{author}{Rangelov, M.}, \& \bibinfo{author}{Vayssilov, G.~N.}
  (\bibinfo{year}{2013}).
\newblock \bibinfo{title}{Ab initio molecular dynamics of {N}a$^+$ and
  {M}g$^{2+}$ countercations at the backbone of {RNA} in water solution}.
\newblock {\it \bibinfo{journal}{ACS Chemical Biology}\/},  {\it
  \bibinfo{volume}{8}\/}, \bibinfo{pages}{1576--1589}.
  \DOIprefix\doi{10.1021/cb300463h}.
\bibitem[{Kolev et~al.(2018)Kolev, Petkov, Rangelov, Trifonov, Milenov \&
  Vayssilov}]{Kolev2018}
\bibinfo{author}{Kolev, S.~K.}, \bibinfo{author}{Petkov, P.~S.},
  \bibinfo{author}{Rangelov, M.~A.}, \bibinfo{author}{Trifonov, D.~V.},
  \bibinfo{author}{Milenov, T.~I.}, \& \bibinfo{author}{Vayssilov, G.~N.}
  (\bibinfo{year}{2018}).
\newblock \bibinfo{title}{Interaction of {N}a$^+$, {K}$^+$, {M}g$^{2+}$ and
  {C}a$^{2+}$ counter cations with {RNA}}.
\newblock {\it \bibinfo{journal}{Metallomics}\/},  {\it
  \bibinfo{volume}{10}\/}, \bibinfo{pages}{659--678}.
  \DOIprefix\doi{10.1039/c8mt00043c}.
\bibitem[{Kubo et~al.(2004)Kubo, Shiomitsu, Odai, Sugimoto, Suzuki \&
  Ito}]{Kubo2004}
\bibinfo{author}{Kubo, M.}, \bibinfo{author}{Shiomitsu, E.},
  \bibinfo{author}{Odai, K.}, \bibinfo{author}{Sugimoto, T.},
  \bibinfo{author}{Suzuki, H.}, \& \bibinfo{author}{Ito, E.}
  (\bibinfo{year}{2004}).
\newblock \bibinfo{title}{Picosecond dynamics of the glutamate receptor in
  response to agonist-induced vibrational excitation}.
\newblock {\it \bibinfo{journal}{Proteins: Structure, Function, and
  Bioinformatics}\/},  {\it \bibinfo{volume}{54}\/}, \bibinfo{pages}{231--236}.
  \DOIprefix\doi{10.1002/prot.10578}.
\bibitem[{Kuno et~al.(1971)Kuno, Turkanis \& Weakly}]{Kuno1971}
\bibinfo{author}{Kuno, M.}, \bibinfo{author}{Turkanis, S.~A.}, \&
  \bibinfo{author}{Weakly, J.~N.} (\bibinfo{year}{1971}).
\newblock \bibinfo{title}{Correlation between nerve terminal size and
  transmitter release at the neuromuscular junction of the frog}.
\newblock {\it \bibinfo{journal}{Journal of Physiology}\/},  {\it
  \bibinfo{volume}{213}\/}, \bibinfo{pages}{545--556}.
  \DOIprefix\doi{10.1113/jphysiol.1971.sp009399}.
\bibitem[{Ladegaard et~al.(2019)Ladegaard, Mulsow, Houser, Jensen, Johnson,
  Madsen \& Finneran}]{Ladegaard2019}
\bibinfo{author}{Ladegaard, M.}, \bibinfo{author}{Mulsow, J.},
  \bibinfo{author}{Houser, D.~S.}, \bibinfo{author}{Jensen, F.~H.},
  \bibinfo{author}{Johnson, M.}, \bibinfo{author}{Madsen, P.~T.}, \&
  \bibinfo{author}{Finneran, J.~J.} (\bibinfo{year}{2019}).
\newblock \bibinfo{title}{Dolphin echolocation behaviour during active
  long-range target approaches}.
\newblock {\it \bibinfo{journal}{Journal of Experimental Biology}\/},  {\it
  \bibinfo{volume}{222}\/}, \bibinfo{pages}{jeb189217}.
  \DOIprefix\doi{10.1242/jeb.189217}.
\bibitem[{Lamme et~al.(1998)Lamme, Zipser \& Spekreijse}]{Lamme1998}
\bibinfo{author}{Lamme, V. A.~F.}, \bibinfo{author}{Zipser, K.}, \&
  \bibinfo{author}{Spekreijse, H.} (\bibinfo{year}{1998}).
\newblock \bibinfo{title}{Figure-ground activity in primary visual cortex is
  suppressed by anesthesia}.
\newblock {\it \bibinfo{journal}{Proceedings of the National Academy of
  Sciences}\/},  {\it \bibinfo{volume}{95}\/}, \bibinfo{pages}{3263--3268}.
\bibitem[{Landau \& Lifshitz(1965)}]{Landau1965}
\bibinfo{author}{Landau, L.~D.}, \& \bibinfo{author}{Lifshitz, E.~M.}
  (\bibinfo{year}{1965}).
\newblock {\it \bibinfo{title}{Quantum Mechanics}\/}.
\newblock Course of Theoretical Physics.
\newblock \bibinfo{address}{Oxford}: \bibinfo{publisher}{Pergamon Press}.
\bibitem[{Lau \& London(2018)}]{Lau2018}
\bibinfo{author}{Lau, L.}, \& \bibinfo{author}{London, K.}
  (\bibinfo{year}{2018}).
\newblock \bibinfo{title}{Cortical blindness and altered mental status
  following routine hemodialysis, a case of iatrogenic cerebral air embolism}.
\newblock {\it \bibinfo{journal}{Case Reports in Emergency Medicine}\/},  {\it
  \bibinfo{volume}{2018}\/}, \bibinfo{pages}{9496818}.
  \DOIprefix\doi{10.1155/2018/9496818}.
\bibitem[{Lewis \& Rosenfeld(2016)}]{Lewis2016}
\bibinfo{author}{Lewis, P.~M.}, \& \bibinfo{author}{Rosenfeld, J.~V.}
  (\bibinfo{year}{2016}).
\newblock \bibinfo{title}{Electrical stimulation of the brain and the
  development of cortical visual prostheses: An historical perspective}.
\newblock {\it \bibinfo{journal}{Brain Research}\/},  {\it
  \bibinfo{volume}{1630}\/}, \bibinfo{pages}{208--224}.
  \DOIprefix\doi{10.1016/j.brainres.2015.08.038}.
\bibitem[{MacIver et~al.(1996)MacIver, Mikulec, Amagasu \&
  Monroe}]{MacIver1996}
\bibinfo{author}{MacIver, M.~B.}, \bibinfo{author}{Mikulec, A.~A.},
  \bibinfo{author}{Amagasu, S.~M.}, \& \bibinfo{author}{Monroe, F.~A.}
  (\bibinfo{year}{1996}).
\newblock \bibinfo{title}{Volatile anesthetics depress glutamate transmission
  via presynaptic actions}.
\newblock {\it \bibinfo{journal}{Anesthesiology}\/},  {\it
  \bibinfo{volume}{85}\/}, \bibinfo{pages}{823--834}.
\bibitem[{Maffeo et~al.(2012)Maffeo, Bhattacharya, Yoo, Wells \&
  Aksimentiev}]{Maffeo2012}
\bibinfo{author}{Maffeo, C.}, \bibinfo{author}{Bhattacharya, S.},
  \bibinfo{author}{Yoo, J.}, \bibinfo{author}{Wells, D.}, \&
  \bibinfo{author}{Aksimentiev, A.} (\bibinfo{year}{2012}).
\newblock \bibinfo{title}{Modeling and simulation of ion channels}.
\newblock {\it \bibinfo{journal}{Chemical Reviews}\/},  {\it
  \bibinfo{volume}{112}\/}, \bibinfo{pages}{6250--6284}.
  \DOIprefix\doi{10.1021/cr3002609}.
\bibitem[{Magistretti et~al.(2015)Magistretti, Spaiardi, Johnson \&
  Masetto}]{Magistretti2015}
\bibinfo{author}{Magistretti, J.}, \bibinfo{author}{Spaiardi, P.},
  \bibinfo{author}{Johnson, S.~L.}, \& \bibinfo{author}{Masetto, S.}
  (\bibinfo{year}{2015}).
\newblock \bibinfo{title}{Elementary properties of {C}a$^{2+}$ channels and
  their influence on multivesicular release and phase-locking at auditory hair
  cell ribbon synapses}.
\newblock {\it \bibinfo{journal}{Frontiers in Cellular Neuroscience}\/},  {\it
  \bibinfo{volume}{9}\/}, \bibinfo{pages}{123}.
\bibitem[{McNeer et~al.(2009)McNeer, Boh\'{o}rquez \& \"{O}zdamar}]{McNeer2009}
\bibinfo{author}{McNeer, R.~R.}, \bibinfo{author}{Boh\'{o}rquez, J.}, \&
  \bibinfo{author}{\"{O}zdamar, O.} (\bibinfo{year}{2009}).
\newblock \bibinfo{title}{Influence of auditory stimulation rates on evoked
  potentials during general anesthesia: Relation between the transient auditory
  middle-latency response and the 40-{H}z auditory steady state response}.
\newblock {\it \bibinfo{journal}{Anesthesiology}\/},  {\it
  \bibinfo{volume}{110}\/}, \bibinfo{pages}{1026--1035}.
  \DOIprefix\doi{10.1097/aln.0b013e31819dad6f}.
\bibitem[{Melkikh(2019)}]{Melkikh2019}
\bibinfo{author}{Melkikh, A.~V.} (\bibinfo{year}{2019}).
\newblock \bibinfo{title}{Thinking as a quantum phenomenon}.
\newblock {\it \bibinfo{journal}{Biosystems}\/},  {\it
  \bibinfo{volume}{176}\/}, \bibinfo{pages}{32--40}.
  \DOIprefix\doi{10.1016/j.biosystems.2018.12.007}.
\bibitem[{Melkikh \& Khrennikov(2015)}]{Melkikh2015}
\bibinfo{author}{Melkikh, A.~V.}, \& \bibinfo{author}{Khrennikov, A.}
  (\bibinfo{year}{2015}).
\newblock \bibinfo{title}{Nontrivial quantum and quantum-like effects in
  biosystems: unsolved questions and paradoxes}.
\newblock {\it \bibinfo{journal}{Progress in Biophysics and Molecular
  Biology}\/},  {\it \bibinfo{volume}{119}\/}, \bibinfo{pages}{137--161}.
  \DOIprefix\doi{10.1016/j.pbiomolbio.2015.07.001}.
\bibitem[{Morris(2017)}]{Morris2017}
\bibinfo{author}{Morris, K.} (\bibinfo{year}{2017}).
\newblock \bibinfo{title}{The combination problem: Subjects and unity}.
\newblock {\it \bibinfo{journal}{Erkenntnis}\/},  {\it \bibinfo{volume}{82}\/},
  \bibinfo{pages}{103--120}. \DOIprefix\doi{10.1007/s10670-016-9808-8}.
\bibitem[{Nagel(1965)}]{Nagel1965}
\bibinfo{author}{Nagel, T.} (\bibinfo{year}{1965}).
\newblock \bibinfo{title}{Physicalism}.
\newblock {\it \bibinfo{journal}{The Philosophical Review}\/},  {\it
  \bibinfo{volume}{74}\/}, \bibinfo{pages}{339--356}.
  \DOIprefix\doi{10.2307/2183358}.
\bibitem[{Nagel(1974)}]{Nagel1974}
\bibinfo{author}{Nagel, T.} (\bibinfo{year}{1974}).
\newblock \bibinfo{title}{What is it like to be a bat?}
\newblock {\it \bibinfo{journal}{The Philosophical Review}\/},  {\it
  \bibinfo{volume}{83}\/}, \bibinfo{pages}{435--450}.
  \DOIprefix\doi{10.2307/2183914}.
\bibitem[{Nagel(1987)}]{Nagel1987}
\bibinfo{author}{Nagel, T.} (\bibinfo{year}{1987}).
\newblock {\it \bibinfo{title}{What Does It All Mean? A Very Short Introduction
  to Philosophy}\/}.
\newblock \bibinfo{address}{New York}: \bibinfo{publisher}{Oxford University
  Press}.
\bibitem[{Nagel(2012)}]{Nagel2012}
\bibinfo{author}{Nagel, T.} (\bibinfo{year}{2012}).
\newblock {\it \bibinfo{title}{Mind and Cosmos: Why the Materialist
  Neo-Darwinian Conception of Nature Is Almost Certainly False}\/}.
\newblock \bibinfo{address}{Oxford}: \bibinfo{publisher}{Oxford University
  Press}.
\bibitem[{Nagele et~al.(2005)Nagele, Mendel, Placzek, Scott, d'Avignon \&
  Crowder}]{Nagele2005}
\bibinfo{author}{Nagele, P.}, \bibinfo{author}{Mendel, J.~B.},
  \bibinfo{author}{Placzek, W.~J.}, \bibinfo{author}{Scott, B.~A.},
  \bibinfo{author}{d'Avignon, D.~A.}, \& \bibinfo{author}{Crowder, C.~M.}
  (\bibinfo{year}{2005}).
\newblock \bibinfo{title}{Volatile anesthetics bind rat synaptic {SNARE}
  proteins}.
\newblock {\it \bibinfo{journal}{Anesthesiology}\/},  {\it
  \bibinfo{volume}{103}\/}, \bibinfo{pages}{768--778}.
\bibitem[{von Neumann(1955)}]{vonNeumann1955}
\bibinfo{author}{von Neumann, J.} (\bibinfo{year}{1955}).
\newblock {\it \bibinfo{title}{Mathematical Foundations of Quantum
  Mechanics}\/}.
\newblock \bibinfo{address}{Princeton}: \bibinfo{publisher}{Princeton
  University Press}.
\bibitem[{Nielsen \& Chuang(2010)}]{Nielsen2010}
\bibinfo{author}{Nielsen, M.~A.}, \& \bibinfo{author}{Chuang, I.~L.}
  (\bibinfo{year}{2010}).
\newblock {\it \bibinfo{title}{Quantum Computation and Quantum Information}\/}.
\newblock \bibinfo{address}{Cambridge}: \bibinfo{publisher}{Cambridge
  University Press}.
\bibitem[{Nourski et~al.(2017)Nourski, Banks, Steinschneider, Rhone, Kawasaki,
  Mueller, Todd \& Howard}]{Nourski2017}
\bibinfo{author}{Nourski, K.~V.}, \bibinfo{author}{Banks, M.~I.},
  \bibinfo{author}{Steinschneider, M.}, \bibinfo{author}{Rhone, A.~E.},
  \bibinfo{author}{Kawasaki, H.}, \bibinfo{author}{Mueller, R.~N.},
  \bibinfo{author}{Todd, M.~M.}, \& \bibinfo{author}{Howard, M.~A.}
  (\bibinfo{year}{2017}).
\newblock \bibinfo{title}{Electrocorticographic delineation of human auditory
  cortical fields based on effects of propofol anesthesia}.
\newblock {\it \bibinfo{journal}{NeuroImage}\/},  {\it
  \bibinfo{volume}{152}\/}, \bibinfo{pages}{78--93}.
  \DOIprefix\doi{10.1016/j.neuroimage.2017.02.061}.
\bibitem[{Nyquist(1928)}]{Nyquist1928}
\bibinfo{author}{Nyquist, H.} (\bibinfo{year}{1928}).
\newblock \bibinfo{title}{Thermal agitation of electric charge in conductors}.
\newblock {\it \bibinfo{journal}{Physical Review}\/},  {\it
  \bibinfo{volume}{32}\/}, \bibinfo{pages}{110--113}.
  \DOIprefix\doi{10.1103/PhysRev.32.110}.
\bibitem[{Ohya \& Volovich(2011)}]{Ohya2011}
\bibinfo{author}{Ohya, M.}, \& \bibinfo{author}{Volovich, I.}
  (\bibinfo{year}{2011}).
\newblock {\it \bibinfo{title}{Mathematical Foundations of Quantum Information
  and Computation and Its Applications to Nano- and Bio-systems}\/}.
\newblock Theoretical and Mathematical Physics.
\newblock \bibinfo{address}{Dordrecht}: \bibinfo{publisher}{Springer}.
\newblock \DOIprefix\doi{10.1007/978-94-007-0171-7}.
\bibitem[{Paciaroni et~al.(2008)Paciaroni, Orecchini, Cornicchi, Marconi,
  Petrillo, Haertlein, Moulin \& Sacchetti}]{Paciaroni2008}
\bibinfo{author}{Paciaroni, A.}, \bibinfo{author}{Orecchini, A.},
  \bibinfo{author}{Cornicchi, E.}, \bibinfo{author}{Marconi, M.},
  \bibinfo{author}{Petrillo, C.}, \bibinfo{author}{Haertlein, M.},
  \bibinfo{author}{Moulin, M.}, \& \bibinfo{author}{Sacchetti, F.}
  (\bibinfo{year}{2008}).
\newblock \bibinfo{title}{Coupled thermal fluctuations of proteins and protein
  hydration water on the picosecond timescale}.
\newblock {\it \bibinfo{journal}{Philosophical Magazine}\/},  {\it
  \bibinfo{volume}{88}\/}, \bibinfo{pages}{4071--4077}.
  \DOIprefix\doi{10.1080/14786430802464263}.
\bibitem[{Pal et~al.(2015)Pal, Jones, Wisidagamage, Meisler \&
  Mashour}]{Pal2015}
\bibinfo{author}{Pal, D.}, \bibinfo{author}{Jones, J.~M.},
  \bibinfo{author}{Wisidagamage, S.}, \bibinfo{author}{Meisler, M.~H.}, \&
  \bibinfo{author}{Mashour, G.~A.} (\bibinfo{year}{2015}).
\newblock \bibinfo{title}{Reduced {N}av1.6 sodium channel activity in mice
  increases in vivo sensitivity to volatile anesthetics}.
\newblock {\it \bibinfo{journal}{PLOS ONE}\/},  {\it \bibinfo{volume}{10}\/},
  \bibinfo{pages}{e0134960}. \DOIprefix\doi{10.1371/journal.pone.0134960}.
\bibitem[{Pathak(2013)}]{Pathak2013}
\bibinfo{author}{Pathak, A.} (\bibinfo{year}{2013}).
\newblock {\it \bibinfo{title}{Elements of Quantum Computation and Quantum
  Communication}\/}.
\newblock \bibinfo{address}{Boca Raton}: \bibinfo{publisher}{CRC Press}.
\newblock \DOIprefix\doi{10.1201/b15007}.
\bibitem[{Pati \& Braunstein(2000)}]{Pati2000}
\bibinfo{author}{Pati, A.~K.}, \& \bibinfo{author}{Braunstein, S.~L.}
  (\bibinfo{year}{2000}).
\newblock \bibinfo{title}{Impossibility of deleting an unknown quantum state}.
\newblock {\it \bibinfo{journal}{Nature}\/},  {\it \bibinfo{volume}{404}\/},
  \bibinfo{pages}{164--165}. \DOIprefix\doi{10.1038/404130b0}.
\bibitem[{Peled et~al.(2020)Peled, Te'eni, Georgiev, Cohen \&
  Carmi}]{Peled2020}
\bibinfo{author}{Peled, B.~Y.}, \bibinfo{author}{Te'eni, A.},
  \bibinfo{author}{Georgiev, D.}, \bibinfo{author}{Cohen, E.}, \&
  \bibinfo{author}{Carmi, A.} (\bibinfo{year}{2020}).
\newblock \bibinfo{title}{Double slit with an {E}instein--{P}odolsky--{R}osen
  pair}.
\newblock {\it \bibinfo{journal}{Applied Sciences}\/},  {\it
  \bibinfo{volume}{10}\/}, \bibinfo{pages}{792}.
  \DOIprefix\doi{10.3390/app10030792}.
\bibitem[{Petrenko et~al.(2007)Petrenko, Tsujita, Kohno, Sakimura \&
  Baba}]{Petrenko2007}
\bibinfo{author}{Petrenko, A.~B.}, \bibinfo{author}{Tsujita, M.},
  \bibinfo{author}{Kohno, T.}, \bibinfo{author}{Sakimura, K.}, \&
  \bibinfo{author}{Baba, H.} (\bibinfo{year}{2007}).
\newblock \bibinfo{title}{Mutation of $\alpha$1{G} {T}-type calcium channels in
  mice does not change anesthetic requirements for loss of the righting reflex
  and minimum alveolar concentration but delays the onset of anesthetic
  induction}.
\newblock {\it \bibinfo{journal}{Anesthesiology}\/},  {\it
  \bibinfo{volume}{106}\/}, \bibinfo{pages}{1177--1185}.
  \DOIprefix\doi{10.1097/01.anes.0000267601.09764.e6}.
\bibitem[{Piccinini(2010)}]{Piccinini2010}
\bibinfo{author}{Piccinini, G.} (\bibinfo{year}{2010}).
\newblock \bibinfo{title}{The mind as neural software? {U}nderstanding
  functionalism, computationalism, and computational functionalism}.
\newblock {\it \bibinfo{journal}{Philosophy and Phenomenological Research}\/},
  {\it \bibinfo{volume}{81}\/}, \bibinfo{pages}{269--311}.
  \DOIprefix\doi{10.1111/j.1933-1592.2010.00356.x}.
\bibitem[{Popper \& Eccles(1983)}]{Popper1983}
\bibinfo{author}{Popper, K.~R.}, \& \bibinfo{author}{Eccles, J.~C.}
  (\bibinfo{year}{1983}).
\newblock {\it \bibinfo{title}{The Self and Its Brain: An Argument for
  Interactionism}\/}.
\newblock \bibinfo{address}{London}: \bibinfo{publisher}{Routledge \& Kegan
  Paul}.
\newblock \DOIprefix\doi{10.4324/9780203537480}.
\bibitem[{Purtell et~al.(2015)Purtell, Gingrich, Ouyang, Herold \&
  Hemmings}]{Purtell2015}
\bibinfo{author}{Purtell, K.}, \bibinfo{author}{Gingrich, K.~J.},
  \bibinfo{author}{Ouyang, W.}, \bibinfo{author}{Herold, K.~F.}, \&
  \bibinfo{author}{Hemmings, H.~C.} (\bibinfo{year}{2015}).
\newblock \bibinfo{title}{Activity-dependent depression of neuronal sodium
  channels by the general anaesthetic isoflurane}.
\newblock {\it \bibinfo{journal}{British Journal of Anaesthesia}\/},  {\it
  \bibinfo{volume}{115}\/}, \bibinfo{pages}{112--121}.
  \DOIprefix\doi{10.1093/bja/aev203}.
\bibitem[{Pusuluk et~al.(2018)Pusuluk, Torun \& Deliduman}]{Pusuluk2018}
\bibinfo{author}{Pusuluk, O.}, \bibinfo{author}{Torun, G.}, \&
  \bibinfo{author}{Deliduman, C.} (\bibinfo{year}{2018}).
\newblock \bibinfo{title}{Quantum entanglement shared in hydrogen bonds and its
  usage as a resource in molecular recognition}.
\newblock {\it \bibinfo{journal}{Modern Physics Letters B}\/},  {\it
  \bibinfo{volume}{32}\/}, \bibinfo{pages}{1850308}.
  \DOIprefix\doi{10.1142/S0217984918503086}.
\bibitem[{Ranaghan \& Mulholland(2017)}]{Ranaghan2017}
\bibinfo{author}{Ranaghan, K.~E.}, \& \bibinfo{author}{Mulholland, A.~J.}
  (\bibinfo{year}{2017}).
\newblock \bibinfo{title}{{QM/MM} methods for simulating enzyme reactions}.
\newblock In {\it \bibinfo{booktitle}{Simulating Enzyme Reactivity:
  Computational Methods in Enzyme Catalysis}\/} (pp.
  \bibinfo{pages}{375--403}).
\newblock \bibinfo{publisher}{The Royal Society of Chemistry}.
\newblock \DOIprefix\doi{10.1039/9781782626831-00375}.
\bibitem[{Richardson et~al.(2016)Richardson, P\'{e}rez, Lobsiger, Reid,
  Temelso, Shields, Kisiel, Wales, Pate \& Althorpe}]{Richardson2016}
\bibinfo{author}{Richardson, J.~O.}, \bibinfo{author}{P\'{e}rez, C.},
  \bibinfo{author}{Lobsiger, S.}, \bibinfo{author}{Reid, A.~A.},
  \bibinfo{author}{Temelso, B.}, \bibinfo{author}{Shields, G.~C.},
  \bibinfo{author}{Kisiel, Z.}, \bibinfo{author}{Wales, D.~J.},
  \bibinfo{author}{Pate, B.~H.}, \& \bibinfo{author}{Althorpe, S.~C.}
  (\bibinfo{year}{2016}).
\newblock \bibinfo{title}{Concerted hydrogen-bond breaking by quantum tunneling
  in the water hexamer prism}.
\newblock {\it \bibinfo{journal}{Science}\/},  {\it \bibinfo{volume}{351}\/},
  \bibinfo{pages}{1310}. \DOIprefix\doi{10.1126/science.aae0012}.
\bibitem[{Ridgway et~al.(2015)Ridgway, Dibble, Van~Alstyne \&
  Price}]{Ridgway2015}
\bibinfo{author}{Ridgway, S.}, \bibinfo{author}{Dibble, D.~S.},
  \bibinfo{author}{Van~Alstyne, K.}, \& \bibinfo{author}{Price, D.}
  (\bibinfo{year}{2015}).
\newblock \bibinfo{title}{On doing two things at once: dolphin brain and nose
  coordinate sonar clicks, buzzes and emotional squeals with social sounds
  during fish capture}.
\newblock {\it \bibinfo{journal}{Journal of Experimental Biology}\/},  {\it
  \bibinfo{volume}{218}\/}, \bibinfo{pages}{3987--3995}.
  \DOIprefix\doi{10.1242/jeb.130559}.
\bibitem[{Risselada \& Grubm\"{u}ller(2012)}]{Risselada2012}
\bibinfo{author}{Risselada, H.~J.}, \& \bibinfo{author}{Grubm\"{u}ller, H.}
  (\bibinfo{year}{2012}).
\newblock \bibinfo{title}{How {SNARE} molecules mediate membrane fusion: Recent
  insights from molecular simulations}.
\newblock {\it \bibinfo{journal}{Current Opinion in Structural Biology}\/},
  {\it \bibinfo{volume}{22}\/}, \bibinfo{pages}{187--196}.
  \DOIprefix\doi{10.1016/j.sbi.2012.01.007}.
\bibitem[{Rizo \& S\"{u}dhof(2012)}]{Rizo2012}
\bibinfo{author}{Rizo, J.}, \& \bibinfo{author}{S\"{u}dhof, T.~C.}
  (\bibinfo{year}{2012}).
\newblock \bibinfo{title}{The membrane fusion enigma: {SNARE}s, {S}ec1/{M}unc18
  proteins, and their accomplices--guilty as charged?}
\newblock {\it \bibinfo{journal}{Annual Review of Cell and Developmental
  Biology}\/},  {\it \bibinfo{volume}{28}\/}, \bibinfo{pages}{279--308}.
  \DOIprefix\doi{10.1146/annurev-cellbio-101011-155818}.
\bibitem[{Robinson(1976)}]{Robinson1976}
\bibinfo{author}{Robinson, H.~M.} (\bibinfo{year}{1976}).
\newblock \bibinfo{title}{The mind-body problem in contemporary philosophy}.
\newblock {\it \bibinfo{journal}{Zygon}\/},  {\it \bibinfo{volume}{11}\/},
  \bibinfo{pages}{346--360}.
  \DOIprefix\doi{10.1111/j.1467-9744.1976.tb00291.x}.
\bibitem[{Rosenfeld et~al.(2017)Rosenfeld, Burchardt, Garthoff, Redeker,
  Ortegel, Rau \& Weinfurter}]{Rosenfeld2017}
\bibinfo{author}{Rosenfeld, W.}, \bibinfo{author}{Burchardt, D.},
  \bibinfo{author}{Garthoff, R.}, \bibinfo{author}{Redeker, K.},
  \bibinfo{author}{Ortegel, N.}, \bibinfo{author}{Rau, M.}, \&
  \bibinfo{author}{Weinfurter, H.} (\bibinfo{year}{2017}).
\newblock \bibinfo{title}{Event-ready {B}ell test using entangled atoms
  simultaneously closing detection and locality loopholes}.
\newblock {\it \bibinfo{journal}{Physical Review Letters}\/},  {\it
  \bibinfo{volume}{119}\/}, \bibinfo{pages}{010402}.
  \DOIprefix\doi{10.1103/PhysRevLett.119.010402}.
\bibitem[{Roy \& Llin\'{a}s(2009)}]{Roy2009}
\bibinfo{author}{Roy, S.}, \& \bibinfo{author}{Llin\'{a}s, R.}
  (\bibinfo{year}{2009}).
\newblock \bibinfo{title}{Relevance of quantum mechanics on some aspects of ion
  channel function}.
\newblock {\it \bibinfo{journal}{Comptes Rendus Biologies}\/},  {\it
  \bibinfo{volume}{332}\/}, \bibinfo{pages}{517--522}.
  \DOIprefix\doi{10.1016/j.crvi.2008.11.009}.
\bibitem[{Rudolph et~al.(2015)Rudolph, Tsai, von Gersdorff \&
  Wadiche}]{Rudolph2015}
\bibinfo{author}{Rudolph, S.}, \bibinfo{author}{Tsai, M.-C.},
  \bibinfo{author}{von Gersdorff, H.}, \& \bibinfo{author}{Wadiche, J.~I.}
  (\bibinfo{year}{2015}).
\newblock \bibinfo{title}{The ubiquitous nature of multivesicular release}.
\newblock {\it \bibinfo{journal}{Trends in Neurosciences}\/},  {\it
  \bibinfo{volume}{38}\/}, \bibinfo{pages}{428--438}.
  \DOIprefix\doi{10.1016/j.tins.2015.05.008}.
\bibitem[{Sajja et~al.(2017)Sajja, Tsering, Mooser, DeFreitas, Carpenter \&
  Magge}]{Sajja2017}
\bibinfo{author}{Sajja, A.}, \bibinfo{author}{Tsering, D.},
  \bibinfo{author}{Mooser, A.~C.}, \bibinfo{author}{DeFreitas, T.~A.},
  \bibinfo{author}{Carpenter, J.}, \& \bibinfo{author}{Magge, S.~N.}
  (\bibinfo{year}{2017}).
\newblock \bibinfo{title}{Patient with severe moyamoya disease who presents
  with acute cortical blindness}.
\newblock {\it \bibinfo{journal}{Stroke}\/},  {\it \bibinfo{volume}{48}\/},
  \bibinfo{pages}{e126}. \DOIprefix\doi{10.1161/strokeaha.116.015548}.
\bibitem[{Sakmann \& Neher(1995)}]{Sakmann1995}
\bibinfo{author}{Sakmann, B.}, \& \bibinfo{author}{Neher, E.}
  (\bibinfo{year}{1995}).
\newblock {\it \bibinfo{title}{Single-Channel Recording}\/}.
\newblock (\bibinfo{edition}{2nd} ed.).
\newblock \bibinfo{address}{New York}: \bibinfo{publisher}{Springer}.
\newblock \DOIprefix\doi{10.1007/978-1-4419-1229-9}.
\bibitem[{Sellers et~al.(2015)Sellers, Bennett, Hutt, Williams \&
  Fr\"{o}hlich}]{Sellers2015}
\bibinfo{author}{Sellers, K.~K.}, \bibinfo{author}{Bennett, D.~V.},
  \bibinfo{author}{Hutt, A.}, \bibinfo{author}{Williams, J.~H.}, \&
  \bibinfo{author}{Fr\"{o}hlich, F.} (\bibinfo{year}{2015}).
\newblock \bibinfo{title}{Awake vs. anesthetized: layer-specific sensory
  processing in visual cortex and functional connectivity between cortical
  areas}.
\newblock {\it \bibinfo{journal}{Journal of Neurophysiology}\/},  {\it
  \bibinfo{volume}{113}\/}, \bibinfo{pages}{3798--3815}.
  \DOIprefix\doi{10.1152/jn.00923.2014}.
\bibitem[{Shen et~al.(2011)Shen, Hao \& Long}]{Shen2011}
\bibinfo{author}{Shen, Y.}, \bibinfo{author}{Hao, L.}, \&
  \bibinfo{author}{Long, G.-L.} (\bibinfo{year}{2011}).
\newblock \bibinfo{title}{Why can we copy classical information?}
\newblock {\it \bibinfo{journal}{Chinese Physics Letters}\/},  {\it
  \bibinfo{volume}{28}\/}, \bibinfo{pages}{010306}.
  \DOIprefix\doi{10.1088/0256-307X/28/1/010306}.
\bibitem[{Sholl \& Steckel(2009)}]{Sholl2009}
\bibinfo{author}{Sholl, D.~S.}, \& \bibinfo{author}{Steckel, J.~A.}
  (\bibinfo{year}{2009}).
\newblock {\it \bibinfo{title}{Density Functional Theory: A Practical
  Introduction}\/}.
\newblock \bibinfo{address}{Hoboken, New Jersey}: \bibinfo{publisher}{John
  Wiley \& Sons}.
\bibitem[{Singer(2007)}]{Singer2007}
\bibinfo{author}{Singer, J.~H.} (\bibinfo{year}{2007}).
\newblock \bibinfo{title}{Multivesicular release and saturation of
  glutamatergic signalling at retinal ribbon synapses}.
\newblock {\it \bibinfo{journal}{Journal of Physiology}\/},  {\it
  \bibinfo{volume}{580}\/}, \bibinfo{pages}{23--29}.
  \DOIprefix\doi{10.1113/jphysiol.2006.125302}.
\bibitem[{Sousa et~al.(2016)Sousa, Ribeiro, Neves, Br\'{a}s, Cerqueira,
  Fernandes \& Ramos}]{Sousa2016}
\bibinfo{author}{Sousa, S.~F.}, \bibinfo{author}{Ribeiro, A. J.~M.},
  \bibinfo{author}{Neves, R. P.~P.}, \bibinfo{author}{Br\'{a}s, N.~F.},
  \bibinfo{author}{Cerqueira, N. M. F. S.~A.}, \bibinfo{author}{Fernandes,
  P.~A.}, \& \bibinfo{author}{Ramos, M.~J.} (\bibinfo{year}{2016}).
\newblock \bibinfo{title}{Application of quantum mechanics/molecular mechanics
  methods in the study of enzymatic reaction mechanisms}.
\newblock {\it \bibinfo{journal}{Wiley Interdisciplinary Reviews: Computational
  Molecular Science}\/},  {\it \bibinfo{volume}{7}\/}, \bibinfo{pages}{e1281}.
  \DOIprefix\doi{10.1002/wcms.1281}.
\bibitem[{Sperry(1982)}]{Sperry1982}
\bibinfo{author}{Sperry, R.} (\bibinfo{year}{1982}).
\newblock \bibinfo{title}{Some effects of disconnecting the cerebral
  hemispheres}.
\newblock {\it \bibinfo{journal}{Science}\/},  {\it \bibinfo{volume}{217}\/},
  \bibinfo{pages}{1223--1226}. \DOIprefix\doi{10.1126/science.7112125}.
\bibitem[{Sperry(1966)}]{Sperry1966}
\bibinfo{author}{Sperry, R.~W.} (\bibinfo{year}{1966}).
\newblock \bibinfo{title}{Brain bisection and mechanisms of consciousness}.
\newblock In \bibinfo{editor}{J.~C. Eccles} (Ed.), {\it
  \bibinfo{booktitle}{Brain and Conscious Experience: Study Week September 28
  to October 4, 1964, of the Pontificia Academia Scientiarum}\/}
  chapter~\bibinfo{chapter}{13}. (pp. \bibinfo{pages}{298--313}).
\newblock \bibinfo{address}{New York}: \bibinfo{publisher}{Springer}.
\bibitem[{Sprigge(1994)}]{Sprigge1994}
\bibinfo{author}{Sprigge, T. L.~S.} (\bibinfo{year}{1994}).
\newblock \bibinfo{title}{Consciousness}.
\newblock {\it \bibinfo{journal}{Synthese}\/},  {\it \bibinfo{volume}{98}\/},
  \bibinfo{pages}{73--93}. \DOIprefix\doi{10.1007/bf01064026}.
\bibitem[{Starkhammar et~al.(2011)Starkhammar, Moore, Talmadge \&
  Houser}]{Starkhammar2011}
\bibinfo{author}{Starkhammar, J.}, \bibinfo{author}{Moore, P.~W.},
  \bibinfo{author}{Talmadge, L.}, \& \bibinfo{author}{Houser, D.~S.}
  (\bibinfo{year}{2011}).
\newblock \bibinfo{title}{Frequency-dependent variation in the two-dimensional
  beam pattern of an echolocating dolphin}.
\newblock {\it \bibinfo{journal}{Biology Letters}\/},  {\it
  \bibinfo{volume}{7}\/}, \bibinfo{pages}{836--839}.
  \DOIprefix\doi{10.1098/rsbl.2011.0396}.
\bibitem[{Stelzer et~al.(2008)Stelzer, Poschner, Stalz, Heck \&
  Langosch}]{Stelzer2008}
\bibinfo{author}{Stelzer, W.}, \bibinfo{author}{Poschner, B.~C.},
  \bibinfo{author}{Stalz, H.}, \bibinfo{author}{Heck, A.~J.}, \&
  \bibinfo{author}{Langosch, D.} (\bibinfo{year}{2008}).
\newblock \bibinfo{title}{Sequence-specific conformational flexibility of
  {SNARE} transmembrane helices probed by hydrogen/deuterium exchange}.
\newblock {\it \bibinfo{journal}{Biophysical Journal}\/},  {\it
  \bibinfo{volume}{95}\/}, \bibinfo{pages}{1326--1335}.
  \DOIprefix\doi{10.1529/biophysj.108.132928}.
\bibitem[{Stevens \& Wang(1995)}]{Stevens1995}
\bibinfo{author}{Stevens, C.~F.}, \& \bibinfo{author}{Wang, Y.}
  (\bibinfo{year}{1995}).
\newblock \bibinfo{title}{Facilitation and depression at single central
  synapses}.
\newblock {\it \bibinfo{journal}{Neuron}\/},  {\it \bibinfo{volume}{14}\/},
  \bibinfo{pages}{795--802}. \DOIprefix\doi{10.1016/0896-6273(95)90223-6}.
\bibitem[{St\"{o}hr \& Tkatchenko(2019)}]{Stohr2019}
\bibinfo{author}{St\"{o}hr, M.}, \& \bibinfo{author}{Tkatchenko, A.}
  (\bibinfo{year}{2019}).
\newblock \bibinfo{title}{Quantum mechanics of proteins in explicit water: the
  role of plasmon-like solute-solvent interactions}.
\newblock {\it \bibinfo{journal}{Science Advances}\/},  {\it
  \bibinfo{volume}{5}\/}, \bibinfo{pages}{eaax0024}.
  \DOIprefix\doi{10.1126/sciadv.aax0024}.
\bibitem[{Strang(2016)}]{Strang2016}
\bibinfo{author}{Strang, W.~G.} (\bibinfo{year}{2016}).
\newblock {\it \bibinfo{title}{Introduction to Linear Algebra}\/}.
\newblock (\bibinfo{edition}{5th} ed.).
\newblock \bibinfo{address}{Wellesley, Massachusetts}:
  \bibinfo{publisher}{Wellesley-Cambridge Press}.
\bibitem[{Stringer \& Galway-Witham(2017)}]{Stringer2017}
\bibinfo{author}{Stringer, C.}, \& \bibinfo{author}{Galway-Witham, J.}
  (\bibinfo{year}{2017}).
\newblock \bibinfo{title}{On the origin of our species}.
\newblock {\it \bibinfo{journal}{Nature}\/},  {\it \bibinfo{volume}{546}\/},
  \bibinfo{pages}{212--214}. \DOIprefix\doi{10.1038/546212a}.
\bibitem[{Su \& Xu(2017)}]{Su2017}
\bibinfo{author}{Su, N.~Q.}, \& \bibinfo{author}{Xu, X.}
  (\bibinfo{year}{2017}).
\newblock \bibinfo{title}{Development of new density functional
  approximations}.
\newblock {\it \bibinfo{journal}{Annual Review of Physical Chemistry}\/},  {\it
  \bibinfo{volume}{68}\/}, \bibinfo{pages}{155--182}.
  \DOIprefix\doi{10.1146/annurev-physchem-052516-044835}.
\bibitem[{S\"{u}dhof(2012)}]{Sudhof2012}
\bibinfo{author}{S\"{u}dhof, T.~C.} (\bibinfo{year}{2012}).
\newblock \bibinfo{title}{The presynaptic active zone}.
\newblock {\it \bibinfo{journal}{Neuron}\/},  {\it \bibinfo{volume}{75}\/},
  \bibinfo{pages}{11--25}. \DOIprefix\doi{10.1016/j.neuron.2012.06.012}.
\bibitem[{S\"{u}dhof(2013)}]{Sudhof2013}
\bibinfo{author}{S\"{u}dhof, T.~C.} (\bibinfo{year}{2013}).
\newblock \bibinfo{title}{Neurotransmitter release: the last millisecond in the
  life of a synaptic vesicle}.
\newblock {\it \bibinfo{journal}{Neuron}\/},  {\it \bibinfo{volume}{80}\/},
  \bibinfo{pages}{675--690}. \DOIprefix\doi{10.1016/j.neuron.2013.10.022}.
\bibitem[{S\"{u}dhof \& Rothman(2009)}]{Sudhof2009}
\bibinfo{author}{S\"{u}dhof, T.~C.}, \& \bibinfo{author}{Rothman, J.~E.}
  (\bibinfo{year}{2009}).
\newblock \bibinfo{title}{Membrane fusion: grappling with {SNARE} and {SM}
  proteins}.
\newblock {\it \bibinfo{journal}{Science}\/},  {\it \bibinfo{volume}{323}\/},
  \bibinfo{pages}{474--477}. \DOIprefix\doi{10.1126/science.1161748}.
\bibitem[{Susskind \& Friedman(2014)}]{Susskind2014}
\bibinfo{author}{Susskind, L.}, \& \bibinfo{author}{Friedman, A.}
  (\bibinfo{year}{2014}).
\newblock {\it \bibinfo{title}{Quantum Mechanics: The Theoretical Minimum. What
  You Need to Know to Start Doing Physics}\/}.
\newblock \bibinfo{address}{New York}: \bibinfo{publisher}{Basic Books}.
\bibitem[{Susskind \& Hrabovsky(2013)}]{Susskind2013}
\bibinfo{author}{Susskind, L.}, \& \bibinfo{author}{Hrabovsky, G.}
  (\bibinfo{year}{2013}).
\newblock {\it \bibinfo{title}{The Theoretical Minimum: What You Need to Know
  to Start Doing Physics}\/}.
\newblock \bibinfo{address}{New York}: \bibinfo{publisher}{Basic Books}.
\bibitem[{van Swinderen et~al.(1999)van Swinderen, Saifee, Shebester, Roberson,
  Nonet \& Crowder}]{vanSwinderen1999}
\bibinfo{author}{van Swinderen, B.}, \bibinfo{author}{Saifee, O.},
  \bibinfo{author}{Shebester, L.}, \bibinfo{author}{Roberson, R.},
  \bibinfo{author}{Nonet, M.~L.}, \& \bibinfo{author}{Crowder, C.~M.}
  (\bibinfo{year}{1999}).
\newblock \bibinfo{title}{A neomorphic syntaxin mutation blocks
  volatile-anesthetic action in {C}aenorhabditis elegans}.
\newblock {\it \bibinfo{journal}{Proceedings of the National Academy of
  Sciences}\/},  {\it \bibinfo{volume}{96}\/}, \bibinfo{pages}{2479--2484}.
  \DOIprefix\doi{10.1073/pnas.96.5.2479}.
\bibitem[{Tang(2014)}]{Tang2014}
\bibinfo{author}{Tang, H.} (\bibinfo{year}{2014}).
\newblock \bibinfo{title}{``it is not a something, but not a nothing
  either!''---{M}c{D}owell on {W}ittgenstein}.
\newblock {\it \bibinfo{journal}{Synthese}\/},  {\it \bibinfo{volume}{191}\/},
  \bibinfo{pages}{557--567}. \DOIprefix\doi{10.1007/s11229-013-0291-3}.
\bibitem[{Tegmark(2000)}]{Tegmark2000}
\bibinfo{author}{Tegmark, M.} (\bibinfo{year}{2000}).
\newblock \bibinfo{title}{Importance of quantum decoherence in brain
  processes}.
\newblock {\it \bibinfo{journal}{Physical Review E}\/},  {\it
  \bibinfo{volume}{61}\/}, \bibinfo{pages}{4194--4206}.
  \DOIprefix\doi{10.1103/PhysRevE.61.4194}.
\bibitem[{Troup et~al.(2019)Troup, Zalucki, Kottler, Karunanithi, Anggono \&
  van Swinderen}]{Troup2019}
\bibinfo{author}{Troup, M.}, \bibinfo{author}{Zalucki, O.~H.},
  \bibinfo{author}{Kottler, B.~D.}, \bibinfo{author}{Karunanithi, S.},
  \bibinfo{author}{Anggono, V.}, \& \bibinfo{author}{van Swinderen, B.}
  (\bibinfo{year}{2019}).
\newblock \bibinfo{title}{Syntaxin1{A} neomorphic mutations promote rapid
  recovery from isoflurane anesthesia in {D}rosophila melanogaster}.
\newblock {\it \bibinfo{journal}{Anesthesiology}\/},  {\it
  \bibinfo{volume}{131}\/}, \bibinfo{pages}{555--568}.
  \DOIprefix\doi{10.1097/aln.0000000000002850}.
\bibitem[{Ullrich(2012)}]{Ullrich2012}
\bibinfo{author}{Ullrich, C.~A.} (\bibinfo{year}{2012}).
\newblock {\it \bibinfo{title}{Time-Dependent Density-Functional Theory:
  Concepts and Applications}\/}.
\newblock \bibinfo{address}{Oxford}: \bibinfo{publisher}{Oxford University
  Press}.
\newblock \DOIprefix\doi{10.1093/acprof:oso/9780199563029.001.0001}.
\bibitem[{Vaziri \& Plenio(2010)}]{Vaziri2010}
\bibinfo{author}{Vaziri, A.}, \& \bibinfo{author}{Plenio, M.~B.}
  (\bibinfo{year}{2010}).
\newblock \bibinfo{title}{Quantum coherence in ion channels: resonances,
  transport and verification}.
\newblock {\it \bibinfo{journal}{New Journal of Physics}\/},  {\it
  \bibinfo{volume}{12}\/}, \bibinfo{pages}{085001}.
  \DOIprefix\doi{10.1088/1367-2630/12/8/085001}.
\bibitem[{Verkhratsky \& Nedergaard(2017)}]{Verkhratsky2017}
\bibinfo{author}{Verkhratsky, A.}, \& \bibinfo{author}{Nedergaard, M.}
  (\bibinfo{year}{2017}).
\newblock \bibinfo{title}{Physiology of astroglia}.
\newblock {\it \bibinfo{journal}{Physiological Reviews}\/},  {\it
  \bibinfo{volume}{98}\/}, \bibinfo{pages}{239--389}.
  \DOIprefix\doi{10.1152/physrev.00042.2016}.
\bibitem[{Weber et~al.(1998)Weber, Zemelman, McNew, Westermann, Gmachl,
  Parlati, S\"{o}llner \& Rothman}]{Weber1998}
\bibinfo{author}{Weber, T.}, \bibinfo{author}{Zemelman, B.~V.},
  \bibinfo{author}{McNew, J.~A.}, \bibinfo{author}{Westermann, B.},
  \bibinfo{author}{Gmachl, M.}, \bibinfo{author}{Parlati, F.},
  \bibinfo{author}{S\"{o}llner, T.~H.}, \& \bibinfo{author}{Rothman, J.~E.}
  (\bibinfo{year}{1998}).
\newblock \bibinfo{title}{{SNARE}pins: minimal machinery for membrane fusion}.
\newblock {\it \bibinfo{journal}{Cell}\/},  {\it \bibinfo{volume}{92}\/},
  \bibinfo{pages}{759--772}. \DOIprefix\doi{10.1016/S0092-8674(00)81404-X}.
\bibitem[{Wenzel et~al.(2019)Wenzel, Han, Smith, Hoel, Greger, House \&
  Yuste}]{Wenzel2019}
\bibinfo{author}{Wenzel, M.}, \bibinfo{author}{Han, S.},
  \bibinfo{author}{Smith, E.~H.}, \bibinfo{author}{Hoel, E.},
  \bibinfo{author}{Greger, B.}, \bibinfo{author}{House, P.~A.}, \&
  \bibinfo{author}{Yuste, R.} (\bibinfo{year}{2019}).
\newblock \bibinfo{title}{Reduced repertoire of cortical microstates and
  neuronal ensembles in medically induced loss of consciousness}.
\newblock {\it \bibinfo{journal}{Cell Systems}\/},  {\it
  \bibinfo{volume}{8}\/}, \bibinfo{pages}{467--474}.
  \DOIprefix\doi{10.1016/j.cels.2019.03.007}.
\bibitem[{Werner et~al.(2011)Werner, Swihart, Rau, Jia, Borghese, McCracken,
  Iyer, Fanselow, Oh, Sonner, Eger, Harrison, Harris \& Homanics}]{Werner2011}
\bibinfo{author}{Werner, D.~F.}, \bibinfo{author}{Swihart, A.},
  \bibinfo{author}{Rau, V.}, \bibinfo{author}{Jia, F.},
  \bibinfo{author}{Borghese, C.~M.}, \bibinfo{author}{McCracken, M.~L.},
  \bibinfo{author}{Iyer, S.}, \bibinfo{author}{Fanselow, M.~S.},
  \bibinfo{author}{Oh, I.}, \bibinfo{author}{Sonner, J.~M.},
  \bibinfo{author}{Eger, E.~I.}, \bibinfo{author}{Harrison, N.~L.},
  \bibinfo{author}{Harris, R.~A.}, \& \bibinfo{author}{Homanics, G.~E.}
  (\bibinfo{year}{2011}).
\newblock \bibinfo{title}{Inhaled anesthetic responses of recombinant receptors
  and knockin mice harboring $\alpha$2({S270H/L277A}) {GABA}$_\textrm{A}$
  receptor subunits that are resistant to isoflurane}.
\newblock {\it \bibinfo{journal}{Journal of Pharmacology and Experimental
  Therapeutics}\/},  {\it \bibinfo{volume}{336}\/}, \bibinfo{pages}{134}.
  \DOIprefix\doi{10.1124/jpet.110.170431}.
\bibitem[{Wittgenstein(2009)}]{Wittgenstein2009}
\bibinfo{author}{Wittgenstein, L.} (\bibinfo{year}{2009}).
\newblock {\it \bibinfo{title}{Philosophical Investigations}\/}.
\newblock \bibinfo{address}{Chichester}: \bibinfo{publisher}{Wiley-Blackwell}.
\bibitem[{Wolman(2012)}]{Wolman2012}
\bibinfo{author}{Wolman, D.} (\bibinfo{year}{2012}).
\newblock \bibinfo{title}{The split brain: a tale of two halves}.
\newblock {\it \bibinfo{journal}{Nature}\/},  {\it \bibinfo{volume}{483}\/},
  \bibinfo{pages}{260--263}. \DOIprefix\doi{10.1038/483260a}.
\bibitem[{Wootters \& Zurek(1982)}]{Wootters1982}
\bibinfo{author}{Wootters, W.~K.}, \& \bibinfo{author}{Zurek, W.~H.}
  (\bibinfo{year}{1982}).
\newblock \bibinfo{title}{A single quantum cannot be cloned}.
\newblock {\it \bibinfo{journal}{Nature}\/},  {\it \bibinfo{volume}{299}\/},
  \bibinfo{pages}{802--803}. \DOIprefix\doi{10.1038/299802a0}.
\bibitem[{Wu et~al.(2004)Wu, Sun, Evers, Crowder \& Wu}]{Wu2004}
\bibinfo{author}{Wu, X.-S.}, \bibinfo{author}{Sun, J.-Y.},
  \bibinfo{author}{Evers, A.~S.}, \bibinfo{author}{Crowder, M.}, \&
  \bibinfo{author}{Wu, L.-G.} (\bibinfo{year}{2004}).
\newblock \bibinfo{title}{Isoflurane inhibits transmitter release and the
  presynaptic action potential}.
\newblock {\it \bibinfo{journal}{Anesthesiology}\/},  {\it
  \bibinfo{volume}{100}\/}, \bibinfo{pages}{663--670}.
  \DOIprefix\doi{10.1097/00000542-200403000-00029}.
\bibitem[{Yablo(1992)}]{Yablo1992}
\bibinfo{author}{Yablo, S.} (\bibinfo{year}{1992}).
\newblock \bibinfo{title}{Mental causation}.
\newblock {\it \bibinfo{journal}{The Philosophical Review}\/},  {\it
  \bibinfo{volume}{101}\/}, \bibinfo{pages}{245--280}.
  \DOIprefix\doi{10.2307/2185535}.
\bibitem[{Yoshor et~al.(2007)Yoshor, Bosking, Ghose \& Maunsell}]{Yoshor2007}
\bibinfo{author}{Yoshor, D.}, \bibinfo{author}{Bosking, W.~H.},
  \bibinfo{author}{Ghose, G.~M.}, \& \bibinfo{author}{Maunsell, J. H.~R.}
  (\bibinfo{year}{2007}).
\newblock \bibinfo{title}{Receptive fields in human visual cortex mapped with
  surface electrodes}.
\newblock {\it \bibinfo{journal}{Cerebral Cortex}\/},  {\it
  \bibinfo{volume}{17}\/}, \bibinfo{pages}{2293--2302}.
  \DOIprefix\doi{10.1093/cercor/bhl138}.
\bibitem[{Zhao(2012)}]{Zhao2012}
\bibinfo{author}{Zhao, Y.} (\bibinfo{year}{2012}).
\newblock \bibinfo{title}{The knowledge argument against physicalism: its
  proponents and its opponents}.
\newblock {\it \bibinfo{journal}{Frontiers of Philosophy in China}\/},  {\it
  \bibinfo{volume}{7}\/}, \bibinfo{pages}{304--316}.
  \DOIprefix\doi{10.3868/s030-001-012-0018-4}.
\bibitem[{Zhou et~al.(2013)Zhou, Bacaj, Yang, Pang \& S\"{u}dhof}]{Zhou2013}
\bibinfo{author}{Zhou, P.}, \bibinfo{author}{Bacaj, T.}, \bibinfo{author}{Yang,
  X.}, \bibinfo{author}{Pang, Z.~P.}, \& \bibinfo{author}{S\"{u}dhof, T.~C.}
  (\bibinfo{year}{2013}).
\newblock \bibinfo{title}{Lipid-anchored {SNARE}s lacking transmembrane regions
  fully support membrane fusion during neurotransmitter release}.
\newblock {\it \bibinfo{journal}{Neuron}\/},  {\it \bibinfo{volume}{80}\/},
  \bibinfo{pages}{470--483}. \DOIprefix\doi{10.1016/j.neuron.2013.09.010}.
\bibitem[{Zhou et~al.(2015)Zhou, Lai, Bacaj, Zhao, Lyubimov, Uervirojnangkoorn,
  Zeldin, Brewster, Sauter, Cohen, Soltis, Alonso-Mori, Chollet, Lemke,
  Pfuetzner, Choi, Weis, Diao, S\"{u}dhof \& Brunger}]{Zhou2015}
\bibinfo{author}{Zhou, Q.}, \bibinfo{author}{Lai, Y.}, \bibinfo{author}{Bacaj,
  T.}, \bibinfo{author}{Zhao, M.}, \bibinfo{author}{Lyubimov, A.~Y.},
  \bibinfo{author}{Uervirojnangkoorn, M.}, \bibinfo{author}{Zeldin, O.~B.},
  \bibinfo{author}{Brewster, A.~S.}, \bibinfo{author}{Sauter, N.~K.},
  \bibinfo{author}{Cohen, A.~E.}, \bibinfo{author}{Soltis, S.~M.},
  \bibinfo{author}{Alonso-Mori, R.}, \bibinfo{author}{Chollet, M.},
  \bibinfo{author}{Lemke, H.~T.}, \bibinfo{author}{Pfuetzner, R.~A.},
  \bibinfo{author}{Choi, U.~B.}, \bibinfo{author}{Weis, W.~I.},
  \bibinfo{author}{Diao, J.}, \bibinfo{author}{S\"{u}dhof, T.~C.}, \&
  \bibinfo{author}{Brunger, A.~T.} (\bibinfo{year}{2015}).
\newblock \bibinfo{title}{Architecture of the synaptotagmin-{SNARE} machinery
  for neuronal exocytosis}.
\newblock {\it \bibinfo{journal}{Nature}\/},  {\it \bibinfo{volume}{525}\/},
  \bibinfo{pages}{62--67}. \DOIprefix\doi{10.1038/nature14975}.

\end{thebibliography}
\end{document}